\documentclass[%
 reprint,
 amsmath,amssymb,
 aps,
]{revtex4-2}

\usepackage{graphicx}
\graphicspath{ {images/} }
\usepackage{dcolumn}
\usepackage{bm}
\usepackage[retainorgcmds]{IEEEtrantools}

\begin{document}

\preprint{APS/123-QED}

\title{Enhanced spectrum of primordial perturbations, galaxy formation 
and small scale structure}

\author{Karim H. Seleim}
 \altaffiliation[Also at ]{Physics Department, Faculty of Science, 
 Alexandria University, Alexandria, Egypt}
\author{Amr A. El-Zant}%
 \email{amr.elzant@bue.edu.eg}

\affiliation{
 Centre for Theoretical Physics, The British University in Egypt,  Sherouk City 11837, Cairo, Egypt
}%
\author{A. M. Abdel-Moneim.}
\affiliation{Physics Department, Faculty of Science, Alexandria University, Alexandria, Egypt}%

\date{\today}

\begin{abstract}
The standard cosmological structure formation scenario is successful on large scales. 
Several apparent problems affect it however at galactic scales, such as the
small scale problems at low redshift and more recent issues involving  early 
massive galaxy and black hole formation. 
As these arise where complex baryonic physics becomes important, 
this is often assumed to be behind the problems. But the same scales are also those 
where the primordial spectrum is relatively unconstrained, and there
are several ways in which it can be modified. 
We focus on that arising from effects possibly 
associated with the crossing of high energy cutoff scale by fluctuation modes 
during inflation.  Elementary arguments 
show that adiabatic evolution cannot modify the near scale invariance,  we thus 
discuss a simple model for the contrary 
extreme of sudden transition. Numerical calculations and simple arguments 
suggest that its predictions, for parameters con2sidered here, are more generic than may be expected, with significant modifications requiring a rapid transition. 
We examine the implications of such a scenario,
in this simplest form of sudden jump as well as gradual variants, on the matter power 
spectrum and halo mass function in light of the limitations imposed by particle production. 
We show the resulting enhancement and oscillation in the power spectrum on currently 
nonlinear scales can potentially simultaneously 
alleviate both the apparent problem of early structure formation 
and,  somewhat counterintuitively, problems at low 
redshift concerning the abundance of dwarf galaxies, including those too big to fail. 
We discuss consequences that can observationally constrain 
the scenario and its parameters,  
including an inflationary Hubble scale $\lesssim 10^{-8} M_{\rm Pl}$, 
while touching on the possibility of simultaneous 
modification of power on the largest scales. 
\end{abstract}

\maketitle


\section{Introduction}

Structure  can condense from  small density perturbations in a nearly homogeneous universe 
through gravitational instability.  In the  context of  contemporary cosmology the 
density perturbations  are  seeded by quantum fluctuations in 
a primordial scalar field driving inflation, which later decays as the 
universe reheats and the standard model particles (and putative dark ones)
arise (e.g., \cite{LiddleLythB, UzanB}).  
The statistical properties  of the primordial perturbations thus leave their  mark
on the  cosmic microwave background (CMB) and large scale structure of the galaxy distribution. 
On these scales, on which they can be inferred with precision, the properties  of 
the perturbations 
are consistent with a nearly scale invariant primordial spectrum essentially 
determining their statistics~\cite{Planck18}.

At a phenomenological level, the simplest models of inflation do predict 
a near scale invariant primordial power spectrum 
if the inflaton potential is specified in such a way that the resulting Hubble parameter 
is nearly constant over a sufficient 
number of e-folds. Nevertheless, this prediction is not unique~\cite{SaloBond89, Dod_Slow_Tilt}; indeed, 
little is known of the microscopic physics of inflation, or the wider particle physics model 
it may be a part of, and the coupling of the inflaton driving inflation to other fields may lead to 
changes in the potential that can ruin the predictions of standard slow 
roll models~(.e.g., \cite{BrokSuperg, KolbResPart00, McGuill_HEC02, Sasaki08} 
and~\cite{PartPhysModInf, SupergRev, InflStr} for reviews).   
Stages of singular or rapid evolution of the potential 
or its derivatives, interrupting slow roll,  leave imprints on the primordial spectrum 
of fluctuations~\cite{Starob92, StarobPol, StaroStepnum, KolbResPart00, 
LesgourguesBSI00,  YamaguchiSuper01, CouplingFieldsBSI03, Hannestad_Res03,
Gong05, DvorHu09, Multi_Sasaki12}. 
Such effects may lead to a variety of 'features' and changes 
that break the scale free spectrum, and may have  observable 
consequences associated with significant enhancement or 
suppression of power on large scales~\cite{StaroCluster98, CDMwStep, PriFeatSup, KofLindSuppress03, 
BrokenScaleCMBLSS05, WMAPPOW06, StaroCMB09}, as well as on smaller 
(galactic and sub-galactic) scales~\cite{KamionLidd, DwarfBSI00, ZentBull02,  WDMBSC03, SSF2019, Superdens10, Baughtherm18},
including the formation of primordial black holes~\cite{Yamaguchi01,Bringmann01}. 
However the anomalous variation in the inflaton potential needs to be localized in order for 
single field slow roll inflation 
to proceed for a sufficient number of e-folds, and is also limited on large 
scales by observations suggesting a nearly scale invariant primordial power spectrum~\cite{PeirisRun06, Planck18}.

In some models, such as DBI inflation, the breaking of scale invariance and associated features in the power spectrum 
are best represented in terms of sharp temporal variations in the sound speed of  
perturbations~\cite{BeanDBI, JapDBI, HuDBI1, HuDBI2, Sud_1, PhenoSound_1, PhenoSound_2}.  
This is still accompanied by corresponding changes in the equation of state of the inflaton and therefore, as in the cases above, by anomalous background evolution. 
On the other hand, 
if inflation proceeds modestly longer than the minimal number of e-folds 
needed to solve problems such as the apparent causal connection in the CMB field  ('horizon problem'),
primordial perturbations are expected to arise  from well within  a high energy cutoff 
scale where new physics may transpire. 
This gives rise to what is often termed as the  trans-Planckian 
problem. The minimal  length scale that it entails 
can be introduced by modifying the commutation relations or 
introducing nonlinear dispersion relations for the propagation of 
fluctuations (early studies include~\cite{BDM00, BDM01, Easther_GUP_01, Niemeyer_Com_Fluid01, BDB_M03,
Easther_BEFT05, Kempf_GUP06}, while a relatively recent review can be found in~\cite{BDMR13}). 

If the Hubble scale of inflation $H$ 
is  smaller than the scale $k_c$, 
where new physics may appear, 
the background evolution is unmodified 
by possible new physics;
modifications to the scale invariant spectrum may then arise 
solely from  anomalous evolution of the fluctuation modes,
rather than of the inflation potential. 
However, if some form of decoupling is assumed --- 
e.g. motivated by the fact that the power spectrum is evaluated 
at horizon exit scales $H \ll k_c$ ---
then the corrections can be quite small; 
of order $(H/k_c)^q$, with $q \gtrsim 1$. 
In the context of local effective field theory $q = 2$~\cite{Kaloper02}; 
if the fluctuation modes are assumed to simply emerge from the quantum foam  
at conformal time $\eta_c (k)$,  depending on the comoving mode wavenumber $k$ 
--- instead of the usual Bunch Davies initial conditions taken at $\eta \rightarrow - \infty$ --- 
then $q=1$~\cite{Danielsson02, Easther02}; if the modes emerge from the local adiabatic vacuum 
then $q = 3$~\cite{Niemeyer02}.   A covariant small scale cutoff, introduced by 
removing field configurations that are off shell by more than a Planck scale leads to 
$q = 1$~\cite{Chatwin-Davies16}.

Much larger corrections are nevertheless possible in principle.  
This is the case,  in particular, if the emergence of the inflating modes 
from the high energy cutoff scale is assumed to be preceded by non-adiabatic 
evolution arising from a nonlinear dispersion relation at 
scales $> k_c$~\cite{BDM00, BDM01, BDM02, Zhu_Disp_Mod16, Ashoor_Disp_Num17}. Modes then do not emerge from the foam in their lowest energy states. 
Excited states arise, providing anomalous initial conditions 
for further evolution and leading to enhancement and oscillations 
in the power spectrum.
The most general parametrization of the effects 
of a non-adiabatic high energy scale exit would therefore appear to 
include both phenomena~\cite{BDB_M03, ArmendarizPicon03}.

Though physics at these scales is 
largely unknown, it can in principle be envisioned that the 
introduction of a cutoff scale in itself can  
modify the effective dynamics of the fluctuations. Indeed, 
the introduction of a minimal length scale and a large variety of phenomenological  descriptions of 'quantum spacetimes',  can be characterized by nonlinear dispersion relations~\cite{Camelia08, Hossenfelder13}. 
At the most intuitive level, 
a simple hydrodynamic analogy suggests such a modification to the  
propagation of fluctuations~\cite{Unruh81, Unruh94, Analog_Rev}. 
This, much in the same way that an 
effective macroscopic 
description of wave propagation through a  fluid or  lattice  may still
be employed at wavelengths approaching the interparticle distances, 
provided this is 
phenomenologicaly taken into account through a modified dispersion relation.  
As,  at scales smaller than the interparticle distances waves cannot 
propagate at all, it is in the transition between such a cutoff scale  
and the scale on which the standard effective 
macroscopic description applies
that a nonlinear dispersion relation may describe the propagation of fluctuations. 
At this simplest intuitive level, 
one may expect the dynamics of a sound wave, initially moving in a medium where
interparticle spacing is large and comparable with its wavelength, 
to keep memory of the anomalous evolution,  even after it crosses 
and propagates into a medium with smaller interparticle spacing,  
where the effective theory is perfectly valid and 'decoupling' is guaranteed. 
The rough analogy here would be with an inflaton fluctuation 
mode inflating from wavenumbers above the 
high energy cutoff scale to ones below it, on its way to the horizon.  
Applications of more sophisticated 'analogue' models 
to the inflationary scenario show that modifications of the dispersion 
relation can indeed lead to significant changes of power spectrum of 
field correlations~\cite{Cha16}.

An important limitation on modifications of the power spectrum 
through the inclusion of 
excited states relates to the fact that these are
necessarily associated with departures from a vacuum state. 
And too much excitation can lead to
departures significant enough to prohibit inflation 
from starting and persisting in the first place~\cite{Tanaka00, Niemeyer_nomod_disp01, Starobinsky_PP01, Giovannini03, Porrati04}.  
However, as has been pointed out, the limitations that arise thus may not be very constraining~\cite{Greene04, BDM04, Daniel04}.    
In the present investigation we wish to examine whether, 
within the limits imposed by particle production, 
excitations of inflaton modes, 
stemming from the presence of high energy cutoff scale, 
can lead to significant and astrophysicaly interesting 
modifications of the primordial power spectrum.

At present, the non-observation of  significant 
departures from  scale invariance
on large (linear)  scales, where the primordial spectrum 
can be rather precisely inferred, seem to embody the main 
evidence against such 
modifications.  Indeed, observations on scales on which 
the density perturbation is linear preclude even relatively 
small modifications~\cite{Goldstein02, Easther_Peir_04, Okamoto04, Martin_Ring04, Martin_Ring04A}. 
Observations are however much less constraining  at smaller scales,  where they are
limited by the Silk damping of the CMB, 
and by nonlinear structure formation erasing the possibility 
of directly mapping the observed power spectrum to the primordial one. 
Precise CMB and large scale structure inference is therefore limited to scales 
$\gtrsim 10~{\rm Mpc}$. For a horizon scale of $\sim 10~{\rm Gpc}$ this spans three orders of magnitudes. 
More model dependent constraints are available from the 
Lyman-$\alpha$ forest down to wavenumbers roughly corresponding 
to  comoving spatial scales  of order of Mpc.  
Beyond that, the spectrum is currently quite weakly constrained~\cite{Chluba_D12, Chluba_S12, EmamiSmoot17}. 
On the other hand, the smallest structures 
that form in the context of the standard cold dark matter scenario have 
earth mass and roughly solar system size $\sim 10^{-4} {\rm pc}$. 
From such scales to the smallest scales at which the linear power spectrum can be directly recovered
one counts 11 orders of magnitude ---  nine  more than those separating the nonlinear scale to the horizon. 

It is not inconceivable that the scale invariance of the primordial power spectrum does not hold 
in some parts of the aforementioned range. 
On the contrary, despite the significant successes of the current model of structure 
formation on large scales~\cite{White_F12},  
through the past couple of decades a variety of problems have arisen on galactic scales. 
There is a group of quite possibly related longstanding issues connected to the central 
densities of dark matter halos,  and 
the abundance and dynamical properties of local dwarf 
galaxies~\cite{delPopolo17, Bullock_B17}; and, in apparent contradiction, more recent issuees related to 
an apparent preponderance of massive old galaxies  and supermassive black holes at redshifts $3  \lesssim z \lesssim 9$  that may pose a challenge the current $\Lambda$CDM-based structure formation paradigm~\cite{DeRosa_SMBH14,Wu_SMBH_Natur15, Dusty_Gal_15, Hirano_Spergel2015,
Imp_earl16,  First_StarsBH_Volont16, Bouwens_halostellar17, Massive_Q_Glaze17,  Dust_quas17, Silk_LCDM_SMBH18, 
SMBH19, Queis_Obs_mod19, Bromm_SMBH_Rev19, Dominant_Yoshi2019}.

Such problems appear in the highly nonlinear regime of structure formation;  
where small density perturbations, born of primordial ones in the presumed inflaton,  
have sufficiently grown under gravity to form gravitationaly bound objects. 
Since it is also at such scales that 
complex baryonic physics becomes important,  
it was natural to suppose that the main determinant lies in complex
baryonic physics  of galaxy formation and evolution.
For example, for the small scale problems at $z= 0$, 
processes involving energy input to dark matter halos    
through dynamical friction with baryons~\cite{Zant2001, Zant2004, 
RomanoDiaz2008, Goerdt2010, Cole2011, Inoue2011,  
Laporte2015, Nipoti2015, delPopolo2018} or through random potential fluctuations
driven by starbursts or active galactic nuclei (AGN)~\cite{Read2005, Mashchenko2008, Peirani2008, Pontzen2012, Teyssier2013, Pontzen2014, EZFC}, were invoked. 
(In addition to suggestions modifying the dark matter particle physics models, as in warm dark matter~\cite{Colin2000, Bode2001,  Maccio2012b, Lovell2014, El-Zant2015}, self interacting dark matter~\cite{Spergel2000,Burkert2000,Kochanek2000,Miralda2002,
Zavala2013, Elbert2015} and fuzzy dark matter~\cite{Goodman2000, Hu2000, Schive2014, Marsh2014, Hui_etal2017}). Similar attempts are ongoing in the case of the 
the more recent early structure formation issue (some are discussed in Section~\ref{sec:earlygals}). 

However, it is also precisely at the nonlinear scales, 
where baryonic physics becomes important, 
that the primordial power spectrum is 
relatively unconstrained. That modifications thereof can be relevant 
to small scale problems associated with galaxy formation 
has long been realized~\cite{KamionLidd}, but not as 
extensively investigated as the baryonic solutions discussed above.
While feedback from starbursts and AGN is 
now  recognized as a central ingredient of 
galaxy formation, independently of its possible role in alleviating the  
aforementioned issues, and whereas massive baryonic clumps,  
proposed to mediate dynamical friction coupling to dark matter, have since been observed in forming galaxies 
(e.g.~\cite{Genzzel_Clumps11, Caval_Clumps18}), it is also important to further examine mechanisms
that address galactic scale problems through 
modification of the primordial spectrum. 
Deriving the consequences of such modifications is 
in itself an interesting tool for understanding the processes 
from which they may arise in an inflationary era.

In this study we investigate the effect on the power spectrum 
from field excitations, stemming from  
non-adiabatic transition through a high energy cutoff regime 
corresponding to currently nonlinear scales, and within 
the limits imposed by particle production. 
We attempt to do this in generic terms, starting 
from well defined initial conditions, with linear dispersion 
relation  (but with sound speed different from unity) and examining the 
effect of the transition.  As this solely affects the fluctuation
modes, the equation of state of the inflaton, and thus the 
background evolution, remains unmodified (unlike  in cases such as DBI 
inflation mentioned above), this helps isolate the 
effect of excitations on the spectrum. 
We then look for associated effects on 
the matter power spectrum and dark matter 
halo mass function. 

In the next section, after illustrating in simplest terms 
how the power spectrum is essentially an adiabatic invariamt of the 
dynamics of inflaton fluctuations, we present 
and discuss a simple model representing the
other extreme of a sudden transition
(in an appendix, we show results that suggest it is generic for a large range of 
parameters; in a second appendix we discuss the situation when the assumption is relaxed).
In Section~\ref{sec:nonadiab} we discuss what this model entails
in more formal terms, evaluating the limits on power spectrum modification 
in terms of particle production.
In Section~\ref{sec:matt_halomass}, we study, within these limits, the possible modifications 
on the matter power spectrum and halo mass function. 
We discuss possible astrophysical consequences and constraints, before presenting 
our conclusions

\section{Adiabaticity, scale invariance and the sudden extreme}
\label{sec:adiab}

\subsection{The evolution of fluctuations}

The general quadratic action for inflationary perturbations with sound speed $c_{s}$
can be expressed in terms of the Mukhanov-Sasaki  (MS) variable $v$ 
as~\cite{mukhanov1992theory,armendariz1999k,garriga1999perturbations}:
\begin{equation}
S^{(2)}=\frac{1}{2} \int \mathrm{d}^{4} x\left(v^{\prime 2}-c_{\mathrm{s}}^{2}(\nabla v)^{2}+\frac{z^{\prime \prime}}{z} v^{2}\right),
\label{eq:MSA}
\end{equation}
where $z =a \frac{\phi_{0}^{\prime}}{\mathcal{H} c_{\mathrm{s}}}$, $\phi_{0}$ is the background inflaton field, $\mathcal{H} =a^{\prime} / a$, and the primes denote derivative with respect to conformal time $\eta$. 
The evolution of each Fourier mode $v_{k}(\eta)$ is governed by 
the MS equation
\begin{equation}
v_{k}^{\prime \prime}+\left(c_{s}^{2} k^{2}-\frac{ z^{\prime \prime}}{z} \right) v_{k}=0.
\label{eq:MS}
\end{equation}

The MS variable is related to the curvature perturbations by $v=z \mathcal{R}$. 
This is a quantity of fundamental interest, as it relates 
primordial quantum fluctuations to the observables,  such as CMB anisotropies; the power spectrum of the large scale  galaxy distribution; and, ultimately 
(more indirectly), the formation of smaller scale structures, such as the dark matter halos hosting galaxies.  
The dimensionless power spectrum of such perturbations is given by
 \begin{equation}
\Delta_{\mathcal{R}}^{2}(k) \equiv \frac{k^{3}}{2 \pi^{2}} \left|\mathcal{R}_{k}\right|^{2},
\label{eq:PPSC}
\end{equation}
where the right-hand side is evaluated at the horizon ($c_{s} k=a H$); as in the absence of isocurvature  perturbations, the comoving curvature perturbations $\mathcal{R}$ are 
conserved on super-horizon scales \cite{salopek1990nonlinear}. Scale-invariant perturbations correspond to $\Delta_{\mathcal{R}}^{2}(k)= {\rm const}$. 
Departures from this can arise if $c_{s}$, or the inflationary Hubble scale $H$, 
depend on time.

In the standard inflationary scenario, 
a massless field,  and quasi de Sitter evolution is assumed (and so $H$ is nearly 
constant throughout the inflationary stage). 
The associated slow roll parameters, defined as 
\begin{equation}
\epsilon \equiv-\frac{\dot{H}}{H^{2}} \quad  \tilde{\eta} \equiv \frac{\dot{\epsilon}}{H \epsilon} \quad \text { and } \quad \kappa \equiv \frac{\dot{c}_{s}}{H c_{s}},
\end{equation}
are always much smaller than unity. Any departure from scale invariance is small, and is 
usually 
quantified by the spectral tilt parameter $n_{s}$:
\begin{equation}
n_{s}-1 \equiv \frac{d \ln \Delta_{\mathcal{R}}^{2}}{d \ln k}=-2 \epsilon-\tilde{\eta}-\kappa.
\end{equation} 
The slow roll parameters being small implies that  $n_{s} \approx 1$.

To first order in the slow roll parameters, and assuming canonical kinetic terms, 
one can also show that 
\begin{equation}
\frac{z^{\prime \prime}}{z}=\frac{1}{\eta^{2}}\left(2+3 \epsilon+\frac{3}{2} \tilde{\eta}\right). 
\end{equation}
In this case, to a good approximation, one can rewrite~(\ref{eq:MS}) as 
\begin{equation}
v_{k}^{\prime \prime}+\left(c_s^2 k^{2}-\frac{ 2}{\eta^2} \right) v_{k}=0.
\label{eq:MSf}
\end{equation}
If no non-standard dispersion relation is invoked then $c_s = 1$, and the 
standard scenario may be fully recovered.

\subsection{Adiabaticity, adiabatic invariants and primordial power spectrum}

\subsubsection{General context}

In the context of equation~(\ref{eq:MS}), setting $c_s = 1$ can be interpreted as the result of 
assuming  a massless field with
linear dispersion relation between the physical frequency $\omega_{\rm phys}$ and the physical wavenumber  $k_{\rm phys}=\frac{k}{a}$, $\omega_{\text {phys }}=k_{\text {phys }}$. 
However, as discussed in the introduction,
this is not a necessity; a  modification of the equation of state of the inflaton (e.g.  such as in DBI inflation), or modification of the dispersion relation due to modes probing a high 
energy cutoff scale, beyond which new physics may arise, can change the situation. 

In the latter case, beyond a cutoff scale  $k_{c}$, one can introduce the relevant modification 
by replacing the square of the comoving wavenumber $k^{2}$ in~(\ref{eq:MS}) with 
\begin{equation}
k^{2} \rightarrow k_{\mathrm{eff}}^{2}(k, \eta) \equiv a^{2}(\eta) \omega_{\mathrm{phys}}^{2}\left[\frac{k}{a(\eta)}\right],
\end{equation}
the main requirement being that the new dispersion 
relation recovers the linear one for scales $k \ll k_{\mathrm{c}}$ \cite{martin2001trans}.This dispersion relation is thus necessarily time dependent, as it must 
transit between two regimes.  
It  can be used to parametrize and reflect the 
effect of a varying sound speed in equations~(\ref{eq:MS}) and~(\ref{eq:MSf}), 
the latter applying 
when the background dynamics is well approximated by standard slow roll.  
Indeed, in this context, Eq.~(\ref{eq:MS}) can be rewritten as 
\begin{equation}
v_{k}^{\prime \prime}+\left[k_{\mathrm{eff}}^{2}(k, \eta)-\frac{z^{\prime \prime}}{z}\right] v_{k}=0.
\label{eq:keff}
\end{equation} 
(A more rigorous derivation, based on a variational principle, can be found here\cite{lemoine2001stress}).

How does the extra time dependence, that thus arises, 
affect the power spectrum derived from the above equation? 
As noticed in several studies, 
mere time dependence in itself is not sufficient to alter the nearly scale invariant 
nature of the primordial spectrum of fluctuations. 
The adiabaticity condition ---  that is, $|\frac{d \omega}{d \eta}| / \omega^{2} < 1$ --- 
must be violated. 
A well known example where this condition is indeed violated invokes
the Corley-Jacobson dispersion relation
\begin{equation}
k_{\mathrm{eff}}^{2}(k, \eta)=k^{2}-k^{2}\left|b_{m}\right|\left[\frac{k}{k_{c} a(\eta)}\right]^{2 m}
\end{equation}
(where $b_m$ is a constant and $m$ an integer).
The studies~\cite{martin2001trans,martin2002corley,BDM01}, 
indeed indicated that 
a modification of the power spectrum, in the form of  a change in the spectral index and superimposed oscillations, was possible. However, several  criticisms were raised, 
including the possibility of complex frequencies arising at early times, rendering the 
quantum field theory ill-defined, and  problems related  to setting the initial conditions in non-adiabatic regime. To circumvent such issues, a new dispersion relation \cite{lemoine2001stress} was proposed, which exhibits linear behaviour in the small and large wavenumbers, but  has intermediate concave region where the adiabaticity is violated locally.

Here we will be considering a simpler scenario, 
which assumes standard Bunch Davies type initial 
conditions, with modified sound speed but still linear dispersion relation. 
The effective sound speed transits to the standard 
relation $\omega = k$ as the boundary around $k_c$ is crossed. 
In this simple controlled context, we wish to estimate the  
rapidity and steepness of the transition 
required in order to produce palpable change 
in the power spectrum. It turns out that such a 
transition must be quite rapid.     

\subsubsection{The power spectrum as an adiabatic invariant}
\label{sec:psadiab}

We now wish to show, in explicit simple terms, that the primordial 
power spectrum is in fact an adiabatic invariant of the evolution 
of inflationary perturbation, and thus cannot be snigficiantly modified by any 
changes in the dispersion relation that keeps the dynamics of 
the perturbations sufficiently adiabatic. 

We will be interested in the case when the dispersion relation is modified 
due to the fluctuation modes probing scales beyond a high energy cutoff, before 
they inflate into lower energy scales on their way to horizon exit. Only the 
effective speed of mode propagation is modified and 
slow roll inflation of a massless field is assumed 
to hold in all stages. So, 
Eq.~(\ref{eq:MSf}) holds to a good approximation. 
However, because of the non-standard dispersion relation assumed, 
$c_s$ in that equation will  not be necessarily unity in all stages. 
In fact its variation would 
incorporate changes parametrized by $k_{\rm eff}$ in  Eq.~(\ref{eq:keff})
above.

In principle $c_s$ in~(\ref{eq:MSf}) 
can be either larger or smaller than unity. Perhaps a
scenario in which modes do not propagate at all for $k_{\rm phys} \gg k_c$, and 
then do so at increasing $c_s \rightarrow 1$ as they emerge from the 'quantum foam'
at $k_{\rm phys}  \sim k_c$, is appealing; it qualitatively connects, for example, 
to waves propagating in a lattice, which are scattered and dispersed to smaller 
speeds as one approaches the interparticle spacing, before ceasing to propagate. 
However 'analogue' models with superluminal speeds $c_s > 1$,
beyond the cutoff scale, have also been proposed~\cite{Analog_Rev}. 
As we discuss in Section~\ref{sec:nonadiab}, 
for our purposes both situations lead to similar results. 

Equation~(\ref{eq:MSf}) refers 
to a simple harmonic oscillator with variable frequency. 
If $c_s$ is constant, the variation solely comes 
from the second term in the bracket. To
separate this effect from that connected 
to possible variation in $c_s$ at a high energy cutoff 
transition, we exploit the fact that $k_c \gg H$.
This enables one neglect the second term in the brackets 
of equation~(\ref{eq:MSf}), at scales ($\sim k_c$) 
around the high energy cutoff 
transition; as, when modes transit from beyond the cutoff scale $k_c$ to below it, 
the conformal time
$\eta_c = \frac{-1}{a_c H} = \frac{-k_c}{H k}$. The term 
in the brackets in the  aforementioned equation is then 
$c_s^2 k^2 \left(1 -  \frac{2}{c_s^2} \frac{H^2}{k_c^2} \right)$.
The second term inside this latter bracket is 
small compared to unity when 
\begin{equation}
 c_s^2 > 2 \left(\frac{H}{k_c}\right)^2.   
\label{eq:noimcon}
\end{equation}
Since we already assume that $k_c \gg H$, this is always the case when $c_s >  1$. 
We will also assume that this condition is satisfied when considering the case 
of $c_s <1$~\footnote{Indeed, as discussed in Section~\ref{sec:jumpc},
in order to get modification of the matter power spectrum 
and halo mass function at scales of the order of a comoving Mpc, 
which are of particular interest, one needs $H/k_c \lesssim 10^{-4}$, much smaller than 
$c_s =0.01$, which is the smallest we consider.}. 

 This leaves us with an equation of a harmonic oscillator with frequency 
 $\omega = c_s k$. 
 The adiabatic  invariant for a standard harmonic oscillator with specific energy
 $E$  and frequency $\omega$ is $J =\frac{E}{\omega}$. 
 Taking the modulus of the amplitude and the 
 velocity $v_{k}^{\prime}\left(\eta\right)=i \omega_{k} v_{k}\left(\eta\right)$, 
 the energy of the oscillator is $E = \omega^2 |v_k|^2$. 
 Whatever the evolution at scales above $k_c$, as long 
 as it is adiabatic $J$ is conserved. Moreover, at 
 scales $< k_c$ one must recover the standard 
 linear dispersion relation, and so $\omega = k$. At such scales, 
 relevant to eventual horizon crossing, one then has
 \begin{equation}
J = k \left|v_{k}\right|^{2}.
\end{equation}
Comparing this with the standard slow roll inflationary power spectrum 
 \begin{equation}
\Delta_{\mathcal{R}}^{2}(k) = \frac{k H^{2} }{2 \pi^{2}} \left|v_{k}\right|^{2},
\label{eq:SPPS}
\end{equation}
one finds that they are equivalent up to a (nearly) constant factor $\frac{ H^{2} }{2 \pi^{2}}$. 

With sufficiently adiabatic evolution through the transition at $k_c$ 
no significant change to the power spectrum can occur. Any appreciable 
effect could then result solely 
from small variations in $H$, or from second term in bracket
of Eq.~(\ref{eq:MSf}), which also turns out to be quite modest, as 
may be expected given the quadratic correction at any physical scale $(k_{\rm phys}/H)^{-2}$. For this implies again that 
this term  
is smaller than the first until modes are quite close  to existing 
the horizon. In Appendix~\ref{app:nuan}, we show numerical calculations that corroborate this contention 
in the context of the simple model described below.

\subsection{A toy model of the sudden extreme}
\label{sec:toysud}

As we have seen,  any adiabatic frequency 
change, due to nonstandard evolution of modes beyond a high 
energy cutoff scale, will not alter the nearly scale 
free form of the resulting power spectrum. 
We thus consider the opposite extreme; that of a sudden change in the  sound speed 
at $k_c$, while employing the same approximation 
of neglecting the second term in the bracket of Eq.~(\ref{eq:MSf}). 
The procedure again separates changes in the power spectrum due to variations 
in $c_s$ at around $k_c \gg H$ from any time dependence 
connected to the second term in the above equation at much smaller physical 
wavenumbers.

The change in sound speed across the transition 
is equivalent to a sudden change in  frequency. 
To illustrate such a situation in simplest terms, 
we consider the effect of such a change on a simple
harmonic oscillator, with initial amplitude $A$, 
frequency $\omega_{\rm in}$ and phase $\phi$. 
Its evolution is given by
\begin{align}
X_{in}(t)=A \cos \left(\omega_{\rm in} t+\phi\right) \\
X^{\prime}_{\rm in}(t)=-A~\omega_{\rm in} \sin \left(\omega_{\rm in} t+\phi\right),
\end{align}
with
\begin{align}
A=\sqrt{X^2_{\rm in}(t)+\frac{V
^2_{in}(t)}{\omega^2_{\rm in}}} \\
\phi= \arccos{\left(\frac{X_{\rm in}(t)}{A}\right)}-\omega_{\rm in} t.
\end{align}
Suppose that at some moment $t = t_s$ the spring constant
is suddenly altered, 
and the corresponding frequency of the oscillator changes to 
$\omega_{\rm out}$. Then, for $t \ge t_s$,
\begin{IEEEeqnarray}{rCl}
X_{\rm out}(t)&=& B \cos({\omega_{\rm out}(t - t_s)})\nonumber \\
&&+C\sin({\omega_{\rm out}(t - t_s)}) \\
X^{\prime}_{\rm out}(t)&=&-B\omega_{\rm out}\sin({\omega_{\rm out}(t - t_s)})\nonumber \\
&&+C\omega_{\rm out}\cos({\omega_{\rm out}(t - t_s)}). 
\end{IEEEeqnarray}
Matching the initial and final states at $t_s$ one obtains
\begin{align}
B=X_{\rm in}(t_s)\\
C=\frac{\dot{X}_{\rm in}(t_s)}{\omega_{\rm out}}.
\end{align}
So the evolution after the jump can be expressed in terms of the initial state at the 
jump as
\begin{IEEEeqnarray}{rCl}
X_{\rm out}(t)&=&X_{\rm in}(t_s)\cos({\omega_{\rm out}(t - t_s)}) \nonumber \\
&&+\frac{\dot{X}_{\rm in}(t_s)}{\omega_{\rm out}}\sin({\omega_{\rm out}(t-t_s)}),
\end{IEEEeqnarray}
with the amplitude and phase changing after the jump. 

We now apply this  toy model 
to attempt to mimic the evolution of fluctuations due to a sudden 
change in frequency (or again, effectively sound speed) of propagation of inflaton fluctuations.  
In our approximation the evolution is effectively 
governed by two independent harmonic oscillators, 
due to the complexity of the mode functions in~(\ref{eq:MSf}). 
Thus, for the real part,
\begin{align}
X_{{\rm in}_{r}}(\eta)=A_{r} \cos({\omega_{\rm in} \eta+\phi_{r}}) \\
\dot{X}_{{\rm in}_{r}}(\eta)=-A_{r} \omega_{\rm in}  \sin({\omega_{\rm in} \eta+\phi_{r}}),
\end{align}
and for the imaginary part we have
\begin{align}
X_{{\rm in}_{i}}(\eta)=A_{i} \cos({\omega_{\rm in} \eta+\phi_{i}}) \\
X^{\prime}_{{\rm in}_{i}}(\eta)=-A_{i} \omega_{\rm in}  \sin({\omega_{\rm in} \eta+\phi_{i}}).
\end{align}
The sudden step will here correspond to conformal time $\eta_c$, 
when an inflating  mode crosses a physical the wavenumber $k_c$, where 'new physics' may arise.
Applying the step condition as previously, for the real part we  find
\begin{IEEEeqnarray}{rCl}
\nonumber
  X_{{\rm out}_{r}} =  X_{{\rm in}_{r}} (\eta_c) \cos[\omega_{\rm out}(\eta-\eta_c)]\\
+ \frac{X^{'}_{{\rm in}_{r}}}{\omega_{\rm out}}  \sin [\omega_{\rm out}(\eta-\eta_c)].
\label{eq:anmoder}
\end{IEEEeqnarray}
Similarly, for the imaginary part 
\begin{IEEEeqnarray}{rCl}
\nonumber
  X_{{\rm out}_{i}} =  X_{{\rm in}_{i}} (\eta_c) \cos[\omega_{\rm out}(\eta-\eta_c)]\\
+ \frac{X^{'}_{{\rm in}_{r}}}{\omega_{\rm out}}  \sin [\omega_{\rm out}(\eta-\eta_c)].
\label{eq:anmodei}
\end{IEEEeqnarray}
The complete solution then is 
\begin{equation}
v_{k}(\eta)=X_{{\rm out}_{r}}(\eta)+ i X_{{\rm out}_{i}}(\eta).
\label{eq:toymodes}
\end{equation}

This can be evaluated for each $k$,
with $\omega_{k_{\rm out}}= k$, given $\omega_{k_{\rm in}} = c_s k$.
As mentioned above (and checked in Appendix~\ref{app:nuan}) 
usage of Eq.~(\ref{eq:toymodes}), 
in order to evaluate the effect on the primordial 
power spectrum of a sudden step in $c_s$ and 
$\omega$ at $k_c$, returns a good approximation. 
The results of Appendix~\ref{app:Logistic} also 
suggest that the sudden jump scenario itself turns out to be much 
more generic to any appreciable change in the power spectrum 
than may seem {\it a priory}. We now discuss how the power spectrum 
is evaluated and the modifications to the standard near scale invariant form 
that arise.

\subsubsection{The power spectrum of primordial fluctuations}
\label{sec:ppseval}

A mode corresponding to comoving 
wavenumber $k$ crosses the high energy cutoff scale at
$\eta_{c}=-\dfrac{k_{c}}{H k}$. At $\eta \ll \eta_c$ we 
assume Bunch Davies type initial conditions but with $\omega_k = c_s k$,
with $c_s \ne 1$. Thus, before the 
transition in sound speed (and frequency),
\begin{equation}
v_{k}(\eta) \rightarrow \frac{1}{\sqrt{\omega_{k_{\rm in}}}} e^{i \omega_{k_{\rm in}} \eta}.
\label{eq:BD}
\end{equation}
For the initial amplitudes 
one then has
\begin{align}
A_{r}=A_{i}=\frac{1}{\sqrt{\omega_{k_{\rm in}}}}, \quad\phi_{r}=0, \quad
\phi_{i}=\frac{\pi}{2}.
\end{align}
Modes with physical wavenumbers larger than $k_c$ at the start of inflation undergo 
a frequency change such that $\omega_{\rm in}/\omega_{\rm out} = c_s$ (where 
$c_s$ refers to the value, different from unity, before the crossing).  
Modes with smaller wavenumbers do not cross the high energy cutoff scale 
and their frequency remains unmodified  (we discuss
how the transition scale is connected to current comoving scales 
in Section~\ref{sec:matt_halomass}).  

All modes eventually cross the horizon.
Using equations~(\ref{eq:toymodes}) and~(\ref{eq:PPSC}), one can 
evaluate the power spectrum in the context of our simplified model 
when $H$ is given. This is done at horizon crossing 
when $\eta = \eta_H$. Alternatively, one can also use equation~(\ref{eq:MSf}) to calculate 
the power spectrum numerically at superhorizon scales, 
as done in the Appendix for purpose of comparison and 
evaluating the relevance of the model.

In de Sitter inflation $H$ is exactly constant, 
and all modes are assumed to exit the horizon 
at time $\eta_H = - 1/k$. Since again 
a standard dispersion relation must reign beyond 
the high energy cutoff scale $k_c$, one expects $\omega_{\rm out} = k$. 
All modes then leave the horizon at the same phase and oscillations implied 
by equations~(\ref{eq:anmoder}) and (\ref{eq:anmodei})
do not appear in the power spectrum; only enhancement is found 
at scales undergoing the jumps. 
Numerically, equation~(\ref{eq:MSf}) can be used to obtain similar results
(cf. Appendix~\ref{app:nuan}).

The Hubble parameter in more realistic models of inflation must vary  slowly with time.
The variations imply that modes do not leave the horizon at the same phase, 
and  oscillations as well as enhancement appear in the primordial power spectrum.
As a simple generic example, we  will adopt  power-law inflation\cite{lucchin1985power,lyth1992curvature} where (in proper time), $a(t)\approx t^p$, with $p>1$. This corresponds to an inflation potential of the form
\begin{equation}
V(\phi)=V_0 \exp \left[-\frac{1}{M_{\rm Pl}}\sqrt{\frac{2}{p}}(\phi-\phi_i)\right],
\end{equation}
where $M_{\rm Pl}$ is the reduced Planck mass. with
 slow-roll parameters  given by
\begin{equation}
\epsilon_v =\frac{M^2_{\rm Pl}}{2} \left(\frac{V_\phi}{V}\right)^2=\frac{1}{p}, \qquad    \tilde{\eta_{v}}=M^2_{\rm Pl}\frac{V_{\phi\phi}}{V}=\frac{2}{p},
\end{equation}
where $V_\phi=\frac{dV}{d\phi}$. The scale factor and the Hubble parameter become
\begin{equation}
a(\eta)=\left(\frac{\eta}{\eta_{\rm in}}\right)^\frac{p}{1-p} , \qquad H(\eta)=-\frac{p}{p-1}\left(\frac{\eta}{\eta_{\rm in}}\right)^\frac{p}{p-1}\frac{1}{\eta},
\end{equation}
with $\eta_{\rm in}=\frac{t_{\rm in}}{1-p}$.
With these forms for the evolution of the scale factor and Hubble parameter one
can again use Eq.~(\ref{eq:toymodes}) in conjunction with (\ref{eq:PPSC}) 
to evaluate the power spectrum, or numerically integrate  Eq.~(\ref{eq:MS}), which 
now takes the following form
\begin{equation}
v''_{k}(\eta)+\left[c_s^2 k^{2} -\frac{2p^2-p}{(1-p)^2}\frac{1}{\eta^2}\right] v_{k}(\eta)=0.
\end{equation}
We will generally use $H = 10^{-4} M_{\rm Pl}$, $p = 55$. and 
$\eta_i = - 10^{4} M^{-1}_{\rm Pl}$, in order to get the correct normalization and tilt 
of the power spectrum on linear scales using power law inflation.
The results, when rescaled accordingly,  
are valid for other values of $H$, since the  
relative enhancement of the power spectrum  
depends only on the in and out frequency ratio 
of oscillations. The 
connection between the cutoff scale $k_c$
and the corresponding comoving scale 
depends on $H/k_c$ rather than the absolute values (Section~\ref{sec:jumpc}). 
As shown in Appendix~\ref{app:nuan}, the step 
enhancement in the power spectrum 
is accompanied by oscillations in this more generic
(as opposed to de Sitter) case. 

\section{Broken invariance and particle production}
\label{sec:nonadiab}

\subsection{General solution in terms of Bogoliubov coefficients}  

We now consider what the simplified sudden step model actually implies in terms 
of quantum fluctuations in an inflaton. For this purpose
we translate it to the language Bogoliubov expansion and coefficients. 
In this context,  the high energy cutoff transition will 
be seen to lead to excitations of the field and particle production. 
The excitations will take place 
as a result of transitions between well defined time independent in and out states.  
They invariably lead to enhancement in the power spectrum,  
generically accompanied by oscillations. 

\subsection{Generic enhancement in power spectrum}

Using~(\ref{eq:MSf}), and again invoking the approximation of neglecting the $\frac{-2}{\eta^{2}}$ term due to $k_c \gg H$, 
we get the mode function differential  equation of a massless free scalar field 
in Minkowski spacetime, with $\omega_{k} = k$. 
Assuming the field to be initially in 
the vacuum state $|0\rangle$, the amplitude of the vacuum fluctuations 
(the square root of the power spectrum) 
are given in terms of the vacuum mode function 
corresponding to the Bunch Davies initial 
conditions~(\ref{eq:BD}).
Then, non-adiabatic evolution (whether sudden or not), 
can transform this initial vacuum state $|0\rangle$ 
to one with excitations, with respect to the old annihilation 
operator $\hat{a}_{\mathrm{k}}^{-}$. 

To find the effect of such excitations on the power spectrum after the 
transition is complete, one can proceed as follows. 
First by writing the mode expansion of the field operator in terms of the annihilation operator $\hat{b}_{\mathrm{k}}^{-}$ of $|0\rangle$ and its complex conjugate
\begin{equation}
\hat{\chi}=\frac{1}{\sqrt{2}} \int\left(e^{i \mathbf{k} \cdot \mathbf{x}} \mu_{k}^{*} \hat{b}_{\mathbf{k}}^{-}+e^{-i \mathbf{k} \cdot \mathbf{x}} \mu_{k} \hat{b}_{\mathbf{k}}^{+}\right) \frac{d^{3} \mathbf{k}}{(2 \pi)^{3 / 2}},
\end{equation}
and then computing the two point correlation function in the state $|0\rangle$ 
using this operator. One can then define 
the amplitude of the quantum fluctuations in terms of the new mode function $\mu_{k}(\eta)$ as
\begin{equation}
\Delta_{\mu}(\eta)= \frac{1}{2 \pi} k^{3 / 2}\left|\mu_{k}(\eta)\right|.
\end{equation} 
This new normalized mode function is a linear combination of the old one and its complex conjugate. Using Bogoliubov coefficients, it can be written as
\begin{equation}
\mu_{k}(\eta)=\alpha_{k} v_{k}(\eta)+\beta_{k} v_{k}^{*}(\eta).
\end{equation}
Thus we have
\begin{equation}
\Delta_{\mu}(\eta)=\frac{1}{2 \pi} \frac{k^{3 / 2}}{\sqrt{\omega_{k}}}\left[\left|\alpha_{k}\right|^{2}+\left|\beta_{k}\right|^{2}+2 \operatorname{Re}\left(\alpha_{k} \beta_{k}^{*} e^{2 i \omega_{k} \eta}\right)\right]^{1 / 2}.
\label{eq:powerat}
\end{equation}
The second coefficient $\beta$ refers to excitations away from the vacuum state; as we will 
see below it directly counts particle production. 
The ratio of the primordial power spectrum after 
and before the sudden change can be expressed  as 
\begin{equation}
\frac{\Delta_{\mu}^{2}(\eta)}{\Delta_{v}^{2}(\eta)}=1+2\left|\beta_{k}\right|^{2}+2 \operatorname{Re}\left(\alpha_{k} \beta_{k}^{*} e^{2 i \omega_{k} \eta}\right).
\end{equation}
Averaging over a period larger the periodic time of the system (or in generic 
inflation models, over horizon exists of the different modes with different $H$), 
eliminates the oscillating term. 
The main result is that excitations away from the  vacuum state lead 
to typically larger RMS fluctuations and power spectrum. 

An important point  to note here is that the generic enhancement in the power spectrum will occur whether the 'jump' in frequency is upward --- that is whether 
$\omega_{\rm in} < \omega_{\rm out}$ --- 
or the downward, with  $\omega_{\rm in} > \omega_{\rm out}$. 
Or, assuming a dispersion relation  $w = c_s k$ to govern the propagation of fluctuations before and after the jump, the power spectrum will be enhanced whether $c_s$ is larger before the jump, or whether it is larger afterwards. This is seen explicitly below. 

\subsection{Relations between the coefficients and the frequencies}
\label{sec:betaom}

The above does not necessarily assume instantaneous transition jumps between the well defined 
in to out states, just that a time dependent transition occurred. 
In our simplified model we have two regions connected by 
a sudden jump, which enables one to calculate the Boguliubov coefficients explicitly in terms of 
the in and out frequencies. 

If one labels the initial vacuum as  $\left|0_{\mathrm{in}}\right\rangle$, and the final vacuum $\left|0_{\mathrm{  out}}\right\rangle$. Before the jump, 
and assumes the scalar field is in the initial vacuum state,  the mode function is
\begin{equation}
v_{k}^{(\mathrm{in})}(\eta)=\frac{1}{\sqrt{\omega_{k_{\mathrm{in}}}}} e^{i \omega_{k_{\mathrm{in}}} \eta},
\end{equation}
for $\eta < \eta_{c}$. 
Before the jump, the frequency
$\omega_{\rm in} = c_s k$ with $c_s \ne 1$.  
In order to connect with standard inflationary scenario, the frequency after the jump 
is  $\omega_{k_{\rm out}} = k$. 
The final frequency $\omega_{k_{\rm out}}$ is therefore necessarily 
different from the initial one $\omega_{k_{\rm in}}$. This causes 
excitations in the field, which modify the power spectrum. 

After the jump, the mode function $v_{k}^{(\mathrm{in})}(\eta)$ evolves 
into the superposition of $v_{k}^{(\text {out})}(\eta)$ and its complex conjugate:
\begin{equation}
v_{k}^{(\mathrm{in})}(\eta)=\frac{1}{\sqrt{\omega_{k_{\mathrm{out}}}}}\left[\alpha_{k}^{*} e^{i \omega_{k_{\rm out}}\left(\eta-\eta_{c}\right)}-\beta_{k} e^{-i \omega_{k_{\mathrm{out}}}\left(\eta-\eta_{c}\right)}\right].
\end{equation}
The Bogoliubov coefficients $\alpha_{k}, \beta_{k}$ are determined by the requirement that the solution and its first derivative must be continuous at the jump, that is at $\eta=\eta_{c}$. The result is
\begin{equation}
\alpha_{k}=\frac{  e^{-i \omega_{k_{\mathrm{in}}} \eta_{c}}}{2}\left(\sqrt{\frac{\omega_{k_{\mathrm{in}}}}{\omega_{k_{\mathrm{out}}}}}+\sqrt{\frac{\omega_{k_{\mathrm{out}}}}{\omega_{k_{\mathrm{in}}}}}\right)
\label{eq:alpha}
\end{equation}
\begin{equation}
\beta_{k}=\frac{  e^{i \omega_{k_{\mathrm{in}}} \eta_{c}}}{2}\left(\sqrt{\frac{\omega_{k_{\mathrm{in}}}}{\omega_{k_{\mathrm{out}}}}}-\sqrt{\frac{\omega_{k_{\mathrm{out}}}}{\omega_{k_{\mathrm{in}}}}}\right)
\label{eq:beta}
\end{equation}
\\
This explicitly shows  that the time dependence introduced by 
assuming a sufficiently rapid transition from in to out states can lead to significant excitation in the 
inflaton field for large enough frequency ratio. 
As is clear, the absolute values of $\beta_k$ and $\alpha_k$ derived above do not depend on whether $\omega_{\rm in} >  \omega_{\rm out}$ or the reverse. This  again shows 
that generic enhancement in the power spectrum is expected, independent of the direction of the jump.

\subsection{Limits from particle production}
\label{sec:pplim}

As we have seen, 
excitations of the inflaton generically lead to enhanced  power spectrum. 
In section~\ref{sec:matt_halomass} below, we will suggest that these may have important consequences 
at galactic scales, at both high and low redshifts, pertaining to such 
apparent problems as the 
dearth of dwarf galaxies, 'too big to fail' and early galaxy formation, while maintining 
a standard spectrum at scales where it is highly constrained. 
But how much excitations 
of the field can one have without ruining the inflationary scenario itself? Indeed, the exponential expansion during inflation hinges on a dark energy equation of state, too much excitation and particle production can turn it instead into a radiation field, with deceleration replacing the exponential expansion. 

The radiation energy density associated with the relativistic particles, 
which can be assumed to be 
produced through excitations of the field,  may be expressed as~\cite{BDM04}
\begin{equation}
\langle \rho \rangle = \int_{k_{\rm phys} = H}^{k_{\rm phys} = k_c}d^3 {\bf k}_{\rm phys}
\omega_{\rm phys} (k_{\rm phys}) n_{k_{\rm phys}},
\label{eq:backen}
\end{equation}
where $k_{\rm phys}$ and $\omega_{\rm phys}$ are the physical wavenumbers and frequencies.
The occupation number of excited states can be expressed in terms of the second Bogoliubov coefficient 
as $n (k) = |\beta_k|^2$.  In the relevant integration range the relation between the wavenumbers and frequencies 
is linear, and the integral is dominated by larger values of $k_{\rm phys}$. In this case,  
$\langle \rho \rangle \approx \beta^2 k_c^4$, where $\beta$ corresponds to 
$\beta_k$ at larger values of $k$ dominating the integral.  

In the context of the sudden step scenario 
$\beta_k$ is some non-zero constant for modes affected  
by the jump (and zero otherwise),
and the above estimate is rigorously justified. 
In  order for inflation to start and proceed then, 
$\beta^2 k_c^4$ must be smaller than the energy density scale of inflation $H^2 M_{\rm Pl}^2$. 
This leads to the 
condition
\begin{equation}
|\beta|  < \frac{M_{\rm Pl} H}{k_c^2}. 
\label{eq:Backlim}
\end{equation}
If $k_c = M_{\rm Pl}$ this is small for $H \ll k_c$. However much smaller cutoff 
scales may in principle be allowed (and claimed all the way down to the TeV scale e.g.~\cite{ArkTeV981, ArkTev982, RanSun99};  also~\cite{StrinScale04} for review). 
For the largest field inflation allowed by recent data, 
with $H \lesssim 3 \times 10^{-5} M_{\rm Pl}$  
and relatively conventional  high scale $k_c \gtrsim 10^{-3} M_{\rm Pl}$, one finds   
$|\beta| \lesssim  30$ as an upper limit.  
In general, one only needs $k_c/M_{\rm Pl}  \approx H/k_c$ to get 
a Bugoliubov coefficient of order one. The backreaction condition 
above may thus in principle allow for large modifications that could be tested and 
constrained observationally, even in the nonlinear regime of structure formation. 

As long as~(\ref{eq:Backlim}) is satisfied inflation can start and proceed, but 
in order to obtain a near invariant spectrum on large scales, the time derivative 
of the backreaction energy must also be small as the large scale modes exit the horizon.  
The limits of integration in Eq.~(\ref{eq:backen}) is an upper limit on backreaction energy, which assumes
that the whole interval between 
$k_c$ and $H$ is filled with excited states corresponding to modes that have already crossed $k_c$. 
In this case the time derivative $\frac{d \langle \rho \rangle}{dt} \sim \beta^2 H^3 \dot{H}$
is much smaller in absolute value 
than the change in energy density of the inflaton
$\sim M_{\rm Pl}^2 H \dot{H}$  for values of $\beta^2$ of interest.  
However, at earlier times, as modes are crossing $k_c$ and filling up the interval down to $H$, 
the integration interval is variable, the time derivative of the backreaction energy
$d \frac{\langle \rho \rangle}{dt} \sim \beta^2  k^3_{\rm phys} (k_c) \dot{k}_{\rm phys} (k_c) 
\approx \beta^2 H k^4_{\rm phys} (k_c)$
can be much larger (here $k_{\rm phys} (k_c)$ refers to the physical wavenumber of the 
first scale that crosses $k_c$; it  decreases as the mode inflates towards the horizon, when $k_{\rm phys} (k_c) = H$). 
This  leads to the constraint 
$\beta^2 \lesssim \epsilon \frac{H^2 M_{\rm Pl}^2}{k_c^4} \left(\frac{k_c}{k_{\rm phys}}\right)^4$, 
where $\epsilon = - \dot{H}/H^2$. A more detailed treatment 
gives a similar constraint, 
$\beta^2 \lesssim 2 (6 \pi)^2 \epsilon \frac{H^2 M_{\rm Pl}^2}{k_c^4} \left(\frac{k_c}{k_{\rm phys}}\right)^4$~\footnote{cf. Ref \cite{Greene04}. 
Their more rigorous formulation using effective field theory invokes an 'earliest time', 
which is defined as 
the time the smallest CMB scale leaves the cutoff scale. 
The earliest time here would correspond to 
that when the first scales 
cross the high energy cutoff. In both situations the origin of the time variation of the backreaction
lies in the same change in number of 
excited states in the interval 
between $k_c$ and $H$.}. 

The effect of the changing  energy density as the excited states are filling up
the interval between $k_c$ and $H$
can be quite complicated, as it would require evaluation of  the 
modified evolution, taking into account the rescaling of the energy density
(which itself can act as vacuum energy~\cite{BDM04}).
Here we just point out that, simply assuming the usual relation 
$\epsilon = \frac{H^2}{8 \pi^2 P_0}$  to hold when the effect is small enough,
leads to the condition 
\begin{equation}
|\beta| \lesssim   \left(\frac{2 \times 10^{-9}}{P_0}\right)^{1/2} \times 6.7 \times 10^{4} \left(\frac{H}{k_{\rm phys} (k_c)}\right)^2, 
\label{eq:backrd}
\end{equation}
with $P_0 \approx 2 \times 10^{-9}$ the standard characteristic value of the standard 
primordial power spectrum of scalar fluctuations, 
This rough estimate suggests that $|\beta|$ can be of order one, without affecting
the power spectrum on larger scales exiting the horizon, if these scales exit
when the spatial physical  
scale that first crosses the high energy threshold has inflated enough 
to be about $0.004$ times the size of the horizon. 
We further discuss the  possible interpretation of this constraint 
in Section~\ref{sec:jumpc}.

As the ratio of the power spectrum modified by excitations 
to the vacuum power spectrum scales as $1 + 2  \beta^2$, 
considerable modifications may be allowed in principle, if $|\beta|$ is 
of order one or larger. 
In the following we consider possible 
consequences of, and constraints on,  such modification on currently
nonlinear scales, where existing observational
constraints are relatively weak and apparent problems 
with galaxy formation at low and high 
redshift arise. 

\section{Matter power spectrum and halo mass function}
\label{sec:matt_halomass}

In this section we examine some possible  astrophysical implication of the 
sudden change of frequency at a high energy cutoff.  For this purpose
we compute the linear matter power spectrum and the dark matter halo mass function. 
The modified halo mass function
will be of interest, particularly in terms of its possible observable 
consequences on the galaxy stellar mass function.
For the actual calculations  we assume a $\lambda$CDM universe 
with $\Omega_m =0.3$, $\Omega_\Lambda = 0.7$,
$h=0.7$, and RMS dispersion in the density field at $8~h^{-1} {\rm Mpc}$
at $z = 0$, $\sigma_{8}=0.8$.

\subsection{The matter power spectrum}

\begin{figure*}
\includegraphics[width=0.49\textwidth]{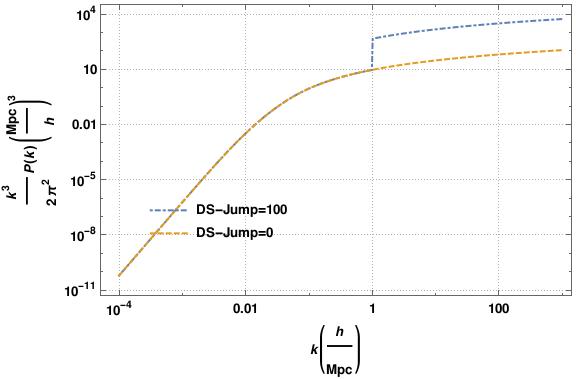}
\includegraphics[width=0.49\textwidth]{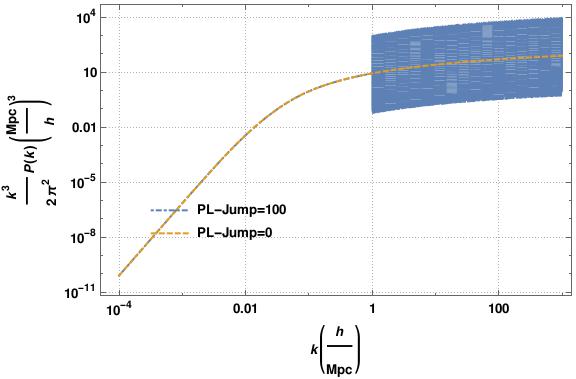}
\caption{Dimensionless matter power spectra at $z= 0$. 
The perturbed cases correspond 
to sudden jumps of a factor of 100 in mode frequency (and therefore sound speed) 
due to shift in dispersion 
relation at the high energy cutoff scale (chosen to correspond to a comoving wavenumber 
$k = 1 h/{\rm Mpc}$ as discussed in the text).
Spectra are shown for a de Sitter background  (left) and 
corresponding power law inflation model as a generic example
(cf. Section~\ref{sec:ppseval}).  The oscillations in the latter case 
are absent in the de Sitter one due to all 
modes leaving the horizon at the same phase. The frequency ratio 
corresponds to a Boguliubov coefficient $|\beta| = 4.95$ 
(Eq.~\ref{eq:beta}). 
Note that, for power law inflation, there is net enhancement 
despite the strong oscillations, which appear symmetric around the uperturbed 
spectrum on the logscale. 
This will result in similar mass dispersions in the de Sitter and power law models, where 
the smoothing also leads to gradual enhancement despite the sudden jump 
at the cutoff scale (Fig.~\ref{fig:Disp}).}
\label{fig:MPS}
\end{figure*}

\subsubsection{Evaluation procedure}

The power spectrum of perturbation in the CDM is evaluated from 
\begin{equation}
P(k,a)=\frac{4}{9} \frac{k^{4} P_{i}(k)}{\Omega_m^{2} H_{0}^{4}} T^{2}(k) D^{2}(a),
\end{equation}
where $P_{i}(k) \equiv  \Delta_{\mathcal{R}}^{2}(k)$ is the primordial power spectrum,
$D(a)$ the linear growth factor, and $H_0$ is the present value of 
the Hubble parameter.

As we will be primarily interested in generic consequences, rather than 
detailed comparison with data, for this purpose we genrerally use 
the BBKS fitting form~\cite{BBKS}  
\begin{equation}
T \left(x\equiv\frac{k}{k_{eq}}\right) = 
\frac{\ln~(1+0.171 x)}{0.171 x} \left[F(x)\right]^{-1/4},
\end{equation}
\\
with $k_{\rm eq} =  0.073~\Omega_m h^2 {\rm Mpc}^{-1}$
and
\begin{equation}
    F(x) = 1+ 0.284 x +    
(1.18 x)^2  + (0.399 x)^3 + (0.490 x)^4.
\end{equation}
For the growth factor, we use\cite{lukic2007halo}
\begin{equation}
D(z)=\frac{D^{+}(z)}{D^{+}(z=0)},
\end{equation}
where 
\begin{equation}
D^{+}(z)=\frac{5 \Omega_{m}}{2} \frac{H(z)}{H_{0}} \int_{z}^{\infty} \frac{\left(1+z^{\prime}\right) d z^{\prime}}{\left[H\left(z^{\prime}\right) / H_{0}\right]^{3}},
\end{equation}
with
\begin{equation}
H(z)=H_{0} \sqrt{\Omega_{m}(1+z)^{3}+\left(1-\Omega_{m}\right)}.
\end{equation}
We have  verified our results against full solution of the perturbation 
equations using the public code CLASS (class-code.net), and show results
using this code in examining the predictions of the enhanced 
primordial spectrum scenario in the more general 
case when the assumption of sudden jump is 
relaxed (Appendix~\ref{app:nosud} and Fig.~\ref{fig:RatG200}).  

\subsubsection{Choice of jump scale}
\label{sec:jumpc}

CMB and large scale structure observations place quite 
tight constraints on the amplitude of the primordial power spectrum 
on scales $\sim 10 {\rm Mpc}$ or larger, 
We now show how smaller scales can be affected by a modified power spectrum, 
while larger scales remain unaffected, if inflation proceeds for approximately the 
number of e-folds needed to solve the horizon problem.

Observable inflation takes place after the comoving spatial 
scale $k^{-1} (H_0) \sim H_{0}^{-1}$ exits the horizon; it is 
characterized by the minimum number of e-folds needed to solve the horizon problem: $N [k(H_0)]= \ln \left[a_{\mathrm{end}} / a_{k (H_0)}\right]$, where the subscripts denote the end of inflation and the epoch of horizon exit of the scale  $k^{-1} (H_0)$. 
The condition can be written as
\begin{equation}
\frac{a_{k(H_0)}}{a_0}=\frac{H_{0}}{H},
\end{equation}
where, $a_{0}$ is the current scale factor.  
$H$ is the Hubble parameter at the horizon exit of the scale $k (H_0)^{-1}$ 
during inflation.

Given a comoving scale $k$,  one may ask when it was equal to a given physical scale $k_c$
during inflation. This gives  the following condition  
\begin{equation}
\frac{k}{a_0 H_0}=\frac{a_c k_c}{a_{k (H_0)} H}.
\end{equation}
As an example, we set $k (k_c)  = 1~{\rm Mpc}^{-1}$, $H = 10^{-4} {M_{\rm Pl}}$, and 
$k_c = {M_{\rm Pl}}$. We then find that $a_{c}\sim a_{k (H_0)}$; that is, 
at the time the current horizon scale $H_{0}^{-1}$ exits the horizon during inflation, the 
comoving spatial scale $\sim 1~\mathrm{Mpc}$ 
is of the order of the Planck length.

This general picture is reproduced even if the cutoff scale 
$k_c$ is not the Planck scale. 
All one needs is $H/k_c \approx 10^{-4}$. If inflation proceeds for a number 
of e-folds
larger than the number $N_{\rm min} = N [k(H_0)]$
required to solve the horizon problem, then the 'jump scale' can still correspond to 
$k (k_c)) = 1  {\rm Mpc^{-1}}$ if $H/k_c < 10^{-4}$. 
In general, the number of e-folds allowed, with  
$k (k_c)$ corresponding to the smallest comoving 
spatial scale affected by the high energy cutoff transition, is 
\begin{equation}
N = N_{\rm min}     + \ln \left[ \left(\frac{k_c}{H}\right) \left(\frac{k (H_0)} {k (k_c)}\right) \right],
\label{eq:cefolds}
\end{equation}

The scale $k (H_0) \approx  10^{-4}~{\rm  Mpc}^{-1}$ 
is fixed by the present size of the horizon, while 
$k (k_c) = 1~{\rm  Mpc^{-1}}$
happens to roughly correspond 
to the largest scale on which
significant modification of the power spectrum would not affect 
its inference from galaxy cluster counts and lensing surveys (but, 
depending on the exact scale,  not necessarily  
Lyman-$\alpha$ bounds, as discussed in  Section~\ref{sec:Lyman}).
Larger values of $k (k_c)$ are in principle 
possible, and in this case the power spectrum can be modified on smaller scales, 
affecting smaller nonlinear structures.  
However, if one takes into 
account our crude estimate of the time variation of the backreaction, this 
may be constrained.  For, as mentioned in relation to Eq.~(\ref{eq:backrd}), 
to maintain $|\beta|$ of order 1, one may need $H/k_{\rm phys} (k_c) \gtrsim 0.004$.
If $k (k_c) \approx 1~{\rm Mpc^{-1}}$, the comoving 
scale exiting the horizon when this is satisfied is $\approx 0.004~{\rm Mpc^{-1}}$. 
Larger scales, 
with smaller wavenumbers, can be affected if 
one insists on $|\beta| \gtrsim 1$. In the context of the simplest scenario 
with constant $\beta_k$ beyond the cutoff regime, 
the power spectrum may be modified on such scales. 
This may be allowed on comoving
scales $k < 0.004~{\rm Mpc}^{-1}$, and may even be relevant to 
supposed anomalies of the CMB on large scales, but
not on smaller spatial scales, where modifications are tightly constrained. 
That changes in the power spectrum on the largest scales may be 
connected with backreaction associated with initial evolution has already
been noted  (e.g.~\cite{NormaS06}), and may be of interest in the present context, 
but its proper examination is beyond our present scope.

Here we will be mainly interested in the enhancement 
of the power spectrum on large nonlinear scales, corresponding 
to $k \approx  {\rm Mpc}^{-1}$, 
because of the  particularly interesting consequences for galaxy formation we discuss  in Section~\ref{sec:intcons}. 
Fig.~\ref{fig:MPS} shows the resulting dimensionless matter power spectrum for a jump corresponding to ratio of sound speeds  (or in and out frequencies) of $100$ on such scales. The ratio is associated with a Boguliubov coefficient $|\beta|$ of about 5.
The large value is chosen as to clearly delineate  phenomena  associated with significant 
excitation on nonlinear structure formation. This fixes our basic fiducial model. 
We will examine, in addition,  the effect of smaller enhancements and comoving spatial cutoff scales
(Section~\ref{sec:Lyman}), as well as the effect of relaxing the sudden jump assumption
(Appendix~\ref{app:nosud}; Section~\ref{sec:Lyman}).

\begin{figure*}
\includegraphics[width=0.49\textwidth]{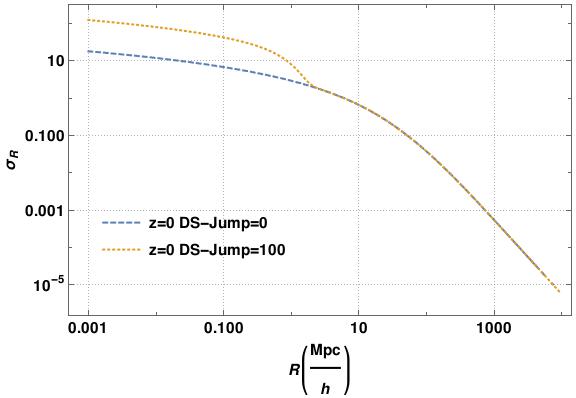}
\includegraphics[width=0.49\textwidth]{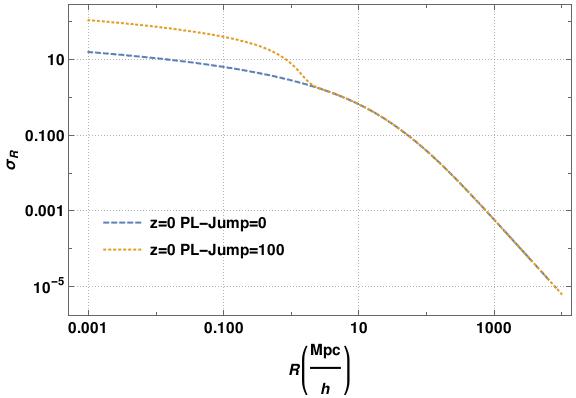}
\caption{RMS mass fluctuations corresponding to power spectra shown in Fig.~\ref{fig:MPS}. 
Note that the strong oscillations in the power spectrum, 
in the case of power law inflation, have little effect 
here, as they are smoothed over and integrated out as the dispersions are extracted
from the power spectra. Despite the sharp jump in linear power spectra, the 
change in the RMS mass fluctuations beyond the cutoff scale is also gradual.} 
\label{fig:Disp}
\end{figure*}

\subsection{Halo mass function}

\subsubsection{Evaluation procedure}

On nonlinear scales, modifications of the primordial power spectrum 
are primarily encoded in the mass function of 
self gravitating dark matter objects, the halos hosting galaxies. 
We evaluate this function using the Press-Schecter formalism, 
which estimates the number of dark matter halos per unit mass and 
comoving volume, given the
linear matter power spectrum {\it via} a spherical collapse 
model~\cite{press1974formation, bond1991excursion}. 
This is given by
\begin{equation}
\frac{dn}{dm}=\frac{\rho_{0}}{M^2} f (\sigma)  \mid \frac{d\ln \sigma}{d\ln M} \mid
\label{eq:HMF}
\end{equation}
where 
$\rho_0$ is the mean matter density at $z=0$ and 
$f(\sigma)$ is given by
\begin{equation}
f (\sigma) = \sqrt{\frac{2}{\pi}} \nu 
\exp \left(\frac{- \nu^2}{2} \right),  
\end{equation}
where $\nu = \delta_c/\sigma$,  with 
$\delta_{c}=1.686$  the critical 
overdensity for spherical collapse and 
$\sigma$  the RMS variance of mass fluctuations 
within a sphere of radius $R$ and containing mass $M=\vartheta_{f} \rho_{0} R^3$, 
where $\vartheta_{f}$ a constant that depends on the filter function $W$.  
For Gaussian filter it is $\vartheta_{f}=(2 \pi)^{3 / 2}$. The filter function is characterized by its size $R$ or mass $M$. In the case of Gaussian filter we use here, 
the relation between them is 
\begin{equation}
M=4.37 \times 10^{12} \Omega_m h^{-1}\left(\frac{R}{h^{-1} \mathrm{Mpc}}\right)^{3} M_\odot.
\end{equation}
As our primary aim is to illustrate generic consequences of 
enhanced small scale power spectrum, we generally kept to the aforementioned
simplest form of the Press-Schecter formalism. 
However we have also verified the 
insensitivity of our results to that choice by comparing with 
an ellipsoidal collapse fitting function, 
which provides better fits to mass functions of halos identified in cosmological 
simulations~\cite{SMT01, ST02, Despali16, Comparat17},  
\begin{equation}
 f  (\sigma) =   A \sqrt{\frac{2 a_s}{\pi}} 
 \left[1+ \left(\frac{\nu^2}{a_s}\right)^{p_s} \right] 
\nu \exp \left(- \frac{a_s \nu^2}{2} \right), 
\label{eq:fST}
\end{equation}
where we set $p_s = 0.3$, $A = 0.3222$ and $a_s =0.707$. 
Results using that form are shown in Appendix~\ref{app:nosud}, 
where we examine implication of an enhanced small scale spectrum when the assumption 
of sudden jump is relaxed, and also in Fig.~\ref{fig:RatG200}.

The mass variance is calculated through the integral,
\begin{equation}
\sigma^2(R)=\frac{1}{2 \pi^2}\int_{0}^{\infty} k^2 P(k) W^2(k R) d k, 
\end{equation}
where $P(k)$ is the linear power spectrum and $W(kR)$ is the Fourier transform of the Gaussian filter function
\begin{equation}
W(k R)=\exp \left(-\frac{(k R)^{2}}{2}\right).
\end{equation}
Fig.~\ref{fig:Disp} shows the thus calculated dispersion for de Sitter and power low 
inflation models. As can be seen the strong oscillations in the power spectrum of the 
latter case are smoothed and integrated over, and the results are quite similar in the two cases. 
Also, despite the sharp jump in the corresponding power spectra, the 
change in the RMS mass fluctuations in the nonlinear regime beyond the cutoff scale is gradual.

\begin{figure}
\includegraphics[width=0.49\textwidth]{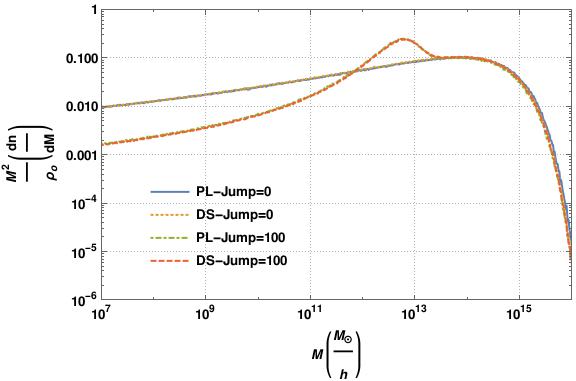}
\caption{Multiplicity function, describing the fraction of mass in dark halos of mass $M$, 
for the power spectra shown in Fig.~\ref{fig:MPS} and dispersions of Fig.~\ref{fig:Disp}. 
As may already be expected from the latter figure, the results are similar 
in de Sitter and power law inflation 
(labelled DS and PL respectively), due to the smoothing and integration over 
the oscillations as the mass function is derived.}
\label{fig:HMFz0}
\end{figure}

\subsubsection{Mass function at redshift zero} 
\label{sec:HMFz=0}

Fig.~\ref{fig:HMFz0} shows the resulting Press-Schecter halo multiplicity function, which 
estimates the fraction of mass in halos of mass $M$, 
corresponding to 
the unperturbed  and perturbed (with jump) matter power spectra 
shown in Fig.~\ref{fig:MPS}. 
As may be expected given the mass dispersions 
shown in Fig.~\ref{fig:Disp}, the results are virtually identical in case of de Sitter and power  
law inflation, despite the strong oscillations in the spectrum in the latter case. 
Perhaps also expected  is the enhancement at larger masses embodied in the bump encompassing a scale around a few $10^{12} {M_\odot}$, 
when the power spectrum is boosted. 
More conterintutively, there is a dearth of small halos when the power spectrum is perturbed. This is due to those smaller halos more rapidly merging into larger ones, as we discuss further below
(next subsection). 

The bump at higher masses 
is not in itself directly observable; as it can be accounted for by changing the galaxy-halo 
occupation numbers. Indeed it roughly corresponds to the highest mass to light ratio inferred when 
fitting galaxies to halos in context of the standard model.
Nevertheless, the compatibility of such enhancement with data can be tested through a combination of abundance 
matching and dynamical modelling.  In fact, 
for galaxies with stellar mass above $5 \times 10^{10} M_\odot$,
abundance matching with 
standard power spectrum seems to overpredict the observed stellar masses for a given dynamical mass~\cite{Maccio_abunmatch20}.  
As the galaxy number density (determined by the
stellar mass function), is a decreasing function of mass, the discrepancy may in principle 
be accommodated in our current context as follow: 
by increasing the abundance of halos with larger dynamical masses
(with the bump at higher masses), the galaxy population
associated with such halos will then be one with correspondingly 
larger number density, and hence smaller masses. 
This is essentially the same effect 
that may help alleviate the apparent problems with early massive galaxy formation as we 
discuss in Section~\ref{sec:earlygals}. Thus the solution to such potential problems does not 
only appear consistent with the distribution of dynamical masses 
at low redshift but may even resolve certain  problems there; 
at both high and low mass scales.

\subsubsection{Enhancement at high $z$ and large $M$, and 
suppression at the opposite ends}

\begin{figure}
\includegraphics[width=0.49 \textwidth]{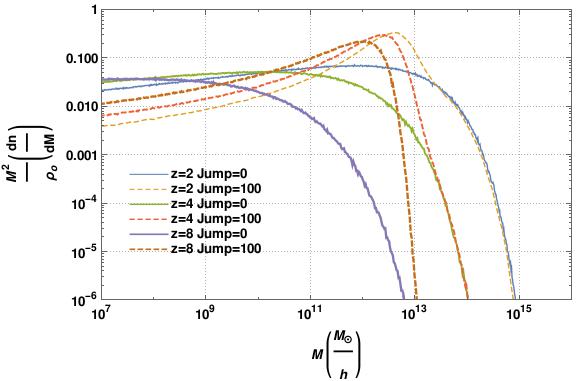}
\caption{
Same as in Fig.~\ref{fig:HMFz0} but at the indicated redshifts (using power law
inflation model). 
Solid lines show the results
with standard (nearly) scale invariant primordial spectrum, 
which are compared to those obtained when the power spectrum is boosted at 
smaller scales, as a result of an imposed step (corresponding to a ratio of hundred fold) in mode frequency and propagation speed (cf. Fig.~\ref{fig:MPS}).}.
\label{fig:multiplic}
\end{figure}

For the same parameters used above, 
Fig.~\ref{fig:multiplic} shows the multiplicity function
at selected redshifts. Four trends are clear.  First, the enhancement in the primordial 
power spectrum leads to  enhancement 
in the number of halos at intermediate masses, culminating approximately at the mass 
scale corresponding to the length scale to where the jump in the power spectrum is placed.  
The second is that the effect is larger, and
is apparent for a larger range in masses, at higher redshifts. Indeed, 
Fig.~\ref{fig:Rat1.1} shows  a maximum enhancement of almost 
four orders of magnitude at $z = 8$, compared to only a factor of a few at $z =0$. 
This is because  high mass halos are rarer at higher redshifts and thus 
the relative increase due to the enhancement resulting from the discontinuity in the power spectrum is larger. It also results 
in the  rate of change in the mass function with redshift, at fixed mass, 
being smaller in the case of enhanced spectrum  
than in the unperturbed case.  

Finally, there is the somewhat counterintuitive effect, mentioned at the conclusion of the previous subsection, 
of significant  {\it suppression} of the multiplicity function contribution of 
halos at smaller masses, below a few $10^{11} {M_\odot}$.  
This interesting result may be understood by recalling that 
enhancement in the power spectrum at comoving scales associated with masses of $\gtrsim 10^{12} {\rm M_\odot}$, 
implies that all smaller scales  are also enhanced. 
And enhancement at smaller (length and mass) scales in turn implies
that smaller halos form at higher redshift and that by the redshifts considered here they have already been typically 
subsumed in larger ones; that is, the typical mass scale, for a given fluctuation level  at a given redshift,  shifts up.  
This leads to a relative decrease in the number of halos with masses $\lesssim 10^9 {M_\odot}$. 
As opposed to the case of the enhancement of the multiplicity of relatively high mass halos,  the de-enhancement is here relatively larger at smaller $z$, as the lower mass halos are now those that are rarer at such redshifts.  

\subsection{Interpretation and possible consequences}
\label{sec:intcons}

As, in the current model of structure formation, galaxies form in seeds provided 
by the potential wells of dark matter halos, the significant modifications to the 
halo mass function  are expected to leave imprints on the associated 
galaxy stellar mass function. Pertinent questions here thus include whether those modifications 
have consequences for problems arising at small scales within the current 
standard scenario of structure formation, as outlined in the introduction; 
or, in contrast,  whether such modifications can constrained also on nonlinear scales. 

\subsubsection{Small scale problems at low redshift and the dearth of dwarf galaxies}

One straightforward consequence of the suppression of halo multiplicity at small scales 
pertains to the longstanding issue of the dearth of dwarf galaxies in the standard scenario: 
a galaxy like the Milky Way is expected, in the context of the $\Lambda$CDM with a standard 
primordial power spectrum, to have hundreds of satellites that are not observed, 
and some of the predicted hosting halos are too 'large' to have 'failed' to form galaxies. 
These are aspects of the so-called small scale problems of the standard 
scenario has given rise to various explanations, e.g in terms of baryonic physics, warm dark matter, 
fuzzy dark matter, as well as direct suppression of the small scale power spectrum.  

Our somewhat counter-intuitive result, on the other hand, is that an enhancement of power 
on small scales can also lead to a suppression in the number of small halos (as these `overmerge' 
into larger entities).  This suppression at $z =0$ at the scales where 
issues such as the dearth of small galaxies ($M_h \lesssim 10^{9} M_\odot$)
and too big to fail ($10^{9}  \lesssim  M_h \lesssim  10^{11} M_\odot$)
problems appear, can therefore be 
of relevance to apparent small scale crises arising 
in the context the standard model of structure formation.
The order of magnitude suppression at smaller masses is  
directly relevant  to resolving the apparent discrepancy between the number of observed small satellite galaxies 
and large number of small halos found in cold dark matter simulations. The suppression on the 
larger mass scales, on the other hand, may help alleviate the too big to fail issue; 
when this is posed as an abundance matching problem, whereby the abundance of simulated
halos is too large at the masses inferred from the dynamics of observed galaxies
in the range $10^{9}  \lesssim M_h/M\odot \lesssim 10^{11}$~(e.g., ref. \cite{Ellis14}).

\subsubsection{The excess of early massive galaxies}
\label{sec:earlygals}
 
The enhancement at higher mass scales may, on the other hand, have consequences for 
the more recently raised issues associated with early galaxy formation. 
These are  extensions of longstanding phenomena related to what is referred to as
'downsizing' (e.g.~\cite{Somerdown}), required to account for 
preponderance of early massive galaxies; 
a phenomenon that does not appear entirely natural in a hierarchical structure formation 
scenario, where the smaller halos embodying the potential wells hosting the galaxies form first.
 
The problem of early galaxy formation has been termed `impossibly early' in the context of the 
standard $\Lambda$CDM scenario of structure formation~\cite{Imp_earl16}. 
In that work, the authors attempt to infer the halo mass function at high-$z$, primarily 
from  stellar mass functions  derived using photometric spectral energy 
distribution  templates and ultraviolet luminosity functions. The halo mass 
is then inferred by assuming a stellar to halo mass of 
$M_*/M_h = 1/70$.  If this local value of $M_*/M_h$ is used, then Fig.~1 of the aforementioned work 
 suggests that the number density of massive galaxies 
can greatly exceed that of the halos they should inhabit for $z \gtrsim 4$ in the standard
$\Lambda$CDM structure formation scenario. 
The discrepancy becomes more severe as one moves up in redshift and mass, reaching 
four orders of magnitude or more.  

The above would seem to rule out the standard scenario of structure formation 
in the context of $\Lambda$CDM cosmology. 
However, a couple of caveats have been pointed out. 
First, regarding the assumption that $M_*/M_h$ does not vary with redshift. 
For, as  can also be seen from Fig.~1 of~\cite{Imp_earl16}, instead of moving the 
points inferred from the observed 
stellar number densities down orders of magnitude to 
fit the corresponding halo number densities, 
one can move the points horizontally to the left by an order of magnitude. 
This fitting procedure in effect invokes a $z$ (and $M_*$) dependent $M_*/M_h$, to replace 
the fiducial local value of  $M_*/M_h = 1/70$ assumed by the authors.  The procedure, 
requiring $M_*/M_h \sim 1/7$, is 
still in principle consistent with a universal baryon fraction of $1/6.3$, associated with 
the standard cosmological  scenario, but only 
just~\cite{Silk_LCDM_SMBH18}~\footnote{Note that in~\cite{Silk_LCDM_SMBH18}
the more conservative cumulative stellar mass function, which averages over 
the increasing $\frac{M_*}{M_h} (M_*)$,  is employed. The data for $M_* = 10^8 {M_\odot}$ to $10^{10} {M_\odot}$ 
are consistent with $\Lambda$CDM.  Tension still arises for the data point shown for 
$M_* \ge 10^{11.7} {M_\odot}$ at $z \approx 5.5$ (their Fig.~2 top-left).
The results in 
Fig.~18 of~\cite{COSMOS15} show data consistent with  $M_*/M_h =  1/6.3$, or even larger, already
at $z \sim 5$, 
for  $M_* \gtrsim 10^{11} {M_\odot}$. 
In general,  it seems that galaxies can be accommodated into halos with 
$M_*/M_h \ll 1/6.3$ for $z \gtrsim 5$ if one keeps to 
$M_*  \lesssim 10^{10.5}  {M_\odot}$ (e.g., \cite{IncFBehrooz13}, and Fig.~9  
of~\cite{UMach}).
On the other hand,  the most extreme cases in the data presented by Steinhardt et. al.~\cite{Imp_earl16} 
(where $M_*/M_h \approx 1/7$ seems required),  are just barely consistent with 
standard $\Lambda$CDM.}.

Another caveat that has been pointed out concerns the extraction of $M_*$ and associated number 
densities from the ultraviolet luminosity function at high $z$, 
which some of the data points of~\cite{Imp_earl16} relies on~\cite{Mancuso16}.
However, a multi-wavelength analysis of a sample of massive galaxies at $z > 3$
also leads to a cumulative mass function that can be consistent (within estimated errors)
with $M_*/M_h$ approaching the universal baryon fraction 
at $z \sim 5.5$ and $M_* > 10^{11} {\rm M_\odot}$~\cite{Balmer19}. 
That work also shows (Fig.~14) that the number densities of massive galaxies 
are very difficult to reproduce in 
hydrodynamic numerical simulations --- with significant underestimate for $z > 3$ ---
which may be expected, as their 
reproduction would seem to  require that all 
available baryons reside inside galaxies, and their near  total conversion to stars over a 
short time ($\sim {\rm Gyr}$).   This would have as consequence the presence of a significantly 
`quenched', quiescent population of massive galaxies already at high redshift. The presence of such a population,
which is indeed observed, poses significant challenges.  Synthesizing
the stellar populations of one such 
object, observed at $z = 3.717$, for example, 
seems to again require prior evolution involving a $M_*/M_h$  
reaching the universal baryon fraction~\cite{NatImp} (see also~\cite{Capak19}).   
There  now appears to be a substantial population of such galaxies, 
observed at increasing redshift~\cite{subqueis14, SchreiSAM18, PALMA19, R+D19, Quiescent1.5Gyr19, GirelliSAM19, XMassQ19}, 
and not easily reproduced by either hydrodynamic
simulations~\cite{ R+D19, Quiescent1.5Gyr19} or semi analytical models~\cite{SchreiSAM18, GirelliSAM19}.

Although questions as to the ultimate severity of these problems will only be settled 
with the next generation surveys  (e.g. with the JWST),  
the situation warrants pointing out that 
they can in principle be alleviated by invoking small scale 
enhancement of the primordial  power spectrum examined here. 

\begin{figure}
\includegraphics[width=0.49 \textwidth]{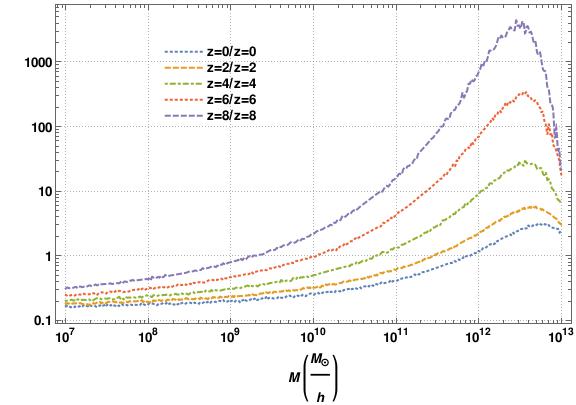}
\caption{
Ratios of mass functions with modified power spectrum 
to those resulting from 
unmodified power spectrum, at the same redshift. 
The results correspond to the ratios of dashed to solid lines  
in Fig.~\ref{fig:multiplic} and Fig.~\ref{fig:HMFz0}.}
\label{fig:Rat1.1}
\end{figure}

\begin{figure*}
\includegraphics[width=0.49\textwidth]{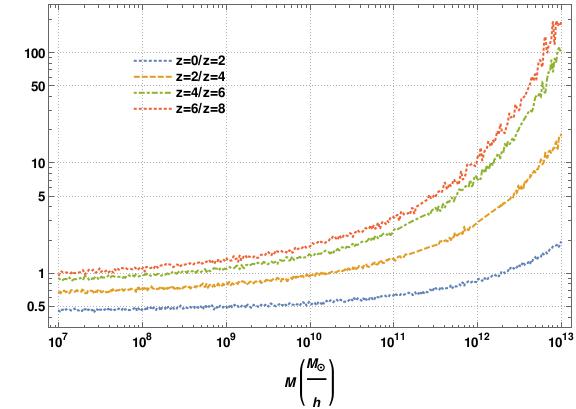}
\includegraphics[width=0.49\textwidth]{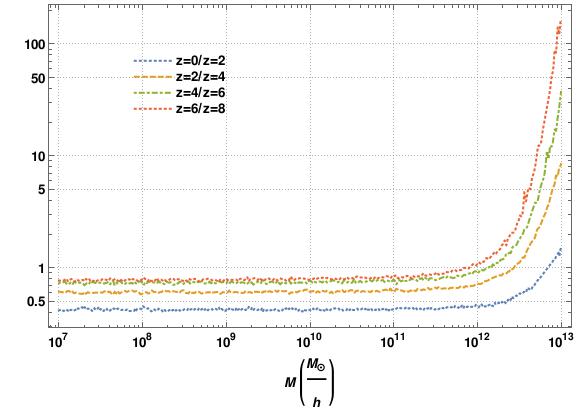}
\caption{Ratios of  the mass functions at different redshifts, for the 
standard case (left) and that with modified power spectrum (right).
Note the slower evolution (reflected in the smaller ratios) 
at higher $z$ for most of the mass range in the modified case.}
\label{fig:Ratz1}
\end{figure*}

Fig.~\ref{fig:Rat1.1} shows that significant enhancements 
can be achieved at mass scales  $10^{12} {M_\odot} \lesssim M\lesssim
10^{13} {M_\odot}$, with a peak
at a scale corresponding to highest dark matter to stellar mass ratio in standard modelling, at which the enhancements reach 
even the `impossibly' large levels claimed in~\cite{Imp_earl16}. 
Perhaps no less important is the slower 
evolution of the mass function for $z \gtrsim 4$, observed 
in Fig.~\ref{fig:Ratz1}, which is more consistent
with the redshift evolution of the inferred stellar mass densities in~\cite{Imp_earl16}
than the much faster evolution in the standard 
case (the slow evolution of the stellar mass function for  $4 \lesssim z \lesssim 7$ was also observed 
for example by Song et. al.~\cite{Song16}).  
This would seem to waive the apparent requirement of a $M_*/M_h$ that is high dependent 
on redshift in order to fit the data. 

With better statistics, and firmer grip on observational systematics, it should be 
possible to distinguish between scenarios involving enhancement in the 
primordial power spectrum, such as the one presented here
and reconciliation with data through improvement of the baryonic model; by invoking 
further `downsizing' physics input, in terms of feedback, quenching and other `subgrid' physics 
(assuming the data remain  consistent with the strict upper bounds placed 
in the context of $\Lambda$CDM~\cite{Silk_LCDM_SMBH18}). As the baryonic 
models become better  
constrained, there may be particular 
consequences that could also constrain (or confirm) the sort of
scenario discussed here. We now discuss some of these. 

\subsubsection{Other observables, constraints and variation on basic model}
\label{sec:Lyman}

\begin{figure*}
\includegraphics[width=0.49 \textwidth]{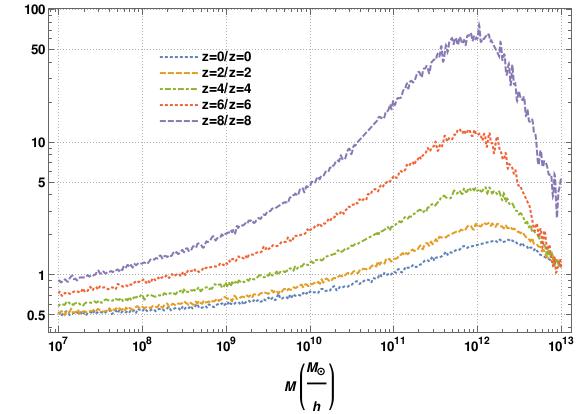}
\includegraphics[width=0.49 \textwidth]{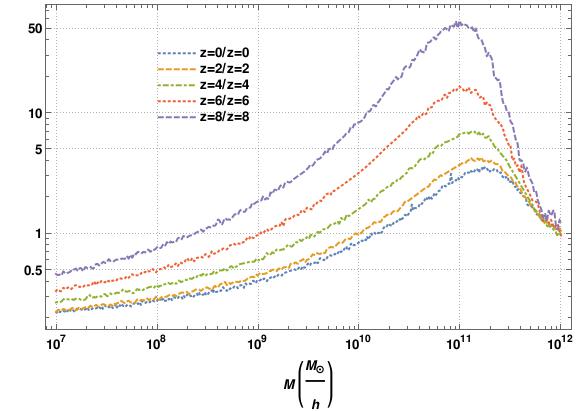}
\caption{Same as in Fig.~\ref{fig:Rat1.1} but with step frequency ratios of 10 instead of 
100, leading about to about an order of magnitude less modification in the power spectrum, with
Bogulibov coefficient $|\beta| = 1.42$ instead of $4.95$ (left); and for power spectrum 
comoving modification scale $k =3  h/{\rm Mpc}$, instead 
of $k = 1 h/{\rm Mpc}$ (right).}
\label{fig:Rat1.23}
\end{figure*}

In the context of the enhanced spectrum scenario presented here, 
the clustering of halos, on mass scales and redshifts where numbers are predicted to be 
significantly enhanced, may be measurably different from the standard case. 
This is because  the biasing with respect to the matter distribution 
would be expected to be different (since they would correspondent 
to less rare density peaks).   
Combined clustering and abundance matching analysis in the context of a `halo model' 
(e.g.,~\cite{Peacock_B}), particularly  at higher redshifts~\cite{Harikane16}, could thus
in principle test, and place constraints on, scenarios involving primordial power spectrum enhancements. 
The galaxy-matter correlation function,  entering into
calculations of galaxy-galaxy weak lensing signals,  should also be different in the 
present scenario from the standard case. The difference should again be especially significant 
at higher redshifts, where the abundance of high mass halos is strongly increased, 
making for a relatively clumpy matter distribution. 
Tests are also possible at low redshift. particularly as regards to the peak 
in abundance of Milky Way sized halos, which seems consistent with observations
(Section~\ref{sec:HMFz=0}). Some observations on the other hand suggest that 
the standard model itself may overpredict the halo mass function at 
scales $10^{13} M_\odot \lesssim M \lesssim 10^{14} M_\odot$~\cite{Leauthaud_Lens17}; further enhancements of the mass function  at such scales may be thus constrained.

Another observable that can potentially place immediate constraints on the 
scenario discussed here is the Lyman-$\alpha$ forest. 
Here, detailed comparison with data involves 
complex simulations that depend on assumptions regarding the state of the intergalactic medium, 
which become less robust at nonlinear scales~\cite{NonlinLyman, Hui_etal2017}. 
In the nonlinear regime the modifications in the power spectrum are primarily 
imprinted in the RMS dispersion and the halo mass function, where the complex 
pattern of enhancement and suppression at different scales and redshifts would 
contribute to the mass fluctuations probed by one dimensional Lyman-$\alpha$ spectra. 
The large enhancement at higher masses and redshifts may also affect the thermal history of the intergalactic medium. 
It may therefore be worth investigating if and how such changes affect the standard 
constraints regarding the power spectrum.  
Pending such investigation, as the modifications to the linear
power spectrum and mass dispersion considered above are large 
and fall within the region relevant to Lyman-$\alpha$ observations, 
it apt to probe what is to be expected if more modest
modifications are made.

In Fig.~\ref{fig:Rat1.23} (left panel)
we show the relative change in the mass function for (about an order of magnitude) 
smaller perturbation in the power spectrum, as well as on scales deeper in the nonlinear regime. 
As can be seen, in the former case, significant enhancement in the mass  function   
can still occur at the right scale at higher redshift 
(where halos are exceedingly rare in the standard scenario), so as to  
alleviate the  apparent early galaxy formation problem. The 
reduction in  number of small halos, relevant to the dearth of small
galaxies and too big to fail problems at low $z$, is smaller however.

When the modification 
in the power spectrum is placed on a smaller spatial scales, 
deeper in the nonlinear regime (Fig.~\ref{fig:Rat1.23}, right panel),
the decrease in number of small halos at $z = 0$ is again significant and relevant 
to the dearth of dwarf galaxies, but 
does not cover all the mass range relevant to the too big to fail problem.   
As may be expected, the enhancement at high redshifts happens at a  
smaller mass scale (they are also smaller 
because halos in the standard scenario are already more abundant at such scales). 
Enhancement at such scales is not directly applicable to the problem of 
high $M_*/M_h$ at $M_* \gtrsim 10^{10.5} M_\odot$ and $z \gtrsim 4$, as discussed 
in previous subsection. It  may nevertheless be relevant at higher redshift, as the 
progenitors of massive quiescent galaxies were assembled (especially if the star formation 
rate density does not steeply decline beyond $z = 8$, as suggested by some authors; c.f.
ref.~\cite{Capak19}, particularly the discussion Section~7.3 and references therein).
As one further increases the comoving wavenumber associated with the high energy cutoff $k_c$, this 
general trend persists.  The suppression on small scales at $z=0$ are found to correspond to
masses of  small halos that are overabundant in CDM up to comoving 
cutoff $k \sim 9~{\rm Mpc^{-1}}$, 
which essentially avoids Lyman-$\alpha$ bounds. At larger modification wavenumbers, however, 
one finds enhancement 
rather than suppression at mass scales $\lesssim 10^{9} M_\odot$, relevant to the dearth of dwarf galaxies issue. Although enhancing the power spectrum 
at such smaller spatial scales would not appear to alleviate any of issues regarding 
galaxy formation discussed here, it could still have consequences 
for early black hole formation and the epoch of ionization (see also~\cite{Hirano_Spergel2015}).

\begin{figure}
\includegraphics[width=0.49 \textwidth]{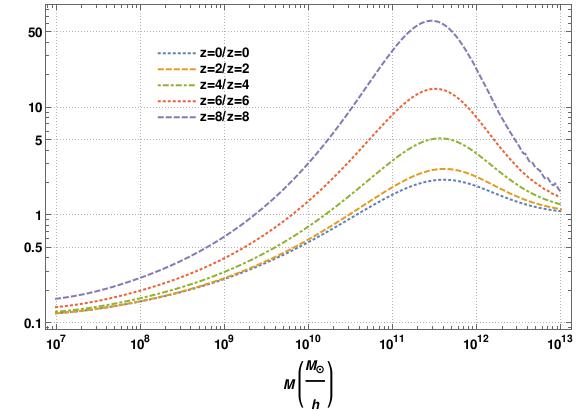}
\caption{
Same as in Figures~\ref{fig:Rat1.1} and~\ref{fig:Rat1.23} but with gradually 
modified power spectrum, rather 
than sudden jump. The spectrum is modified using Eqs.~(\ref{eq:parfun}) and 
(\ref{eq:gradual}) with 
$S = 200$, $b = 2$ and $k_c = 3 h~{\rm Mpc}^{-1}$  comoving. 
The results correspond to ratios 
 of dashed and solid lines in Fig.~\ref{fig: gradualHMF}
(taking the dashed line at $b=2$ for the left hand panel).
The associated modifications to the power spectra are those shown in
Fig.~\ref{fig: gradualPS}} 
\label{fig:RatG200}
\end{figure}

On the other hand, relaxing the assumption of a sudden jump transition leads to 
suppression  and enhancement of halo abundances on larger mass scales,
This allows for retaining the advantages of 
a sudden cutoff at comoving wavenumber 
$1 h / {\rm Mpc}$, while  keeping the mass function at $10^{13} M_\odot$ at $z = 0$ unchanged,  
and  modifying the power spectrum 
much more modestly at $1 h /{\rm Mpc}$ 
(cf. Appendix~\ref{app:nosud}, Fig.~\ref{fig: gradualPS}). 
Fig.~\ref{fig:RatG200} shows the relative change in the mass function for such a gradual transition 
in the power spectrum around a characteristic wavenumber of $3 h/{\rm Mpc}$ comoving.  
As can be seen, significant suppression at $z =0$ is again recovered at mass scales of order $10^{11} M_\odot$, 
relevant to both the dearth of dwarf galaxies and the too big to fail problems. Enhancement  
at higher $z$ also occurs at scales significantly larger than the corresponding sudden 
jump case (shown on right panel of  Fig.~\ref{fig:Rat1.23}), 
which renders the enhancement more directly relevant to early massive galaxy formation issues. 
The mass function is unmodified at scales $10^{13} M_\odot$ at $z =0$. 

Thus, potential resolution of all or some of the galactic and sub-galactic scale 
issues through modification  of the power spectrum, rather than (or in addition to)
baryonic physics input, may in principle be  tested and constrained through distinctive predictions.
This is true in general and is not confined to our particular simple model of a sudden jump; such tests 
will thus become more relevant if the small scale issues connected to the 
standard structure formation scenario are confirmed to persist with incoming observations. 
In the context of the present scenario, 
such observations can potentially probe  imprints (or lack thereof) 
of high energy cutoff physics on the relevant astrophysical scales, and
place constraints on the duration of inflation, 
as the ratio of the Hubble scale of inflation to the high energy 
cutoff scale and the number of inflationary e-folds fix the scale at 
which the matter power spectrum and 
halo mass function is modified (cf. Section~\ref{sec:jumpc}).
For the minimal number of e-folds required to solve the 
horizon problem for example, $H/k_c \approx 10^{-4}$ is 
required to address the galactic scale issues discussed.
Significant power spectrum modification also require $k_c \lesssim H/k_c$ 
(Section~\ref{sec:pplim}). 
These tests can be stringent;  as, given this scale, and 
the level of excitation determined by the  Boguliubov $\beta_k$, 
the predictions of the simplest scenario of sudden jump
through a high energy transition scale,
are unique in terms of the expected effect on the power spectrum.

\section{Conclusion}
\label{sec:conc}

Slow roll inflation predicts a nearly scale invariant  spectrum 
of primordial fluctuations, which is borne out by precise observations of the cosmic microwave 
background and large scale structure in the universe. Nevertheless, that prediction is not 
unique, a variety of effects invoking discontinuous or phased evolution between slow rolls, 
for example, can lead to anomalous 'features' in the spectrum. Excited states arising
from modes crossing  a high energy cutoff scale can also lead
to significant modifications to the scale free spectrum. Although these are essentially ruled out at the 
scales where the aforementioned observations are effective, the primordial spectrum
is relatively unconstrained on smaller, currently nonlinear scales, where the matter distribution
has collapsed into bound self gravitating objects, washing out the primordial signature 
by largely encoding it in the halo mass function. 

On the contrary, at such scales ---  which span many more octaves of  
observable structure than the three 
that are probed in the linear regime ---  
a variety of issues arise in the context of the standard 
model of structure formation; such as the  'small scale problems' at low redshift 
and the apparent problems involving early galaxy and supermassive black hole formation at higher
$z$, which can be seen as extension of longstanding phenomena requiring 'downsizing' in galaxy formation. 
As these issues arise precisely at the scales where complex baryonic 
physics comes to play a central role in the 
standard scenario of structure formation,  it was natural that extensive investigation of 
solutions in these terms have been pursued. However, as these are also the scales where 
the primordial spectrum of fluctuations is relatively weakly constrained, this aspect,
with its effects and consequences, may also warrant further investigation. 

Here we considered 
the effects of excited states arising from the transiting of fluctuation modes
through a high energy cutoff scale. 
As the power spectrum of primordial fluctuations is effectively an adiabatic 
invariant of inflaton dynamics (Section~\ref{sec:psadiab}), 
adiabatic evolution necessarily leaves the 
nearly scale free spectrum intact. We next considered a simple model of the opposite extreme; of 
a sudden jump across the transition. The initial conditions for the fluctuations 
before the jump are well defined, 
taking the Bunch Davies form, but with propagation speed $c_s \ne 1$. 
An intuitive, simple 'analogue' model approximated by such a transition
corresponds to the case of a gas or lattice where 
sound waves do not propagate at all below the interparticle distance, then 
propagate at an anomalous  speed in an effective macroscopic approximation, before
finally propagating with the standard sound speed and dispersion relation  
as the wavelength become progressively larger than the interparticle distance. 

In this context, the primordial spectrum is invariably enhanced rather than suppressed 
(whether the initial $c_s > 1$ or is  $ < 1$), for all scales undergoing the transition 
through the high energy cutoff (Section~\ref{sec:nonadiab}). 
This is accompanied by strong, tightly spaced oscillations in the power spectrum of generic (as opposed to pure de Sitter) models of inflation, where modes exit the horizon at different phases.  
Numerical calculations
 suggest that sufficiently 
non-adiabatic evolution, leading to significant modification of the power 
spectrum implies
an effectively sudden transition for all $c_s > 1$ and for $ 0.01 \lesssim c_s < 1$
(Appendix~\ref{app:Logistic}). 
The simple model of sudden jump, and its
predictions, are in this range thus generic.   We also considered 
the possibility of a more gradual transition when the aforementioned 
conditions are not satisfied (Appendix~\ref{app:nosud}).  

Given the excitation level induced in the inflaton field, 
and the current comoving scale corresponding to the jump across the high energy 
cutoff scale during inflation, 
the predictions of the simple sudden jump models are essentially unique (in terms of its effect on the 
matter power spectrum, mass variance and the dark matter halo mass function). 
The level of excitation can be quantified through a
Bogoliubov coefficient $\beta_k \ne 0$ for scales that undergo the jump, and is easily evaluated 
in terms of the in and out frequency ratio (or equivalently $c_s$ ratio; Section~\ref{sec:betaom}). 
If assumed to be within a few orders of magnitude of the Planck scale, the jump scale 
corresponds to currently nonlinear scales if inflation proceeds for approximately the number of e-folds necessary to solve the horizon problem. In general, the comoving 
jump scale corresponds to currently nonlinear scales for minimal inflation  
if $H/k_c \sim 10^{-4}$, with smaller ratios allowing for larger e-folds  (Section~\ref{sec:jumpc}).
In this context, the nonlinear scales can be modified, while leaving the standard spectrum intact on linear ones. 

Backreaction bounds on $|\beta|$  must be imposed, as 'over-excitation' of the inflaton 
would result in radiation domination rather than inflation;  these may however 
still  allow for major enhancements of the power spectrum $\sim 1 + 2 \beta^2$
(and oscillations in the generic inflation case). As we discuss in Section~\ref{sec:pplim}, 
this would be generally the case if $k_c \lesssim \left(H/k_c\right) M_{\rm Pl}$. 
Such enhancements can have observable consequences, confirming 
or constraining the effect of excitations on structure formation on nonlinear scales.
In order to impose modifications
on such scales in particular, and still keep the excitations 
from overwhelming the inflaton vacuum state, one thus requires 
$H/k_c \lesssim 10^{-4}$ and $k_c \lesssim \left(H/k_c\right) M_{\rm Pl}$. 
This implies $k_c \lesssim 10^{-4} M_{\rm Pl}$ and 
$H \lesssim 10^{-8} M_{\rm Pl}$. 
A rough estimate of the derivative of the backreaction suggests 
possible modification of the power spectrum on the largest scales,
and may place tight constraints on the comoving 
scale at which enhancement of the small scale power spectrum can occur 
(to about a comoving Mpc; Section~\ref{sec:pplim}). 
That modification on the largest scales
can accompany the changes on small, nonlinear ones,
is an interesting possibility that may be worth studying in detail.

To probe for possible characteristic signatures of the 
modifications on nonlinear scales, 
we evaluate (in Section~\ref{sec:matt_halomass}) the dark halo multiplicity function, 
quantifying the fraction of mass in halos of mass $M$.  
In our fiducial example, the peak,
resulting from power spectrum enhancement, is chosen
to correspond to a few times $10^{12} {M_\odot}$. 
This is the mass scale where the highest 
mass to light ratio is inferred when associating galaxies with halos in 
the context of halo models derived within the standard scenario. 
It is also the scale where issues related to the apparent preponderance 
of early massive galaxies, particularly quiescent ones, appear
(Section~\ref{sec:earlygals}). For relatively small enhancements at small 
redshifts $z$,  the enhancement at larger $z \sim 8$ is dramatic, as such massive 
halos are very rare at these redshifts in the standard scenario. 
The change in the number densities of massive galaxy-hosting halos 
with redshift is also much smaller than in the standard case. 
Combined, these effects  may alleviate the apparent 'impossibly' 
early galaxy formation problem, even in the most extreme form claimed. 

Perhaps more surprisingly,  an enhancement of the spectrum at these intermediate 
nonlinear scales leads to {\it suppression} of small halos at low $z$, thus 
potentially alleviating longstanding issues related to the dearth of small galaxies, including 
those 'too big to fail', in the standard structure formation scenario. This is due to 
the enhanced spectrum leading to overmerging of small mass objects at high $z$, so as 
to lead to a suppression of such objects at low $z$. 

The halo mass function, in itself, cannot place strong constraints on 
enhancements of the primordial power spectrum on currently nonlinear scales, 
as one can  vary the galaxy halo 
occupation number to match the data (in the standard scenario, the early galaxy formation 
issues at high $M$ and $z$ arises because this seems to sometimes require very 
large stellar mass fraction, which the enhanced halo mass function here may resolve; 
Section~\ref{sec:earlygals} ). 
However, combined abundance matching and dynamical 
analysis at low-$z$ can. Halo abundance enhancement at the scales 
considered here appears consistent with such analyses; it  may in fact 
alleviate the apparent overprediction of stellar 
masses for given dynamical mass for galaxies with stellar mass 
$\gtrsim 5 \times 10^{10} M_\odot$ 
(\cite{Maccio_abunmatch20};Section~\ref{sec:HMFz=0}). 

Major modifications in the spectrum of primordial fluctuation are eventually 
encoded in more minor modifications to the nonlinear matter power spectrum, 
as these enter primarily through the modified halo mass function rather than the 
statistics of the spatial distribution. 
Nevertheless, the scenario of an enhanced primordial power spectrum 
at scales corresponding to currently nonlinear ones, can 
also be tested through its signature on halo biasing.  
The fact that more massive halos would be less rare may be 
expected to particularly impact such observables as  
galaxy-mass correlations and 
leasing signals (especially at higher redshift where the  
effects of enhancement at higher mass scales are more prominent).  
To address both aforementioned issues 
 --- of massive high-$z$ massive galaxies and small local  ones --- 
simultaneously in the most severe forms claimed, 
through sudden transitions, also 
entails significant modifications at scales probed by 
Lyman-$\alpha$ observations
(the required modifications are more modest, or at 
scales that may be less constrained, if only  partial 
resolution of both issues is sought or if the assumption 
of sudden transition is relaxed; Section~\ref{sec:Lyman}).

Thus, observations, coupled with modelling and simulations 
with modified spectrum, may place constraints on scenarios invoking enhanced power on currently 
nonlinear scales, distinguishing them from 
baryonic solutions to the same problems. In the context of the 
analytical sudden step model primarily considered here, this includes 
constraints on $H/k_c$, $k_c$, $|\beta|$, and the number of inflationary e-folds, 
as discussed above. Given the field excitation levels (i.e.$|\beta|$) and the comoving 
scale of the high energy transition, the consequences for the matter 
spectrum and halo mass function are essentially unique.  
Variants that could also be tested include those involving phased or discontinuous stages of inflation with 
relatively localized peaks in the primordial spectrum. This will become perhaps more pressing 
if next generation surveys (e.g. employing the JWST) confirm problems related to early galaxy formation. 
On smaller (sub-galactic) scales still, primordial power spectrum enhancement may 
be relevant to early supermassive black hole formation, 
and the formation of the first dark matter objects, and
may be tested through such effects as CMB spectral distortions.

\begin{acknowledgments}
We wish to thank Adel Awad, 
Alexey Golovnev, Sergio E. Joras,  Giovanni Marozzi,
Jerome Martin and Joe Silk for useful discussions, and the referee 
for constructive comments and suggestions. 
This project was supported financially
by the Science and Technology Development Fund (STDF),
Egypt. Grant No. 25859.
\end{acknowledgments}

\appendix

\section{Comparison of simplified model with numerical solution, and the effect of relaxing the sudden jump condition}

In this appendix  we test the approximation of the simplified model of 
Section~\ref{sec:toysud}, introduced  
to evaluate the effect of non-adiabatic transition 
at a high energy cutoff scale 
$k_c$ on the primordial power spectrum.  There are two approximations 
that were invoked; the sudden step and the neglect the term proportional to $-\frac{1}{\eta^{2}}$
in the Mukhanov-Sasaki  equations~(\ref{eq:MS}) and (\ref{eq:MSf}).
We start by examining the latter, then we discuss the former. 

\subsection{Model versus numerical solution of Mukhanov-Sasaki equation with step}
\label{app:nuan}

\begin{figure*}
\includegraphics[width=0.49\textwidth]{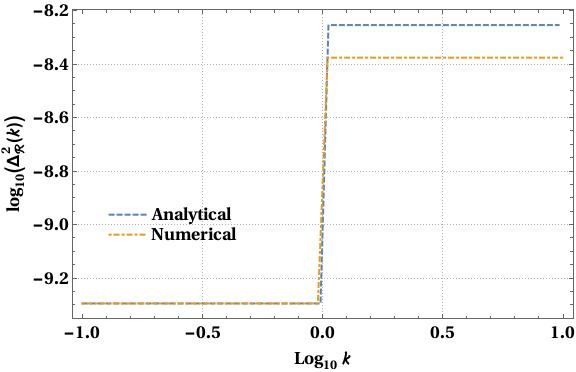}
\includegraphics[width=0.49\textwidth]{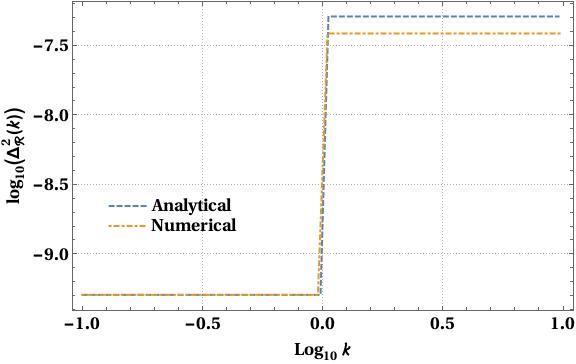}
\caption{Comparison of primordial power spectrum obtained from simplified analytical model 
with full numerical calculation, for sudden jumps corresponding to ratios of the 
in and out sound speed, or equivalently frequencies, of $10$ (left) and $100$ (right).  
This is done for a de Sitter background with $H/k_c = 10^{-4}$ and 
$k$ is shown in units of $k_c$.}
\label{fig: canumdS}
\end{figure*}

\begin{figure*}
\includegraphics[width=0.49\textwidth]{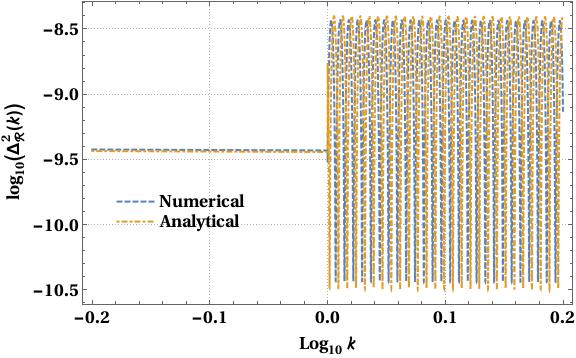}
\includegraphics[width=0.49\textwidth]{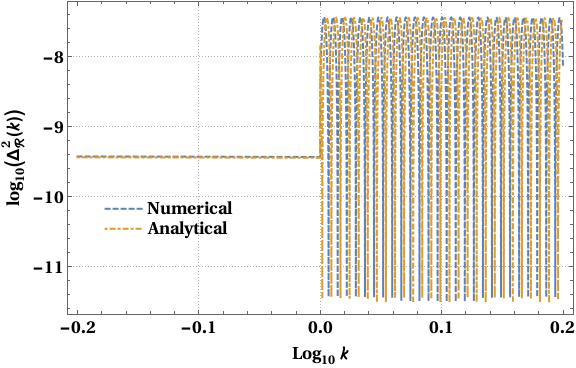}
\caption{Same as in Fig.~\ref{fig: canumdS},  
but for power law inflation model (discussed in Section~\ref{sec:ppseval}).}
\label{fig: canumPL}
\end{figure*}

We evolve the dynamics of fluctuation modes numerically, using the MS equation~(\ref{eq:MS}) for 
de Sitter and power law inflationary backgrounds, while replacing the term $k^{2}$ with
\begin{equation}
k^{2} \rightarrow k_{\mathrm{eff}}^{2}(k, \eta) \equiv a^{2}(\eta) \omega_{\mathrm{phys}}^{2}\left[\frac{k}{a(\eta)}\right].
\end{equation}
where
\begin{equation}
\omega_{\mathrm{phys}}^{2}\left[\frac{k}{a(\eta)}\right]= \left(
\frac{k}{a(\eta )}+\frac{\delta k}{a(\eta)} \mathcal{H}
\left[\frac{k}{k_{c} a(\eta )}-1\right] \right)^{2},
\end{equation} 
and $\mathcal{H}$ is the Heaviside step function. The 
parameter $\delta$ quantifies the size of the step. such that the 
sound speed past the step given by $c_s = 1 + \delta$; it can 
be positive or negative, corresponding to an upward and downward 
jump in sound speed respectively. 
As discussed in Section~\ref{sec:nonadiab}, 
in the context of the sudden step scenario they are equivalent 
in terms of the effect on the power spectrum.

The primordial spectra are evaluated, for both the analytical model and numerical calculations, 
as described in Section~\ref{sec:ppseval}. 
The results are shown and compared 
in figures~\ref{fig: canumdS} and~\ref{fig: canumPL}, for the 
case of de Sitter and power law inflation respectively. As noted in Section~\ref{sec:ppseval}, 
in the de Sitter case all modes exit the horizon 
at exactly the same phase. And 
any initial shift in phase, due to change in effective frequency 
related to the second term in bracket of the MS equation, 
leads to corresponding constant difference in the final power spectrum. This leads to a difference 
between the numerical and analytical results, where 
the aforementioned term is neglected. Nevertheless, the relative 
error in the ratio of the perturbed to unperturbed power spectrum is still of 
order $25 \%$ when the  step frequency ratio is 
$10$. It is an order of magnitude 
lower still when the change of frequency is hundred folds. 

In the case of power law inflation the Hubble scale $H$ is not exactly constant. 
The modes exit the horizon at different phases, and this leads to the oscillations, which accompany 
the enhancement in Fig.~\ref{fig: canumPL}. The corresponding error is then primarily in phases, 
with the maxima and minima of the oscillations practically equal in the 
simplified model and the numerical calculations. The change in phase is 
generally unimportant for calculating quantities with observable consequences; such as the mass 
fluctuations at a given spatial or mass scale, and halo mass multiplicity function. For
these depend on integrals of the matter power spectrum
(as discussed in Section~\ref{sec:matt_halomass}).
The simple analytical model --- with its simple interpretation in terms 
of well defined in and out states; Section~\ref{sec:nonadiab} ---
thus turns out to be a good approximation.

\begin{figure*}
\includegraphics[width=0.49\textwidth]{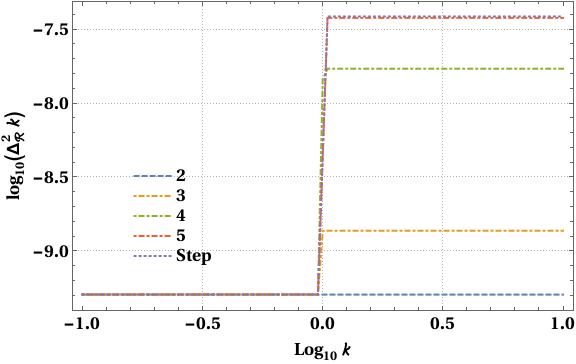}
\includegraphics[width=0.49\textwidth]{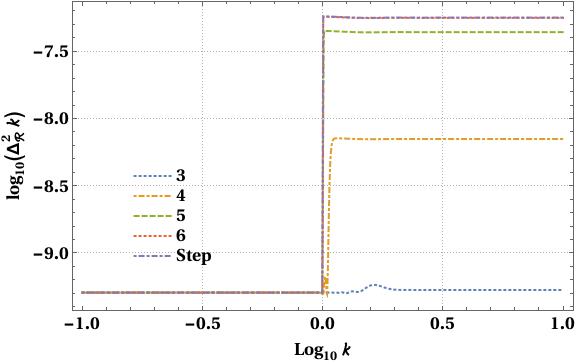}
\caption{The primordial power spectrum evaluated at different levels of violations of the adiabaticty condition~(\ref{eq:adiabcon}),  
when transition across the high energy cutoff scale is interpolated using a logistic function 
(Eq.~\ref{eq:Logistic}), and numerically integrated. 
The numbers in the legend keys refer to the order of magnitude above the critical value of $\gamma$ 
required to violate the adiabaticity condition (Eq.~\ref{eq:gammal}). 
Left panel: interpolation between high energy sound speed $c_s =100$ and standard regime ($c_s =1$).
Right panel: interpolation between high energy sound speed $c_s =0.01$ and standard regime. 
The results are shown for de Sitter inflation, and wavenumbers on the horizontal axis 
are expressed in terms of the high energy cutoff scale $k_c$, with $k_c/H = 300$.}
\label{fig:Logistic}
\end{figure*}

\subsection{Non-adiabaticity versus sudden jump condition}
\label{app:Logistic}

Oscillations can, in general, be considered adiabatic 
if $\omega(\eta)$  changes only slightly  over  a characteristic time $\Delta \eta$ 
of order of one oscillation period. 
If the frequency $\omega$ changes from a value $\omega_{1}$  to another  value $\omega_{2}$, on a  characteristic 
timescale  $\Delta \eta$,  the change  may thus be adiabatic if
\begin{equation}
|\omega_{1}-\omega_{2}|  < \omega^2 \Delta \eta.
\label{eq:adiabcon}
\end{equation}
In the context of our sudden step model, $\omega_1$ and $\omega_2$ will correspond 
to $\omega_{\rm in}$ and $\omega_{\rm out}$, respectively. 
We take the 'typical' $\omega$ on the right hand side to correspond to the minimal frequency; 
supposing that the dynamics may be affected non-adibaticaly if the change in frequency 
is larger than this. 
To examine to what extent that model may describe a more general situation, where 
change may be more gradual, we use the logistic function to parametrize the transition:
\begin{equation}
k_{\rm eff} \equiv  \omega(\eta)=k+\frac{\delta k}{1+\exp \left[-\gamma \left(\frac{\eta}{\eta_{c}}-1\right)\right]}.
\label{eq:Logistic}
\end{equation}
Here the parameter $\gamma$ describes the stiffness of the transition, 
this being steep and steplike for $\gamma \gg 1$, and $\delta$ (which may be positive or negative)
the scale of the step in the transition. Thus, in the high energy regime limit, the sound 
speed $c_s = 1 + \delta$, 
while $c_s =1$  when the transition to the standard low-energy physics regime is complete. 
In these terms,  the characteristic time over which $\omega$
changes between its initial and final value 
is $\frac{\eta_{c}}{\gamma}$. The adiabaticity violation
 condition can then be written as 
 \begin{equation}
\gamma > \frac{{\rm Min}(c_s^2)}{|c_s - 1|} \left(\frac{k_c}{H}\right),
\label{eq:gammal}
\end{equation}
 where we have used $\eta_c = \frac{-k_c}{H k}$, and $c_s \ne 1$
corresponds to the high energy limit sound speed.
 Since, as we have seen in Section~\ref{sec:psadiab}, the power spectrum 
is essentially an adiabatic invariant of the dynamics of inflationary 
perturbations, it is necessary to satisfy this condition in order 
to modify the standard power spectrum. 

Two cases are of particular interest in the context of the 
present study: $c_s \gg 1$, so that ${\rm Min} (c_s) = 1$ and 
$c_c \ll 1$, when ${\rm Min} (c_s) = c_s$.
For sound speeds considered here, the adiaticity condition
(\ref{eq:gammal}) is violated  at smallest possible 
when the sound speed is minimal, that is $c_s =0.01$. 
Still, even in this case,  $\gamma$ is of order one 
or larger if $k_c/H \ge 10^4$, as required to keep the 
significant changes in power spectrum in the nonlinear regime of structure 
formation (Section~\ref{sec:jumpc}). For our parameters, the transition
is thus necessarily stiff.

\begin{figure*}
\includegraphics[width=8.5 cm, height = 5.8 cm]{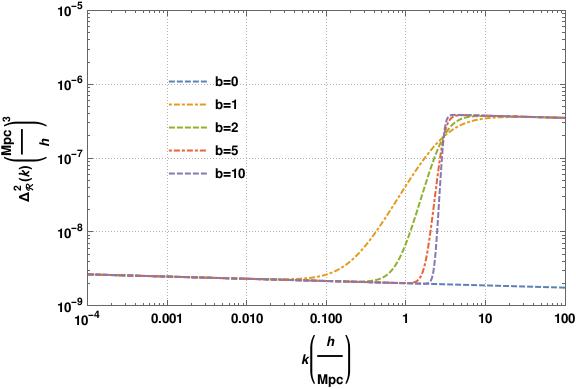}
\includegraphics[width=8.5 cm, height = 5.8 cm]{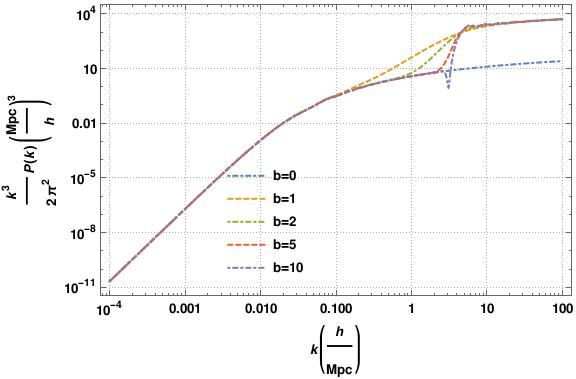}
\caption{Primordial (left) and matter (right)  power spectra, modified at 
small scales using Eqs.~(\ref{eq:parfun}) and (\ref{eq:gradual}), showing 
progressively  steeper transitions with increasing $b$. 
The characteristic high energy transition scale is taken as 
$k_c = 3 h {\rm Mpc}^{-1}$ comoving, and  $S = 200$.}
\label{fig: gradualPS}
\end{figure*}

We now show that the transition is stiff even form much smaller $k_c/H = 300$, which is 
a minimal value, in the sense that with smaller values 
the effect of the second term in brackets 
of Eq.~(\ref{eq:MSf}) becomes important at $\eta \sim \eta_c$
for $c_s =0.01$
(see discussion relating to inequality ~\ref{eq:noimcon}). 
For it turns out 
that the adiabaticty condition needs to be quite strongly violated for sufficient 
change in the dynamics significantly affects 
the power spectrum (that significant changes occurs well beyond the 
adiabaticity breaking condition is common in dynamical systems~\cite{BT2008}). 
This can be seen from Fig.~\ref{fig:Logistic}, where we show that 
large changes only occur 
when $\gamma$ is orders of magnitudes above the value
estimated from~(\ref{eq:gammal}). This is the case for
both the $c_s = 100$ and $c_s =0.01$, with the former being stiffer still as 
expected from~(\ref{eq:gammal})~\footnote{Note also, in the latter case of $c_s =0.01$, 
the small bump on the right, 
visible in the low $\gamma$ spectra beyond the cutoff scale. These are due to non-adibaticity at
the crossing of the high energy transition at $\eta_c$. They are much less prominent 
for smaller $k_c/H \ge 10^{4}$ values, used in the rest of this work. 
They nevertheless represent another potentially interesting effect associated with non-adiabatic evolution}. 
The transition is stiffer still for smaller
values of $c_s > 1$ and larger $c_s < 1$. 
Thus, for sound speed ratios considered here, 
our simple model of a sudden jump, and accompanying signature of a 
sudden break in the power spectrum, 
appears much more generic than may seem {\it a priory}.

\section{Modified mass function from non-sudden spectrum enhancements}
\label{app:nosud}

\begin{figure*}
\includegraphics[width=0.49\textwidth]{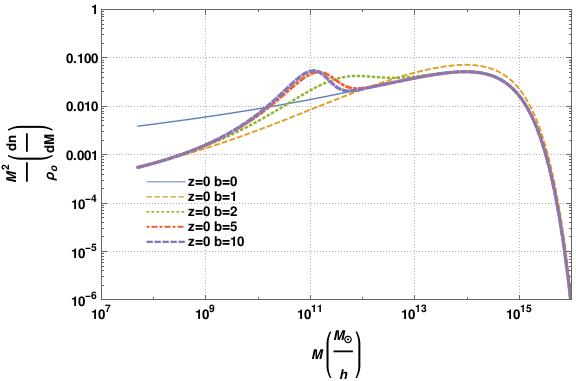}
\includegraphics[width=0.49\textwidth]{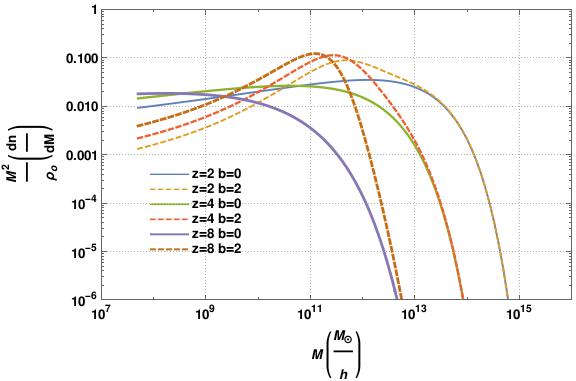}
\caption{Left panel: the halo multiplicity functions 
at $z=0$ for the spectra shown in Fig.~\ref{fig: gradualPS}.
Right panel: the multiplicity functions at shown redshifts 
for the case $b= 2$.}
\label{fig: gradualHMF}
\end{figure*}

As we have seen in the previous appendix, the sudden jump transition in the power 
spectrum at smaller scales is a good approximation for the parameters primarily considered 
in this paper, namely for initial $c_s \ge 0.01$; for such values, the sufficient violation 
of the adiabaticity condition,  required for significant modification 
of the power spectrum, practically implies a sudden transition. Nevertheless, 
as is already apparent in Fig.~\ref{fig:Logistic},  for $c_s = 0,01$ (right hand panel), the 
sudden jump approximation becomes less accurate as a predictor of 
significant change at smaller $c_s$. If one envisages a transition starting at significantly  
smaller  sound speed still, it may then take place more gradually while still imparting 
a palpable effect on the power spectrum.

We consider potentially observable consequences of this effect  by 
examining a series of progressively steeper transitions. We do this by 
modifying the primordial power spectrum in a parametric manner, such 
that 
\begin{equation}
    \Delta_{\mathcal{R}}^{2}(k) =  \Delta_{\mathcal{R} {\rm St}}^{2}(k) 
     \left[(S-1)~G \left(\frac{k}{k_c}\right) + 1\right],  
\label{eq:parfun}
\end{equation}
where $\Delta_{\mathcal{R} {\rm St}}^{2}(k)$ is the standard, nearly scale invariant, 
power spectrum of scalar perturbations. The transition function $G$ 
tends to unity as $k \gg k_c$ and vanishes 
as $k/k_c \rightarrow 0$, and $S\ge 1$ is the enhancement factor
(it determines the ratio of the asymptotic values of the power spectrum 
at small and large scales). 
We have tried several forms for $G$, and the resulting trends were 
verified to be generic. Here we show results for the following form 
(also used in~\cite{JensKam17} for the purpose of suppression of the spectrum rather 
than enhancement): 
\begin{equation}
 G (x) = \frac{1}{2} \left[\tanh~(b ~\log x) + 1\right]
 =     \frac{1}{2} 
\left[ \frac{x^{2 b}  -1}{x^{2 b}  +1} + 1\right],
\label{eq:gradual}
\end{equation}
where $b > 0$ determines the steepness of the transition around 
$x = k/k_c$. In the following we will take $k_c$ to correspond to 
a comoving scale of $3 h~{\rm Mpc}^{-1}$.  
As equations~(\ref{eq:powerat} )and~(\ref{eq:beta}) show
$S$ to be about $50$ for a sudden step scenario with 
initial $c_s = 0.01$, 
the requirement that the initial 
$c_s < 0.01$ implies 
$S > 50$ (otherwise, in line with the aforementioned considerations, 
a gradual transition  may not have 
a significant effect on the power spectrum). 
In what follows we take $S = 200$. 

Fig.~\ref{fig: gradualPS}  shows the primordial power spectrum, as well as the dimensionless 
matter power spectrum calculated using publicly available 
CLASS code (class-code.net), for several values 
of $b$. The corresponding multiplicity functions, calculated using (\ref{eq:HMF}) 
and~(\ref{eq:fST}), are shown in Fig.~\ref{fig: gradualHMF}, 
where the left hand panel displays results at  $z =0$.  
For large $b$, those results are similar to the sudden jump case. 
For $b = 1$ the effect is smeared out, with increase at 
halo masses  $\gtrsim 10^{14} M_\odot$, which would increase 
tension with cluster counts, which is already present in thee standard model. 
The suppression at small mass scales is enhanced by the gradual 
transition. 

Of particular interest is the intermediate, $b = 2$, case. 
The enhancement takes place at larger masses than the corresponding
case with sudden jump (at $3 h {\rm Mpc}^{-1}$). The suppression at $z=0$ takes place at larger masses 
as well. This allows for largely retaining the advantages of the sharp cutoff 
at $1 h {\rm Mpc}^{-1}$  --- in terms of simultaneously alleviating both the dearth of dearth of 
dwarf galaxy and too big to fail problems at $z =0$, as well as 
accounting for early galaxy formation at higher redshifts ---
while avoiding any enhancement at scales of order $\gtrsim 10^{13} M_\odot$ at $z = 0$, and 
relatively mildly perturbing the matter power spectrum at $1 h {\rm Mpc}^{-1}$ 
(Fig.~\ref{fig: gradualPS}; see also Section~\ref{sec:Lyman}). 
The enhancement of the mass function at $z=0$  
may also  be relevant for explaining the overestimation of abundance matching
within the standard model of stellar masses of massive galaxies 
(\cite{Maccio_abunmatch20}; Section~\ref{sec:HMFz=0}).


\bibliography{bibliokk}

\begin{thebibliography}{186}%
\makeatletter
\providecommand \@ifxundefined [1]{%
 \@ifx{#1\undefined}
}%
\providecommand \@ifnum [1]{%
 \ifnum #1\expandafter \@firstoftwo
 \else \expandafter \@secondoftwo
 \fi
}%
\providecommand \@ifx [1]{%
 \ifx #1\expandafter \@firstoftwo
 \else \expandafter \@secondoftwo
 \fi
}%
\providecommand \natexlab [1]{#1}%
\providecommand \enquote  [1]{``#1''}%
\providecommand \bibnamefont  [1]{#1}%
\providecommand \bibfnamefont [1]{#1}%
\providecommand \citenamefont [1]{#1}%
\providecommand \href@noop [0]{\@secondoftwo}%
\providecommand \href [0]{\begingroup \@sanitize@url \@href}%
\providecommand \@href[1]{\@@startlink{#1}\@@href}%
\providecommand \@@href[1]{\endgroup#1\@@endlink}%
\providecommand \@sanitize@url [0]{\catcode `\\12\catcode `\$12\catcode
  `\&12\catcode `\#12\catcode `\^12\catcode `\_12\catcode `\%12\relax}%
\providecommand \@@startlink[1]{}%
\providecommand \@@endlink[0]{}%
\providecommand \url  [0]{\begingroup\@sanitize@url \@url }%
\providecommand \@url [1]{\endgroup\@href {#1}{\urlprefix }}%
\providecommand \urlprefix  [0]{URL }%
\providecommand \Eprint [0]{\href }%
\providecommand \doibase [0]{https://doi.org/}%
\providecommand \selectlanguage [0]{\@gobble}%
\providecommand \bibinfo  [0]{\@secondoftwo}%
\providecommand \bibfield  [0]{\@secondoftwo}%
\providecommand \translation [1]{[#1]}%
\providecommand \BibitemOpen [0]{}%
\providecommand \bibitemStop [0]{}%
\providecommand \bibitemNoStop [0]{.\EOS\space}%
\providecommand \EOS [0]{\spacefactor3000\relax}%
\providecommand \BibitemShut  [1]{\csname bibitem#1\endcsname}%
\let\auto@bib@innerbib\@empty
\bibitem [{\citenamefont {Lyth}\ and\ \citenamefont
  {Liddle}(2009)}]{LiddleLythB}%
  \BibitemOpen
  \bibfield  {author} {\bibinfo {author} {\bibfnamefont {D.~H.}\ \bibnamefont
  {Lyth}}\ and\ \bibinfo {author} {\bibfnamefont {A.~R.}\ \bibnamefont
  {Liddle}},\ }\href@noop {} {\emph {\bibinfo {title} {{The primordial density
  perturbation: cosmology, inflation and the origin of structure; rev.
  version}}}}\ (\bibinfo  {publisher} {Cambridge Univ. Press},\ \bibinfo
  {address} {Cambridge},\ \bibinfo {year} {2009})\BibitemShut {NoStop}%
\bibitem [{\citenamefont {Peter}\ and\ \citenamefont {Uzan}(2009)}]{UzanB}%
  \BibitemOpen
  \bibfield  {author} {\bibinfo {author} {\bibfnamefont {P.}~\bibnamefont
  {Peter}}\ and\ \bibinfo {author} {\bibfnamefont {J.-P.}\ \bibnamefont
  {Uzan}},\ }\href {https://doi.org/019966515X} {\emph {\bibinfo {title}
  {{Primordial cosmology}}}},\ Oxford Graduate Texts\ (\bibinfo  {publisher}
  {Oxford Univ. Press},\ \bibinfo {address} {Oxford},\ \bibinfo {year}
  {2009})\BibitemShut {NoStop}%
\bibitem [{\citenamefont {{Planck Collaboration}}\ \emph
  {et~al.}(2018)\citenamefont {{Planck Collaboration}}, \citenamefont
  {{Akrami}} \emph {et~al.}}]{Planck18}%
  \BibitemOpen
  \bibfield  {author} {\bibinfo {author} {\bibnamefont {{Planck
  Collaboration}}}, \bibinfo {author} {\bibfnamefont {Y.}~\bibnamefont
  {{Akrami}}}, \emph {et~al.},\ }\href@noop {} {\bibfield  {journal} {\bibinfo
  {journal} {arXiv e-prints}\ ,\ \bibinfo {eid} {arXiv:1807.06211}} (\bibinfo
  {year} {2018})},\ \Eprint {https://arxiv.org/abs/1807.06211}
  {arXiv:1807.06211 [astro-ph.CO]} \BibitemShut {NoStop}%
\bibitem [{\citenamefont {Salopek}\ \emph {et~al.}(1989)\citenamefont
  {Salopek}, \citenamefont {Bond},\ and\ \citenamefont {Bardeen}}]{SaloBond89}%
  \BibitemOpen
  \bibfield  {author} {\bibinfo {author} {\bibfnamefont {D.~S.}\ \bibnamefont
  {Salopek}}, \bibinfo {author} {\bibfnamefont {J.~R.}\ \bibnamefont {Bond}},\
  and\ \bibinfo {author} {\bibfnamefont {J.~M.}\ \bibnamefont {Bardeen}},\
  }\href {https://doi.org/10.1103/PhysRevD.40.1753} {\bibfield  {journal}
  {\bibinfo  {journal} {Phys. Rev.}\ }\textbf {\bibinfo {volume} {D40}},\
  \bibinfo {pages} {1753} (\bibinfo {year} {1989})}\BibitemShut {NoStop}%
\bibitem [{\citenamefont {Dodelson}\ and\ \citenamefont
  {Stewart}(2002)}]{Dod_Slow_Tilt}%
  \BibitemOpen
  \bibfield  {author} {\bibinfo {author} {\bibfnamefont {S.}~\bibnamefont
  {Dodelson}}\ and\ \bibinfo {author} {\bibfnamefont {E.}~\bibnamefont
  {Stewart}},\ }\href {https://doi.org/10.1103/PhysRevD.65.101301} {\bibfield
  {journal} {\bibinfo  {journal} {Phys. Rev.}\ }\textbf {\bibinfo {volume}
  {D65}},\ \bibinfo {pages} {101301} (\bibinfo {year} {2002})},\ \Eprint
  {https://arxiv.org/abs/astro-ph/0109354} {arXiv:astro-ph/0109354 [astro-ph]}
  \BibitemShut {NoStop}%
\bibitem [{\citenamefont {{Adams}}\ \emph {et~al.}(1997)\citenamefont
  {{Adams}}, \citenamefont {{Ross}},\ and\ \citenamefont
  {{Sarkar}}}]{BrokSuperg}%
  \BibitemOpen
  \bibfield  {author} {\bibinfo {author} {\bibfnamefont {J.~A.}\ \bibnamefont
  {{Adams}}}, \bibinfo {author} {\bibfnamefont {G.~G.}\ \bibnamefont
  {{Ross}}},\ and\ \bibinfo {author} {\bibfnamefont {S.}~\bibnamefont
  {{Sarkar}}},\ }\href {https://doi.org/10.1016/S0370-2693(96)01484-0}
  {\bibfield  {journal} {\bibinfo  {journal} {Physics Letters B}\ }\textbf
  {\bibinfo {volume} {391}},\ \bibinfo {pages} {271} (\bibinfo {year}
  {1997})},\ \Eprint {https://arxiv.org/abs/hep-ph/9608336}
  {arXiv:hep-ph/9608336 [hep-ph]} \BibitemShut {NoStop}%
\bibitem [{\citenamefont {Chung}\ \emph {et~al.}(2000)\citenamefont {Chung},
  \citenamefont {Kolb}, \citenamefont {Riotto},\ and\ \citenamefont
  {Tkachev}}]{KolbResPart00}%
  \BibitemOpen
  \bibfield  {author} {\bibinfo {author} {\bibfnamefont {D.~J.~H.}\
  \bibnamefont {Chung}}, \bibinfo {author} {\bibfnamefont {E.~W.}\ \bibnamefont
  {Kolb}}, \bibinfo {author} {\bibfnamefont {A.}~\bibnamefont {Riotto}},\ and\
  \bibinfo {author} {\bibfnamefont {I.~I.}\ \bibnamefont {Tkachev}},\ }\href
  {https://doi.org/10.1103/PhysRevD.62.043508} {\bibfield  {journal} {\bibinfo
  {journal} {Phys. Rev.}\ }\textbf {\bibinfo {volume} {D62}},\ \bibinfo {pages}
  {043508} (\bibinfo {year} {2000})},\ \Eprint
  {https://arxiv.org/abs/hep-ph/9910437} {arXiv:hep-ph/9910437 [hep-ph]}
  \BibitemShut {NoStop}%
\bibitem [{\citenamefont {Burgess}\ \emph {et~al.}(2003)\citenamefont
  {Burgess}, \citenamefont {Cline}, \citenamefont {Lemieux},\ and\
  \citenamefont {Holman}}]{McGuill_HEC02}%
  \BibitemOpen
  \bibfield  {author} {\bibinfo {author} {\bibfnamefont {C.~P.}\ \bibnamefont
  {Burgess}}, \bibinfo {author} {\bibfnamefont {J.~M.}\ \bibnamefont {Cline}},
  \bibinfo {author} {\bibfnamefont {F.}~\bibnamefont {Lemieux}},\ and\ \bibinfo
  {author} {\bibfnamefont {R.}~\bibnamefont {Holman}},\ }\href
  {https://doi.org/10.1088/1126-6708/2003/02/048} {\bibfield  {journal}
  {\bibinfo  {journal} {JHEP}\ }\textbf {\bibinfo {volume} {02}},\ \bibinfo
  {pages} {048}},\ \Eprint {https://arxiv.org/abs/hep-th/0210233}
  {arXiv:hep-th/0210233 [hep-th]} \BibitemShut {NoStop}%
\bibitem [{\citenamefont {Romano}\ and\ \citenamefont
  {Sasaki}(2008)}]{Sasaki08}%
  \BibitemOpen
  \bibfield  {author} {\bibinfo {author} {\bibfnamefont {A.~E.}\ \bibnamefont
  {Romano}}\ and\ \bibinfo {author} {\bibfnamefont {M.}~\bibnamefont
  {Sasaki}},\ }\href {https://doi.org/10.1103/PhysRevD.78.103522} {\bibfield
  {journal} {\bibinfo  {journal} {Phys. Rev.}\ }\textbf {\bibinfo {volume}
  {D78}},\ \bibinfo {pages} {103522} (\bibinfo {year} {2008})},\ \Eprint
  {https://arxiv.org/abs/0809.5142} {arXiv:0809.5142 [gr-qc]} \BibitemShut
  {NoStop}%
\bibitem [{\citenamefont {{Lyth}}\ and\ \citenamefont
  {{Riotto}}(1999)}]{PartPhysModInf}%
  \BibitemOpen
  \bibfield  {author} {\bibinfo {author} {\bibfnamefont {D.~H.~D.~H.}\
  \bibnamefont {{Lyth}}}\ and\ \bibinfo {author} {\bibfnamefont {A.~A.}\
  \bibnamefont {{Riotto}}},\ }\href
  {https://doi.org/10.1016/S0370-1573(98)00128-8} {\bibfield  {journal}
  {\bibinfo  {journal} {Phys. Rep.}\ }\textbf {\bibinfo {volume} {314}},\
  \bibinfo {pages} {1} (\bibinfo {year} {1999})},\ \Eprint
  {https://arxiv.org/abs/hep-ph/9807278} {arXiv:hep-ph/9807278 [hep-ph]}
  \BibitemShut {NoStop}%
\bibitem [{\citenamefont {{Yamaguchi}}(2011)}]{SupergRev}%
  \BibitemOpen
  \bibfield  {author} {\bibinfo {author} {\bibfnamefont {M.}~\bibnamefont
  {{Yamaguchi}}},\ }\href {https://doi.org/10.1088/0264-9381/28/10/103001}
  {\bibfield  {journal} {\bibinfo  {journal} {Classical and Quantum Gravity}\
  }\textbf {\bibinfo {volume} {28}},\ \bibinfo {eid} {103001} (\bibinfo {year}
  {2011})},\ \Eprint {https://arxiv.org/abs/1101.2488} {arXiv:1101.2488
  [astro-ph.CO]} \BibitemShut {NoStop}%
\bibitem [{\citenamefont {{Baumann}}\ and\ \citenamefont
  {{McAllister}}(2015)}]{InflStr}%
  \BibitemOpen
  \bibfield  {author} {\bibinfo {author} {\bibfnamefont {D.}~\bibnamefont
  {{Baumann}}}\ and\ \bibinfo {author} {\bibfnamefont {L.}~\bibnamefont
  {{McAllister}}},\ }\href@noop {} {\emph {\bibinfo {title} {{Inflation and
  String Theory}}}}\ (\bibinfo {year} {2015})\BibitemShut {NoStop}%
\bibitem [{\citenamefont {{Starobinskij}}(1992)}]{Starob92}%
  \BibitemOpen
  \bibfield  {author} {\bibinfo {author} {\bibfnamefont {A.~A.}\ \bibnamefont
  {{Starobinskij}}},\ }\href@noop {} {\bibfield  {journal} {\bibinfo  {journal}
  {Soviet Journal of Experimental and Theoretical Physics Letters}\ }\textbf
  {\bibinfo {volume} {55}},\ \bibinfo {pages} {489} (\bibinfo {year}
  {1992})}\BibitemShut {NoStop}%
\bibitem [{\citenamefont {{Polarski}}\ and\ \citenamefont
  {{Starobinsky}}(1992)}]{StarobPol}%
  \BibitemOpen
  \bibfield  {author} {\bibinfo {author} {\bibfnamefont {D.}~\bibnamefont
  {{Polarski}}}\ and\ \bibinfo {author} {\bibfnamefont {A.~A.}\ \bibnamefont
  {{Starobinsky}}},\ }\href {https://doi.org/10.1016/0550-3213(92)90062-G}
  {\bibfield  {journal} {\bibinfo  {journal} {Nuclear Physics B}\ }\textbf
  {\bibinfo {volume} {385}},\ \bibinfo {pages} {623} (\bibinfo {year}
  {1992})}\BibitemShut {NoStop}%
\bibitem [{\citenamefont {{Adams}}\ \emph {et~al.}(2001)\citenamefont
  {{Adams}}, \citenamefont {{Cresswell}},\ and\ \citenamefont
  {{Easther}}}]{StaroStepnum}%
  \BibitemOpen
  \bibfield  {author} {\bibinfo {author} {\bibfnamefont {J.}~\bibnamefont
  {{Adams}}}, \bibinfo {author} {\bibfnamefont {B.}~\bibnamefont
  {{Cresswell}}},\ and\ \bibinfo {author} {\bibfnamefont {R.}~\bibnamefont
  {{Easther}}},\ }\href {https://doi.org/10.1103/PhysRevD.64.123514} {\bibfield
   {journal} {\bibinfo  {journal} {\prd}\ }\textbf {\bibinfo {volume} {64}},\
  \bibinfo {pages} {123514} (\bibinfo {year} {2001})},\ \Eprint
  {https://arxiv.org/abs/astro-ph/0102236} {arXiv:astro-ph/0102236 [astro-ph]}
  \BibitemShut {NoStop}%
\bibitem [{\citenamefont {Lesgourgues}(2000)}]{LesgourguesBSI00}%
  \BibitemOpen
  \bibfield  {author} {\bibinfo {author} {\bibfnamefont {J.}~\bibnamefont
  {Lesgourgues}},\ }\href {https://doi.org/10.1016/S0550-3213(00)00301-1}
  {\bibfield  {journal} {\bibinfo  {journal} {Nucl. Phys.}\ }\textbf {\bibinfo
  {volume} {B582}},\ \bibinfo {pages} {593} (\bibinfo {year} {2000})},\ \Eprint
  {https://arxiv.org/abs/hep-ph/9911447} {arXiv:hep-ph/9911447 [hep-ph]}
  \BibitemShut {NoStop}%
\bibitem [{\citenamefont {Yamaguchi}(2001{\natexlab{a}})}]{YamaguchiSuper01}%
  \BibitemOpen
  \bibfield  {author} {\bibinfo {author} {\bibfnamefont {M.}~\bibnamefont
  {Yamaguchi}},\ }\href {https://doi.org/10.1103/PhysRevD.64.063502} {\bibfield
   {journal} {\bibinfo  {journal} {Phys. Rev.}\ }\textbf {\bibinfo {volume}
  {D64}},\ \bibinfo {pages} {063502} (\bibinfo {year} {2001}{\natexlab{a}})},\
  \Eprint {https://arxiv.org/abs/hep-ph/0103045} {arXiv:hep-ph/0103045
  [hep-ph]} \BibitemShut {NoStop}%
\bibitem [{\citenamefont {Feng}\ \emph {et~al.}(2003)\citenamefont {Feng},
  \citenamefont {Li}, \citenamefont {Zhang},\ and\ \citenamefont
  {Zhang}}]{CouplingFieldsBSI03}%
  \BibitemOpen
  \bibfield  {author} {\bibinfo {author} {\bibfnamefont {B.}~\bibnamefont
  {Feng}}, \bibinfo {author} {\bibfnamefont {M.-z.}\ \bibnamefont {Li}},
  \bibinfo {author} {\bibfnamefont {R.-J.}\ \bibnamefont {Zhang}},\ and\
  \bibinfo {author} {\bibfnamefont {X.-m.}\ \bibnamefont {Zhang}},\ }\href
  {https://doi.org/10.1103/PhysRevD.68.103511} {\bibfield  {journal} {\bibinfo
  {journal} {Phys. Rev.}\ }\textbf {\bibinfo {volume} {D68}},\ \bibinfo {pages}
  {103511} (\bibinfo {year} {2003})},\ \Eprint
  {https://arxiv.org/abs/astro-ph/0302479} {arXiv:astro-ph/0302479 [astro-ph]}
  \BibitemShut {NoStop}%
\bibitem [{\citenamefont {Elgaroy}\ \emph {et~al.}(2003)\citenamefont
  {Elgaroy}, \citenamefont {Hannestad},\ and\ \citenamefont
  {Haugboelle}}]{Hannestad_Res03}%
  \BibitemOpen
  \bibfield  {author} {\bibinfo {author} {\bibfnamefont {O.}~\bibnamefont
  {Elgaroy}}, \bibinfo {author} {\bibfnamefont {S.}~\bibnamefont {Hannestad}},\
  and\ \bibinfo {author} {\bibfnamefont {T.}~\bibnamefont {Haugboelle}},\
  }\href {https://doi.org/10.1088/1475-7516/2003/09/008} {\bibfield  {journal}
  {\bibinfo  {journal} {J. Cosmol. Astropart. Phys.}\ }\textbf {\bibinfo
  {volume} {0309}},\ \bibinfo {pages} {008}},\ \Eprint
  {https://arxiv.org/abs/astro-ph/0306229} {arXiv:astro-ph/0306229 [astro-ph]}
  \BibitemShut {NoStop}%
\bibitem [{\citenamefont {Gong}(2005)}]{Gong05}%
  \BibitemOpen
  \bibfield  {author} {\bibinfo {author} {\bibfnamefont {J.-O.}\ \bibnamefont
  {Gong}},\ }\href {https://doi.org/10.1088/1475-7516/2005/07/015} {\bibfield
  {journal} {\bibinfo  {journal} {J. Cosmol. Astropart. Phys.}\ }\textbf
  {\bibinfo {volume} {0507}},\ \bibinfo {pages} {015}},\ \Eprint
  {https://arxiv.org/abs/astro-ph/0504383} {arXiv:astro-ph/0504383 [astro-ph]}
  \BibitemShut {NoStop}%
\bibitem [{\citenamefont {Dvorkin}\ and\ \citenamefont {Hu}(2010)}]{DvorHu09}%
  \BibitemOpen
  \bibfield  {author} {\bibinfo {author} {\bibfnamefont {C.}~\bibnamefont
  {Dvorkin}}\ and\ \bibinfo {author} {\bibfnamefont {W.}~\bibnamefont {Hu}},\
  }\href {https://doi.org/10.1103/PhysRevD.81.023518} {\bibfield  {journal}
  {\bibinfo  {journal} {Phys. Rev.}\ }\textbf {\bibinfo {volume} {D81}},\
  \bibinfo {pages} {023518} (\bibinfo {year} {2010})},\ \Eprint
  {https://arxiv.org/abs/0910.2237} {arXiv:0910.2237 [astro-ph.CO]}
  \BibitemShut {NoStop}%
\bibitem [{\citenamefont {Namjoo}\ \emph {et~al.}(2012)\citenamefont {Namjoo},
  \citenamefont {Firouzjahi},\ and\ \citenamefont {Sasaki}}]{Multi_Sasaki12}%
  \BibitemOpen
  \bibfield  {author} {\bibinfo {author} {\bibfnamefont {M.~H.}\ \bibnamefont
  {Namjoo}}, \bibinfo {author} {\bibfnamefont {H.}~\bibnamefont {Firouzjahi}},\
  and\ \bibinfo {author} {\bibfnamefont {M.}~\bibnamefont {Sasaki}},\ }\href
  {https://doi.org/10.1088/1475-7516/2012/12/018} {\bibfield  {journal}
  {\bibinfo  {journal} {J. Cosmol. Astropart. Phys.}\ }\textbf {\bibinfo
  {volume} {1212}},\ \bibinfo {pages} {018}},\ \Eprint
  {https://arxiv.org/abs/1207.3638} {arXiv:1207.3638 [hep-th]} \BibitemShut
  {NoStop}%
\bibitem [{\citenamefont {{Lesgourgues}}\ \emph {et~al.}(1998)\citenamefont
  {{Lesgourgues}}, \citenamefont {{Polarski}},\ and\ \citenamefont
  {{Starobinsky}}}]{StaroCluster98}%
  \BibitemOpen
  \bibfield  {author} {\bibinfo {author} {\bibfnamefont {J.}~\bibnamefont
  {{Lesgourgues}}}, \bibinfo {author} {\bibfnamefont {D.}~\bibnamefont
  {{Polarski}}},\ and\ \bibinfo {author} {\bibfnamefont {A.~A.}\ \bibnamefont
  {{Starobinsky}}},\ }\href {https://doi.org/10.1046/j.1365-8711.1998.01537.x}
  {\bibfield  {journal} {\bibinfo  {journal} {Mon. Not. R. Ast. Soc.}\ }\textbf
  {\bibinfo {volume} {297}},\ \bibinfo {pages} {769} (\bibinfo {year}
  {1998})},\ \Eprint {https://arxiv.org/abs/astro-ph/9711139}
  {arXiv:astro-ph/9711139 [astro-ph]} \BibitemShut {NoStop}%
\bibitem [{\citenamefont {{Gramann}}\ and\ \citenamefont
  {{H{\"u}tsi}}(2000)}]{CDMwStep}%
  \BibitemOpen
  \bibfield  {author} {\bibinfo {author} {\bibfnamefont {M.}~\bibnamefont
  {{Gramann}}}\ and\ \bibinfo {author} {\bibfnamefont {G.}~\bibnamefont
  {{H{\"u}tsi}}},\ }\href {https://doi.org/10.1046/j.1365-8711.2000.03576.x}
  {\bibfield  {journal} {\bibinfo  {journal} {Mon. Not. R. Ast. Soc.}\ }\textbf
  {\bibinfo {volume} {316}},\ \bibinfo {pages} {631} (\bibinfo {year}
  {2000})},\ \Eprint {https://arxiv.org/abs/astro-ph/0003455}
  {arXiv:astro-ph/0003455 [astro-ph]} \BibitemShut {NoStop}%
\bibitem [{\citenamefont {{Gramann}}\ and\ \citenamefont
  {{H{\"u}tsi}}(2001)}]{PriFeatSup}%
  \BibitemOpen
  \bibfield  {author} {\bibinfo {author} {\bibfnamefont {M.}~\bibnamefont
  {{Gramann}}}\ and\ \bibinfo {author} {\bibfnamefont {G.}~\bibnamefont
  {{H{\"u}tsi}}},\ }\href {https://doi.org/10.1046/j.1365-8711.2001.04733.x}
  {\bibfield  {journal} {\bibinfo  {journal} {Mon. Not. R. Ast. Soc.}\ }\textbf
  {\bibinfo {volume} {327}},\ \bibinfo {pages} {538} (\bibinfo {year}
  {2001})},\ \Eprint {https://arxiv.org/abs/astro-ph/0102466}
  {arXiv:astro-ph/0102466 [astro-ph]} \BibitemShut {NoStop}%
\bibitem [{\citenamefont {{Contaldi}}\ \emph {et~al.}(2003)\citenamefont
  {{Contaldi}}, \citenamefont {{Peloso}}, \citenamefont {{Kofman}},\ and\
  \citenamefont {{Linde}}}]{KofLindSuppress03}%
  \BibitemOpen
  \bibfield  {author} {\bibinfo {author} {\bibfnamefont {C.~R.}\ \bibnamefont
  {{Contaldi}}}, \bibinfo {author} {\bibfnamefont {M.}~\bibnamefont
  {{Peloso}}}, \bibinfo {author} {\bibfnamefont {L.}~\bibnamefont {{Kofman}}},\
  and\ \bibinfo {author} {\bibfnamefont {A.}~\bibnamefont {{Linde}}},\ }\href
  {https://doi.org/10.1088/1475-7516/2003/07/002} {\bibfield  {journal}
  {\bibinfo  {journal} {J. Cosmol. Astropart. Phys.}\ }\textbf {\bibinfo
  {volume} {2003}}\bibfield  {number} {\bibinfo  {number} { (7)},\ \bibinfo
  {eid} {002}},\ }\Eprint {https://arxiv.org/abs/astro-ph/0303636}
  {arXiv:astro-ph/0303636 [astro-ph]} \BibitemShut {NoStop}%
\bibitem [{\citenamefont {{Gong}}(2005)}]{BrokenScaleCMBLSS05}%
  \BibitemOpen
  \bibfield  {author} {\bibinfo {author} {\bibfnamefont {J.-O.}\ \bibnamefont
  {{Gong}}},\ }\href {https://doi.org/10.1088/1475-7516/2005/07/015} {\bibfield
   {journal} {\bibinfo  {journal} {J. Cosmolog. Astropart. Phys.}\ }\textbf
  {\bibinfo {volume} {2005}},\ \bibinfo {eid} {015} (\bibinfo {year} {2005})},\
  \Eprint {https://arxiv.org/abs/astro-ph/0504383} {arXiv:astro-ph/0504383
  [astro-ph]} \BibitemShut {NoStop}%
\bibitem [{\citenamefont {{Sinha}}\ and\ \citenamefont
  {{Souradeep}}(2006)}]{WMAPPOW06}%
  \BibitemOpen
  \bibfield  {author} {\bibinfo {author} {\bibfnamefont {R.}~\bibnamefont
  {{Sinha}}}\ and\ \bibinfo {author} {\bibfnamefont {T.}~\bibnamefont
  {{Souradeep}}},\ }\href {https://doi.org/10.1103/PhysRevD.74.043518}
  {\bibfield  {journal} {\bibinfo  {journal} {\prd}\ }\textbf {\bibinfo
  {volume} {74}},\ \bibinfo {eid} {043518} (\bibinfo {year} {2006})},\ \Eprint
  {https://arxiv.org/abs/astro-ph/0511808} {arXiv:astro-ph/0511808 [astro-ph]}
  \BibitemShut {NoStop}%
\bibitem [{\citenamefont {{Joy}}\ \emph {et~al.}(2009)\citenamefont {{Joy}},
  \citenamefont {{Shafieloo}}, \citenamefont {{Sahni}},\ and\ \citenamefont
  {{Starobinsky}}}]{StaroCMB09}%
  \BibitemOpen
  \bibfield  {author} {\bibinfo {author} {\bibfnamefont {M.}~\bibnamefont
  {{Joy}}}, \bibinfo {author} {\bibfnamefont {A.}~\bibnamefont {{Shafieloo}}},
  \bibinfo {author} {\bibfnamefont {V.}~\bibnamefont {{Sahni}}},\ and\ \bibinfo
  {author} {\bibfnamefont {A.~A.}\ \bibnamefont {{Starobinsky}}},\ }\href
  {https://doi.org/10.1088/1475-7516/2009/06/028} {\bibfield  {journal}
  {\bibinfo  {journal} {J. Cosmol. Astropart. Phys.}\ }\textbf {\bibinfo
  {volume} {2009}}\bibfield  {number} {\bibinfo  {number} { (6)},\ \bibinfo
  {eid} {028}},\ }\Eprint {https://arxiv.org/abs/0807.3334} {arXiv:0807.3334
  [astro-ph]} \BibitemShut {NoStop}%
\bibitem [{\citenamefont {Kamionkowski}\ and\ \citenamefont
  {Liddle}(2000)}]{KamionLidd}%
  \BibitemOpen
  \bibfield  {author} {\bibinfo {author} {\bibfnamefont {M.}~\bibnamefont
  {Kamionkowski}}\ and\ \bibinfo {author} {\bibfnamefont {A.~R.}\ \bibnamefont
  {Liddle}},\ }\href {https://doi.org/10.1103/PhysRevLett.84.4525} {\bibfield
  {journal} {\bibinfo  {journal} {Phys. Rev. Lett.}\ }\textbf {\bibinfo
  {volume} {84}},\ \bibinfo {pages} {4525} (\bibinfo {year} {2000})},\ \Eprint
  {https://arxiv.org/abs/astro-ph/9911103} {arXiv:astro-ph/9911103 [astro-ph]}
  \BibitemShut {NoStop}%
\bibitem [{\citenamefont {{Yokoyama}}(2000)}]{DwarfBSI00}%
  \BibitemOpen
  \bibfield  {author} {\bibinfo {author} {\bibfnamefont {J.}~\bibnamefont
  {{Yokoyama}}},\ }\href {https://doi.org/10.1103/PhysRevD.62.123509}
  {\bibfield  {journal} {\bibinfo  {journal} {\prd}\ }\textbf {\bibinfo
  {volume} {62}},\ \bibinfo {eid} {123509} (\bibinfo {year} {2000})},\ \Eprint
  {https://arxiv.org/abs/astro-ph/0009127} {arXiv:astro-ph/0009127 [astro-ph]}
  \BibitemShut {NoStop}%
\bibitem [{\citenamefont {{Zentner}}\ and\ \citenamefont
  {{Bullock}}(2002)}]{ZentBull02}%
  \BibitemOpen
  \bibfield  {author} {\bibinfo {author} {\bibfnamefont {A.~R.}\ \bibnamefont
  {{Zentner}}}\ and\ \bibinfo {author} {\bibfnamefont {J.~S.}\ \bibnamefont
  {{Bullock}}},\ }\href {https://doi.org/10.1103/PhysRevD.66.043003} {\bibfield
   {journal} {\bibinfo  {journal} {\prd}\ }\textbf {\bibinfo {volume} {66}},\
  \bibinfo {eid} {043003} (\bibinfo {year} {2002})},\ \Eprint
  {https://arxiv.org/abs/astro-ph/0205216} {arXiv:astro-ph/0205216 [astro-ph]}
  \BibitemShut {NoStop}%
\bibitem [{\citenamefont {{Little}}\ \emph {et~al.}(2003)\citenamefont
  {{Little}}, \citenamefont {{Knebe}},\ and\ \citenamefont
  {{Islam}}}]{WDMBSC03}%
  \BibitemOpen
  \bibfield  {author} {\bibinfo {author} {\bibfnamefont {B.}~\bibnamefont
  {{Little}}}, \bibinfo {author} {\bibfnamefont {A.}~\bibnamefont {{Knebe}}},\
  and\ \bibinfo {author} {\bibfnamefont {R.~R.}\ \bibnamefont {{Islam}}},\
  }\href {https://doi.org/10.1046/j.1365-8711.2003.06437.x} {\bibfield
  {journal} {\bibinfo  {journal} {Mon. Not. R. Ast. Soc.}\ }\textbf {\bibinfo
  {volume} {341}},\ \bibinfo {pages} {617} (\bibinfo {year} {2003})},\ \Eprint
  {https://arxiv.org/abs/astro-ph/0205420} {arXiv:astro-ph/0205420 [astro-ph]}
  \BibitemShut {NoStop}%
\bibitem [{\citenamefont {{Enqvist}}\ \emph {et~al.}(2019)\citenamefont
  {{Enqvist}}, \citenamefont {{Sawala}},\ and\ \citenamefont
  {{Takahashi}}}]{SSF2019}%
  \BibitemOpen
  \bibfield  {author} {\bibinfo {author} {\bibfnamefont {K.}~\bibnamefont
  {{Enqvist}}}, \bibinfo {author} {\bibfnamefont {T.}~\bibnamefont
  {{Sawala}}},\ and\ \bibinfo {author} {\bibfnamefont {T.}~\bibnamefont
  {{Takahashi}}},\ }\href@noop {} {\bibfield  {journal} {\bibinfo  {journal}
  {arXiv e-prints}\ ,\ \bibinfo {eid} {arXiv:1905.13580}} (\bibinfo {year}
  {2019})},\ \Eprint {https://arxiv.org/abs/1905.13580} {arXiv:1905.13580
  [astro-ph.CO]} \BibitemShut {NoStop}%
\bibitem [{\citenamefont {{Berezinsky}}\ \emph {et~al.}(2010)\citenamefont
  {{Berezinsky}}, \citenamefont {{Dokuchaev}}, \citenamefont {{Eroshenko}},
  \citenamefont {{Kachelrie{\ss}}},\ and\ \citenamefont
  {{Solberg}}}]{Superdens10}%
  \BibitemOpen
  \bibfield  {author} {\bibinfo {author} {\bibfnamefont {V.}~\bibnamefont
  {{Berezinsky}}}, \bibinfo {author} {\bibfnamefont {V.}~\bibnamefont
  {{Dokuchaev}}}, \bibinfo {author} {\bibfnamefont {Y.}~\bibnamefont
  {{Eroshenko}}}, \bibinfo {author} {\bibfnamefont {M.}~\bibnamefont
  {{Kachelrie{\ss}}}},\ and\ \bibinfo {author} {\bibfnamefont {M.~A.}\
  \bibnamefont {{Solberg}}},\ }\href
  {https://doi.org/10.1103/PhysRevD.81.103529} {\bibfield  {journal} {\bibinfo
  {journal} {\prd}\ }\textbf {\bibinfo {volume} {81}},\ \bibinfo {eid} {103529}
  (\bibinfo {year} {2010})},\ \Eprint {https://arxiv.org/abs/1002.3444}
  {arXiv:1002.3444 [astro-ph.CO]} \BibitemShut {NoStop}%
\bibitem [{\citenamefont {Leo}\ \emph {et~al.}(2018)\citenamefont {Leo},
  \citenamefont {Baugh}, \citenamefont {Li},\ and\ \citenamefont
  {Pascoli}}]{Baughtherm18}%
  \BibitemOpen
  \bibfield  {author} {\bibinfo {author} {\bibfnamefont {M.}~\bibnamefont
  {Leo}}, \bibinfo {author} {\bibfnamefont {C.~M.}\ \bibnamefont {Baugh}},
  \bibinfo {author} {\bibfnamefont {B.}~\bibnamefont {Li}},\ and\ \bibinfo
  {author} {\bibfnamefont {S.}~\bibnamefont {Pascoli}},\ }\href
  {https://doi.org/10.1088/1475-7516/2018/12/010} {\bibfield  {journal}
  {\bibinfo  {journal} {J. Cosmol. Astropart. Phys.}\ }\textbf {\bibinfo
  {volume} {1812}}\bibfield  {number} {\bibinfo  {number} { (12)},\ \bibinfo
  {pages} {010}},\ }\Eprint {https://arxiv.org/abs/1807.04980}
  {arXiv:1807.04980 [astro-ph.CO]} \BibitemShut {NoStop}%
\bibitem [{\citenamefont {Yamaguchi}(2001{\natexlab{b}})}]{Yamaguchi01}%
  \BibitemOpen
  \bibfield  {author} {\bibinfo {author} {\bibfnamefont {M.}~\bibnamefont
  {Yamaguchi}},\ }\href {https://doi.org/10.1103/PhysRevD.64.063503} {\bibfield
   {journal} {\bibinfo  {journal} {Phys. Rev.}\ }\textbf {\bibinfo {volume}
  {D64}},\ \bibinfo {pages} {063503} (\bibinfo {year} {2001}{\natexlab{b}})},\
  \Eprint {https://arxiv.org/abs/hep-ph/0105001} {arXiv:hep-ph/0105001
  [hep-ph]} \BibitemShut {NoStop}%
\bibitem [{\citenamefont {Bringmann}\ \emph {et~al.}(2002)\citenamefont
  {Bringmann}, \citenamefont {Kiefer},\ and\ \citenamefont
  {Polarski}}]{Bringmann01}%
  \BibitemOpen
  \bibfield  {author} {\bibinfo {author} {\bibfnamefont {T.}~\bibnamefont
  {Bringmann}}, \bibinfo {author} {\bibfnamefont {C.}~\bibnamefont {Kiefer}},\
  and\ \bibinfo {author} {\bibfnamefont {D.}~\bibnamefont {Polarski}},\ }\href
  {https://doi.org/10.1103/PhysRevD.65.024008} {\bibfield  {journal} {\bibinfo
  {journal} {Phys. Rev.}\ }\textbf {\bibinfo {volume} {D65}},\ \bibinfo {pages}
  {024008} (\bibinfo {year} {2002})},\ \Eprint
  {https://arxiv.org/abs/astro-ph/0109404} {arXiv:astro-ph/0109404 [astro-ph]}
  \BibitemShut {NoStop}%
\bibitem [{\citenamefont {Easther}\ and\ \citenamefont
  {Peiris}(2006)}]{PeirisRun06}%
  \BibitemOpen
  \bibfield  {author} {\bibinfo {author} {\bibfnamefont {R.}~\bibnamefont
  {Easther}}\ and\ \bibinfo {author} {\bibfnamefont {H.}~\bibnamefont
  {Peiris}},\ }\href {https://doi.org/10.1088/1475-7516/2006/09/010} {\bibfield
   {journal} {\bibinfo  {journal} {J. Cosmol. Astropart. Phys.}\ }\textbf
  {\bibinfo {volume} {0609}},\ \bibinfo {pages} {010}},\ \Eprint
  {https://arxiv.org/abs/astro-ph/0604214} {arXiv:astro-ph/0604214 [astro-ph]}
  \BibitemShut {NoStop}%
\bibitem [{\citenamefont {Bean}\ \emph {et~al.}(2008)\citenamefont {Bean},
  \citenamefont {Chen}, \citenamefont {Hailu}, \citenamefont {Tye},\ and\
  \citenamefont {Xu}}]{BeanDBI}%
  \BibitemOpen
  \bibfield  {author} {\bibinfo {author} {\bibfnamefont {R.}~\bibnamefont
  {Bean}}, \bibinfo {author} {\bibfnamefont {X.}~\bibnamefont {Chen}}, \bibinfo
  {author} {\bibfnamefont {G.}~\bibnamefont {Hailu}}, \bibinfo {author}
  {\bibfnamefont {S.~H.~H.}\ \bibnamefont {Tye}},\ and\ \bibinfo {author}
  {\bibfnamefont {J.}~\bibnamefont {Xu}},\ }\href
  {https://doi.org/10.1088/1475-7516/2008/03/026} {\bibfield  {journal}
  {\bibinfo  {journal} {J. Cosmol. Astropart. Phys.}\ }\textbf {\bibinfo
  {volume} {0803}},\ \bibinfo {pages} {026}},\ \Eprint
  {https://arxiv.org/abs/0802.0491} {arXiv:0802.0491 [hep-th]} \BibitemShut
  {NoStop}%
\bibitem [{\citenamefont {{Nakashima}}\ \emph {et~al.}(2011)\citenamefont
  {{Nakashima}}, \citenamefont {{Saito}}, \citenamefont {{Takamizu}},\ and\
  \citenamefont {{Yokoyama}}}]{JapDBI}%
  \BibitemOpen
  \bibfield  {author} {\bibinfo {author} {\bibfnamefont {M.}~\bibnamefont
  {{Nakashima}}}, \bibinfo {author} {\bibfnamefont {R.}~\bibnamefont
  {{Saito}}}, \bibinfo {author} {\bibfnamefont {Y.}~\bibnamefont
  {{Takamizu}}},\ and\ \bibinfo {author} {\bibfnamefont {J.}~\bibnamefont
  {{Yokoyama}}},\ }\href {https://doi.org/10.1143/PTP.125.1035} {\bibfield
  {journal} {\bibinfo  {journal} {Progress of Theoretical Physics}\ }\textbf
  {\bibinfo {volume} {125}},\ \bibinfo {pages} {1035} (\bibinfo {year}
  {2011})},\ \Eprint {https://arxiv.org/abs/1009.4394} {arXiv:1009.4394
  [astro-ph.CO]} \BibitemShut {NoStop}%
\bibitem [{\citenamefont {Miranda}\ \emph {et~al.}(2012)\citenamefont
  {Miranda}, \citenamefont {Hu},\ and\ \citenamefont {Adshead}}]{HuDBI1}%
  \BibitemOpen
  \bibfield  {author} {\bibinfo {author} {\bibfnamefont {V.}~\bibnamefont
  {Miranda}}, \bibinfo {author} {\bibfnamefont {W.}~\bibnamefont {Hu}},\ and\
  \bibinfo {author} {\bibfnamefont {P.}~\bibnamefont {Adshead}},\ }\href
  {https://doi.org/10.1103/PhysRevD.86.063529} {\bibfield  {journal} {\bibinfo
  {journal} {Phys. Rev.}\ }\textbf {\bibinfo {volume} {D86}},\ \bibinfo {pages}
  {063529} (\bibinfo {year} {2012})},\ \Eprint
  {https://arxiv.org/abs/1207.2186} {arXiv:1207.2186 [astro-ph.CO]}
  \BibitemShut {NoStop}%
\bibitem [{\citenamefont {Miranda}\ and\ \citenamefont {Hu}(2014)}]{HuDBI2}%
  \BibitemOpen
  \bibfield  {author} {\bibinfo {author} {\bibfnamefont {V.}~\bibnamefont
  {Miranda}}\ and\ \bibinfo {author} {\bibfnamefont {W.}~\bibnamefont {Hu}},\
  }\href {https://doi.org/10.1103/PhysRevD.89.083529} {\bibfield  {journal}
  {\bibinfo  {journal} {Phys. Rev.}\ }\textbf {\bibinfo {volume} {D89}},\
  \bibinfo {pages} {083529} (\bibinfo {year} {2014})},\ \Eprint
  {https://arxiv.org/abs/1312.0946} {arXiv:1312.0946 [astro-ph.CO]}
  \BibitemShut {NoStop}%
\bibitem [{\citenamefont {{Park}}\ and\ \citenamefont {{Sorbo}}(2012)}]{Sud_1}%
  \BibitemOpen
  \bibfield  {author} {\bibinfo {author} {\bibfnamefont {M.}~\bibnamefont
  {{Park}}}\ and\ \bibinfo {author} {\bibfnamefont {L.}~\bibnamefont
  {{Sorbo}}},\ }\href {https://doi.org/10.1103/PhysRevD.85.083520} {\bibfield
  {journal} {\bibinfo  {journal} {\prd}\ }\textbf {\bibinfo {volume} {85}},\
  \bibinfo {eid} {083520} (\bibinfo {year} {2012})},\ \Eprint
  {https://arxiv.org/abs/1201.2903} {arXiv:1201.2903 [astro-ph.CO]}
  \BibitemShut {NoStop}%
\bibitem [{\citenamefont {{Gariazzo}}\ \emph {et~al.}(2017)\citenamefont
  {{Gariazzo}}, \citenamefont {{Mena}}, \citenamefont {{Ram{\'\i}rez}},\ and\
  \citenamefont {{Boubekeur}}}]{PhenoSound_1}%
  \BibitemOpen
  \bibfield  {author} {\bibinfo {author} {\bibfnamefont {S.}~\bibnamefont
  {{Gariazzo}}}, \bibinfo {author} {\bibfnamefont {O.}~\bibnamefont {{Mena}}},
  \bibinfo {author} {\bibfnamefont {H.}~\bibnamefont {{Ram{\'\i}rez}}},\ and\
  \bibinfo {author} {\bibfnamefont {L.}~\bibnamefont {{Boubekeur}}},\ }\href
  {https://doi.org/10.1016/j.dark.2017.07.003} {\bibfield  {journal} {\bibinfo
  {journal} {Physics of the Dark Universe}\ }\textbf {\bibinfo {volume} {17}},\
  \bibinfo {pages} {38} (\bibinfo {year} {2017})},\ \Eprint
  {https://arxiv.org/abs/1606.00842} {arXiv:1606.00842 [astro-ph.CO]}
  \BibitemShut {NoStop}%
\bibitem [{\citenamefont {Ballesteros}\ \emph {et~al.}(2019)\citenamefont
  {Ballesteros}, \citenamefont {Beltran~Jimenez},\ and\ \citenamefont
  {Pieroni}}]{PhenoSound_2}%
  \BibitemOpen
  \bibfield  {author} {\bibinfo {author} {\bibfnamefont {G.}~\bibnamefont
  {Ballesteros}}, \bibinfo {author} {\bibfnamefont {J.}~\bibnamefont
  {Beltran~Jimenez}},\ and\ \bibinfo {author} {\bibfnamefont {M.}~\bibnamefont
  {Pieroni}},\ }\href {https://doi.org/10.1088/1475-7516/2019/06/016}
  {\bibfield  {journal} {\bibinfo  {journal} {J. Cosmol. Astropart. Phys.}\
  }\textbf {\bibinfo {volume} {1906}}\bibfield  {number} {\bibinfo  {number} {
  (06)},\ \bibinfo {pages} {016}},\ }\Eprint {https://arxiv.org/abs/1811.03065}
  {arXiv:1811.03065 [astro-ph.CO]} \BibitemShut {NoStop}%
\bibitem [{\citenamefont {Martin}\ and\ \citenamefont
  {Brandenberger}(2001{\natexlab{a}})}]{BDM00}%
  \BibitemOpen
  \bibfield  {author} {\bibinfo {author} {\bibfnamefont {J.}~\bibnamefont
  {Martin}}\ and\ \bibinfo {author} {\bibfnamefont {R.~H.}\ \bibnamefont
  {Brandenberger}},\ }\href {https://doi.org/10.1103/PhysRevD.63.123501}
  {\bibfield  {journal} {\bibinfo  {journal} {Phys. Rev.}\ }\textbf {\bibinfo
  {volume} {D63}},\ \bibinfo {pages} {123501} (\bibinfo {year}
  {2001}{\natexlab{a}})},\ \Eprint {https://arxiv.org/abs/hep-th/0005209}
  {arXiv:hep-th/0005209 [hep-th]} \BibitemShut {NoStop}%
\bibitem [{\citenamefont {Brandenberger}\ and\ \citenamefont
  {Martin}(2001)}]{BDM01}%
  \BibitemOpen
  \bibfield  {author} {\bibinfo {author} {\bibfnamefont {R.~H.}\ \bibnamefont
  {Brandenberger}}\ and\ \bibinfo {author} {\bibfnamefont {J.}~\bibnamefont
  {Martin}},\ }\href {https://doi.org/10.1142/S0217732301004170} {\bibfield
  {journal} {\bibinfo  {journal} {Mod. Phys. Lett.}\ }\textbf {\bibinfo
  {volume} {A16}},\ \bibinfo {pages} {999} (\bibinfo {year} {2001})},\ \Eprint
  {https://arxiv.org/abs/astro-ph/0005432} {arXiv:astro-ph/0005432 [astro-ph]}
  \BibitemShut {NoStop}%
\bibitem [{\citenamefont {Easther}\ \emph {et~al.}(2001)\citenamefont
  {Easther}, \citenamefont {Greene}, \citenamefont {Kinney},\ and\
  \citenamefont {Shiu}}]{Easther_GUP_01}%
  \BibitemOpen
  \bibfield  {author} {\bibinfo {author} {\bibfnamefont {R.}~\bibnamefont
  {Easther}}, \bibinfo {author} {\bibfnamefont {B.~R.}\ \bibnamefont {Greene}},
  \bibinfo {author} {\bibfnamefont {W.~H.}\ \bibnamefont {Kinney}},\ and\
  \bibinfo {author} {\bibfnamefont {G.}~\bibnamefont {Shiu}},\ }\href
  {https://doi.org/10.1103/PhysRevD.64.103502} {\bibfield  {journal} {\bibinfo
  {journal} {Phys. Rev.}\ }\textbf {\bibinfo {volume} {D64}},\ \bibinfo {pages}
  {103502} (\bibinfo {year} {2001})},\ \Eprint
  {https://arxiv.org/abs/hep-th/0104102} {arXiv:hep-th/0104102 [hep-th]}
  \BibitemShut {NoStop}%
\bibitem [{\citenamefont {Niemeyer}(2001)}]{Niemeyer_Com_Fluid01}%
  \BibitemOpen
  \bibfield  {author} {\bibinfo {author} {\bibfnamefont {J.~C.}\ \bibnamefont
  {Niemeyer}},\ }\href {https://doi.org/10.1103/PhysRevD.63.123502} {\bibfield
  {journal} {\bibinfo  {journal} {Phys. Rev.}\ }\textbf {\bibinfo {volume}
  {D63}},\ \bibinfo {pages} {123502} (\bibinfo {year} {2001})},\ \Eprint
  {https://arxiv.org/abs/astro-ph/0005533} {arXiv:astro-ph/0005533 [astro-ph]}
  \BibitemShut {NoStop}%
\bibitem [{\citenamefont {Martin}\ and\ \citenamefont
  {Brandenberger}(2003)}]{BDB_M03}%
  \BibitemOpen
  \bibfield  {author} {\bibinfo {author} {\bibfnamefont {J.}~\bibnamefont
  {Martin}}\ and\ \bibinfo {author} {\bibfnamefont {R.}~\bibnamefont
  {Brandenberger}},\ }\href {https://doi.org/10.1103/PhysRevD.68.063513}
  {\bibfield  {journal} {\bibinfo  {journal} {Phys. Rev.}\ }\textbf {\bibinfo
  {volume} {D68}},\ \bibinfo {pages} {063513} (\bibinfo {year} {2003})},\
  \Eprint {https://arxiv.org/abs/hep-th/0305161} {arXiv:hep-th/0305161
  [hep-th]} \BibitemShut {NoStop}%
\bibitem [{\citenamefont {Easther}\ \emph
  {et~al.}(2005{\natexlab{a}})\citenamefont {Easther}, \citenamefont {Kinney},\
  and\ \citenamefont {Peiris}}]{Easther_BEFT05}%
  \BibitemOpen
  \bibfield  {author} {\bibinfo {author} {\bibfnamefont {R.}~\bibnamefont
  {Easther}}, \bibinfo {author} {\bibfnamefont {W.~H.}\ \bibnamefont
  {Kinney}},\ and\ \bibinfo {author} {\bibfnamefont {H.}~\bibnamefont
  {Peiris}},\ }\href {https://doi.org/10.1088/1475-7516/2005/08/001} {\bibfield
   {journal} {\bibinfo  {journal} {J. Cosmol. Astropart. Phys.}\ }\textbf
  {\bibinfo {volume} {0508}},\ \bibinfo {pages} {001}},\ \Eprint
  {https://arxiv.org/abs/astro-ph/0505426} {arXiv:astro-ph/0505426 [astro-ph]}
  \BibitemShut {NoStop}%
\bibitem [{\citenamefont {Kempf}\ and\ \citenamefont
  {Lorenz}(2006)}]{Kempf_GUP06}%
  \BibitemOpen
  \bibfield  {author} {\bibinfo {author} {\bibfnamefont {A.}~\bibnamefont
  {Kempf}}\ and\ \bibinfo {author} {\bibfnamefont {L.}~\bibnamefont {Lorenz}},\
  }\href {https://doi.org/10.1103/PhysRevD.74.103517} {\bibfield  {journal}
  {\bibinfo  {journal} {Phys. Rev.}\ }\textbf {\bibinfo {volume} {D74}},\
  \bibinfo {pages} {103517} (\bibinfo {year} {2006})},\ \Eprint
  {https://arxiv.org/abs/gr-qc/0609123} {arXiv:gr-qc/0609123 [gr-qc]}
  \BibitemShut {NoStop}%
\bibitem [{\citenamefont {Brandenberger}\ and\ \citenamefont
  {Martin}(2013)}]{BDMR13}%
  \BibitemOpen
  \bibfield  {author} {\bibinfo {author} {\bibfnamefont {R.~H.}\ \bibnamefont
  {Brandenberger}}\ and\ \bibinfo {author} {\bibfnamefont {J.}~\bibnamefont
  {Martin}},\ }\href {https://doi.org/10.1088/0264-9381/30/11/113001}
  {\bibfield  {journal} {\bibinfo  {journal} {Class. Quant. Grav.}\ }\textbf
  {\bibinfo {volume} {30}},\ \bibinfo {pages} {113001} (\bibinfo {year}
  {2013})},\ \Eprint {https://arxiv.org/abs/1211.6753} {arXiv:1211.6753
  [astro-ph.CO]} \BibitemShut {NoStop}%
\bibitem [{\citenamefont {Kaloper}\ \emph {et~al.}(2002)\citenamefont
  {Kaloper}, \citenamefont {Kleban}, \citenamefont {Lawrence},\ and\
  \citenamefont {Shenker}}]{Kaloper02}%
  \BibitemOpen
  \bibfield  {author} {\bibinfo {author} {\bibfnamefont {N.}~\bibnamefont
  {Kaloper}}, \bibinfo {author} {\bibfnamefont {M.}~\bibnamefont {Kleban}},
  \bibinfo {author} {\bibfnamefont {A.~E.}\ \bibnamefont {Lawrence}},\ and\
  \bibinfo {author} {\bibfnamefont {S.}~\bibnamefont {Shenker}},\ }\href
  {https://doi.org/10.1103/PhysRevD.66.123510} {\bibfield  {journal} {\bibinfo
  {journal} {Phys. Rev.}\ }\textbf {\bibinfo {volume} {D66}},\ \bibinfo {pages}
  {123510} (\bibinfo {year} {2002})},\ \Eprint
  {https://arxiv.org/abs/hep-th/0201158} {arXiv:hep-th/0201158 [hep-th]}
  \BibitemShut {NoStop}%
\bibitem [{\citenamefont {Danielsson}(2002)}]{Danielsson02}%
  \BibitemOpen
  \bibfield  {author} {\bibinfo {author} {\bibfnamefont {U.~H.}\ \bibnamefont
  {Danielsson}},\ }\href {https://doi.org/10.1103/PhysRevD.66.023511}
  {\bibfield  {journal} {\bibinfo  {journal} {Phys. Rev.}\ }\textbf {\bibinfo
  {volume} {D66}},\ \bibinfo {pages} {023511} (\bibinfo {year} {2002})},\
  \Eprint {https://arxiv.org/abs/hep-th/0203198} {arXiv:hep-th/0203198
  [hep-th]} \BibitemShut {NoStop}%
\bibitem [{\citenamefont {Easther}\ \emph {et~al.}(2002)\citenamefont
  {Easther}, \citenamefont {Greene}, \citenamefont {Kinney},\ and\
  \citenamefont {Shiu}}]{Easther02}%
  \BibitemOpen
  \bibfield  {author} {\bibinfo {author} {\bibfnamefont {R.}~\bibnamefont
  {Easther}}, \bibinfo {author} {\bibfnamefont {B.~R.}\ \bibnamefont {Greene}},
  \bibinfo {author} {\bibfnamefont {W.~H.}\ \bibnamefont {Kinney}},\ and\
  \bibinfo {author} {\bibfnamefont {G.}~\bibnamefont {Shiu}},\ }\href
  {https://doi.org/10.1103/PhysRevD.66.023518} {\bibfield  {journal} {\bibinfo
  {journal} {Phys. Rev.}\ }\textbf {\bibinfo {volume} {D66}},\ \bibinfo {pages}
  {023518} (\bibinfo {year} {2002})},\ \Eprint
  {https://arxiv.org/abs/hep-th/0204129} {arXiv:hep-th/0204129 [hep-th]}
  \BibitemShut {NoStop}%
\bibitem [{\citenamefont {Niemeyer}\ \emph {et~al.}(2002)\citenamefont
  {Niemeyer}, \citenamefont {Parentani},\ and\ \citenamefont
  {Campo}}]{Niemeyer02}%
  \BibitemOpen
  \bibfield  {author} {\bibinfo {author} {\bibfnamefont {J.~C.}\ \bibnamefont
  {Niemeyer}}, \bibinfo {author} {\bibfnamefont {R.}~\bibnamefont
  {Parentani}},\ and\ \bibinfo {author} {\bibfnamefont {D.}~\bibnamefont
  {Campo}},\ }\href {https://doi.org/10.1103/PhysRevD.66.083510} {\bibfield
  {journal} {\bibinfo  {journal} {Phys. Rev.}\ }\textbf {\bibinfo {volume}
  {D66}},\ \bibinfo {pages} {083510} (\bibinfo {year} {2002})},\ \Eprint
  {https://arxiv.org/abs/hep-th/0206149} {arXiv:hep-th/0206149 [hep-th]}
  \BibitemShut {NoStop}%
\bibitem [{\citenamefont {Chatwin-Davies}\ \emph {et~al.}(2017)\citenamefont
  {Chatwin-Davies}, \citenamefont {Kempf},\ and\ \citenamefont
  {Martin}}]{Chatwin-Davies16}%
  \BibitemOpen
  \bibfield  {author} {\bibinfo {author} {\bibfnamefont {A.}~\bibnamefont
  {Chatwin-Davies}}, \bibinfo {author} {\bibfnamefont {A.}~\bibnamefont
  {Kempf}},\ and\ \bibinfo {author} {\bibfnamefont {R.~T.~W.}\ \bibnamefont
  {Martin}},\ }\href {https://doi.org/10.1103/PhysRevLett.119.031301}
  {\bibfield  {journal} {\bibinfo  {journal} {Phys. Rev. Lett.}\ }\textbf
  {\bibinfo {volume} {119}},\ \bibinfo {pages} {031301} (\bibinfo {year}
  {2017})},\ \Eprint {https://arxiv.org/abs/1612.06445} {arXiv:1612.06445
  [gr-qc]} \BibitemShut {NoStop}%
\bibitem [{\citenamefont {Brandenberger}\ and\ \citenamefont
  {Martin}(2002)}]{BDM02}%
  \BibitemOpen
  \bibfield  {author} {\bibinfo {author} {\bibfnamefont {R.~H.}\ \bibnamefont
  {Brandenberger}}\ and\ \bibinfo {author} {\bibfnamefont {J.}~\bibnamefont
  {Martin}},\ }\href {https://doi.org/10.1142/S0217751X02010765} {\bibfield
  {journal} {\bibinfo  {journal} {Int. J. Mod. Phys.}\ }\textbf {\bibinfo
  {volume} {A17}},\ \bibinfo {pages} {3663} (\bibinfo {year} {2002})},\ \Eprint
  {https://arxiv.org/abs/hep-th/0202142} {arXiv:hep-th/0202142 [hep-th]}
  \BibitemShut {NoStop}%
\bibitem [{\citenamefont {Zhu}\ \emph {et~al.}(2016)\citenamefont {Zhu},
  \citenamefont {Wang}, \citenamefont {Kirsten}, \citenamefont {Cleaver},\ and\
  \citenamefont {Sheng}}]{Zhu_Disp_Mod16}%
  \BibitemOpen
  \bibfield  {author} {\bibinfo {author} {\bibfnamefont {T.}~\bibnamefont
  {Zhu}}, \bibinfo {author} {\bibfnamefont {A.}~\bibnamefont {Wang}}, \bibinfo
  {author} {\bibfnamefont {K.}~\bibnamefont {Kirsten}}, \bibinfo {author}
  {\bibfnamefont {G.}~\bibnamefont {Cleaver}},\ and\ \bibinfo {author}
  {\bibfnamefont {Q.}~\bibnamefont {Sheng}},\ }\href
  {https://doi.org/10.1103/PhysRevD.93.123525} {\bibfield  {journal} {\bibinfo
  {journal} {Phys. Rev.}\ }\textbf {\bibinfo {volume} {D93}},\ \bibinfo {pages}
  {123525} (\bibinfo {year} {2016})},\ \Eprint
  {https://arxiv.org/abs/1604.05739} {arXiv:1604.05739 [gr-qc]} \BibitemShut
  {NoStop}%
\bibitem [{\citenamefont {Ashoorioon}\ \emph {et~al.}(2017)\citenamefont
  {Ashoorioon}, \citenamefont {Casadio}, \citenamefont {Geshnizjani},\ and\
  \citenamefont {Kim}}]{Ashoor_Disp_Num17}%
  \BibitemOpen
  \bibfield  {author} {\bibinfo {author} {\bibfnamefont {A.}~\bibnamefont
  {Ashoorioon}}, \bibinfo {author} {\bibfnamefont {R.}~\bibnamefont {Casadio}},
  \bibinfo {author} {\bibfnamefont {G.}~\bibnamefont {Geshnizjani}},\ and\
  \bibinfo {author} {\bibfnamefont {H.~J.}\ \bibnamefont {Kim}},\ }\href
  {https://doi.org/10.1088/1475-7516/2017/09/008} {\bibfield  {journal}
  {\bibinfo  {journal} {J. Cosmol. Astropart. Phys.}\ }\textbf {\bibinfo
  {volume} {1709}}\bibfield  {number} {\bibinfo  {number} { (09)},\ \bibinfo
  {pages} {008}},\ }\Eprint {https://arxiv.org/abs/1702.06101}
  {arXiv:1702.06101 [hep-th]} \BibitemShut {NoStop}%
\bibitem [{\citenamefont {Armendariz-Picon}\ and\ \citenamefont
  {Lim}(2003)}]{ArmendarizPicon03}%
  \BibitemOpen
  \bibfield  {author} {\bibinfo {author} {\bibfnamefont {C.}~\bibnamefont
  {Armendariz-Picon}}\ and\ \bibinfo {author} {\bibfnamefont {E.~A.}\
  \bibnamefont {Lim}},\ }\href {https://doi.org/10.1088/1475-7516/2003/12/006}
  {\bibfield  {journal} {\bibinfo  {journal} {J. Cosmol. Astropart. Phys.}\
  }\textbf {\bibinfo {volume} {0312}},\ \bibinfo {pages} {006}},\ \Eprint
  {https://arxiv.org/abs/hep-th/0303103} {arXiv:hep-th/0303103 [hep-th]}
  \BibitemShut {NoStop}%
\bibitem [{\citenamefont {Amelino-Camelia}(2013)}]{Camelia08}%
  \BibitemOpen
  \bibfield  {author} {\bibinfo {author} {\bibfnamefont {G.}~\bibnamefont
  {Amelino-Camelia}},\ }\href {https://doi.org/10.12942/lrr-2013-5} {\bibfield
  {journal} {\bibinfo  {journal} {Living Rev. Rel.}\ }\textbf {\bibinfo
  {volume} {16}},\ \bibinfo {pages} {5} (\bibinfo {year} {2013})},\ \Eprint
  {https://arxiv.org/abs/0806.0339} {arXiv:0806.0339 [gr-qc]} \BibitemShut
  {NoStop}%
\bibitem [{\citenamefont {Hossenfelder}(2013)}]{Hossenfelder13}%
  \BibitemOpen
  \bibfield  {author} {\bibinfo {author} {\bibfnamefont {S.}~\bibnamefont
  {Hossenfelder}},\ }\href {https://doi.org/10.12942/lrr-2013-2} {\bibfield
  {journal} {\bibinfo  {journal} {Living Rev. Rel.}\ }\textbf {\bibinfo
  {volume} {16}},\ \bibinfo {pages} {2} (\bibinfo {year} {2013})},\ \Eprint
  {https://arxiv.org/abs/1203.6191} {arXiv:1203.6191 [gr-qc]} \BibitemShut
  {NoStop}%
\bibitem [{\citenamefont {Unruh}(1981)}]{Unruh81}%
  \BibitemOpen
  \bibfield  {author} {\bibinfo {author} {\bibfnamefont {W.~G.}\ \bibnamefont
  {Unruh}},\ }\href {https://doi.org/10.1103/PhysRevLett.46.1351} {\bibfield
  {journal} {\bibinfo  {journal} {Phys. Rev. Lett.}\ }\textbf {\bibinfo
  {volume} {46}},\ \bibinfo {pages} {1351} (\bibinfo {year}
  {1981})}\BibitemShut {NoStop}%
\bibitem [{\citenamefont {Unruh}(1995)}]{Unruh94}%
  \BibitemOpen
  \bibfield  {author} {\bibinfo {author} {\bibfnamefont {W.~G.}\ \bibnamefont
  {Unruh}},\ }\href {https://doi.org/10.1103/PhysRevD.51.2827} {\bibfield
  {journal} {\bibinfo  {journal} {Phys. Rev.}\ }\textbf {\bibinfo {volume}
  {D51}},\ \bibinfo {pages} {2827} (\bibinfo {year} {1995})}\BibitemShut
  {NoStop}%
\bibitem [{\citenamefont {Barcelo}\ \emph {et~al.}(2005)\citenamefont
  {Barcelo}, \citenamefont {Liberati},\ and\ \citenamefont
  {Visser}}]{Analog_Rev}%
  \BibitemOpen
  \bibfield  {author} {\bibinfo {author} {\bibfnamefont {C.}~\bibnamefont
  {Barcelo}}, \bibinfo {author} {\bibfnamefont {S.}~\bibnamefont {Liberati}},\
  and\ \bibinfo {author} {\bibfnamefont {M.}~\bibnamefont {Visser}},\ }\href
  {https://doi.org/10.12942/lrr-2005-12} {\bibfield  {journal} {\bibinfo
  {journal} {Living Rev. Rel.}\ }\textbf {\bibinfo {volume} {8}},\ \bibinfo
  {pages} {12} (\bibinfo {year} {2005})},\ \bibinfo {note} {[Living Rev.
  Rel.14,3(2011)]},\ \Eprint {https://arxiv.org/abs/gr-qc/0505065}
  {arXiv:gr-qc/0505065 [gr-qc]} \BibitemShut {NoStop}%
\bibitem [{\citenamefont {Chä}\ and\ \citenamefont {Fischer}(2017)}]{Cha16}%
  \BibitemOpen
  \bibfield  {author} {\bibinfo {author} {\bibfnamefont {S.-Y.}\ \bibnamefont
  {Chä}}\ and\ \bibinfo {author} {\bibfnamefont {U.~R.}\ \bibnamefont
  {Fischer}},\ }\href {https://doi.org/10.1103/PhysRevLett.118.179901,
  10.1103/PhysRevLett.118.130404} {\bibfield  {journal} {\bibinfo  {journal}
  {Phys. Rev. Lett.}\ }\textbf {\bibinfo {volume} {118}},\ \bibinfo {pages}
  {130404} (\bibinfo {year} {2017})},\ \bibinfo {note} {[Addendum: Phys. Rev.
  Lett.118,no.17,179901(2017)]},\ \Eprint {https://arxiv.org/abs/1609.06155}
  {arXiv:1609.06155 [cond-mat.quant-gas]} \BibitemShut {NoStop}%
\bibitem [{\citenamefont {{Tanaka}}(2000)}]{Tanaka00}%
  \BibitemOpen
  \bibfield  {author} {\bibinfo {author} {\bibfnamefont {T.}~\bibnamefont
  {{Tanaka}}},\ }\href@noop {} {\bibfield  {journal} {\bibinfo  {journal}
  {arXiv e-prints}\ ,\ \bibinfo {eid} {astro-ph/0012431}} (\bibinfo {year}
  {2000})},\ \Eprint {https://arxiv.org/abs/astro-ph/0012431}
  {arXiv:astro-ph/0012431 [astro-ph]} \BibitemShut {NoStop}%
\bibitem [{\citenamefont {Niemeyer}\ and\ \citenamefont
  {Parentani}(2001)}]{Niemeyer_nomod_disp01}%
  \BibitemOpen
  \bibfield  {author} {\bibinfo {author} {\bibfnamefont {J.~C.}\ \bibnamefont
  {Niemeyer}}\ and\ \bibinfo {author} {\bibfnamefont {R.}~\bibnamefont
  {Parentani}},\ }\href {https://doi.org/10.1103/PhysRevD.64.101301} {\bibfield
   {journal} {\bibinfo  {journal} {Phys. Rev.}\ }\textbf {\bibinfo {volume}
  {D64}},\ \bibinfo {pages} {101301} (\bibinfo {year} {2001})},\ \Eprint
  {https://arxiv.org/abs/astro-ph/0101451} {arXiv:astro-ph/0101451 [astro-ph]}
  \BibitemShut {NoStop}%
\bibitem [{\citenamefont {Starobinsky}(2001)}]{Starobinsky_PP01}%
  \BibitemOpen
  \bibfield  {author} {\bibinfo {author} {\bibfnamefont {A.~A.}\ \bibnamefont
  {Starobinsky}},\ }\href {https://doi.org/10.1134/1.1381588} {\bibfield
  {journal} {\bibinfo  {journal} {Pisma Zh. Eksp. Teor. Fiz.}\ }\textbf
  {\bibinfo {volume} {73}},\ \bibinfo {pages} {415} (\bibinfo {year} {2001})},\
  \bibinfo {note} {[JETP Lett.73,371(2001)]},\ \Eprint
  {https://arxiv.org/abs/astro-ph/0104043} {arXiv:astro-ph/0104043 [astro-ph]}
  \BibitemShut {NoStop}%
\bibitem [{\citenamefont {Giovannini}(2003)}]{Giovannini03}%
  \BibitemOpen
  \bibfield  {author} {\bibinfo {author} {\bibfnamefont {M.}~\bibnamefont
  {Giovannini}},\ }\href {https://doi.org/10.1088/0264-9381/20/24/016}
  {\bibfield  {journal} {\bibinfo  {journal} {Class. Quant. Grav.}\ }\textbf
  {\bibinfo {volume} {20}},\ \bibinfo {pages} {5455} (\bibinfo {year}
  {2003})},\ \Eprint {https://arxiv.org/abs/hep-th/0308066}
  {arXiv:hep-th/0308066 [hep-th]} \BibitemShut {NoStop}%
\bibitem [{\citenamefont {Porrati}(2004)}]{Porrati04}%
  \BibitemOpen
  \bibfield  {author} {\bibinfo {author} {\bibfnamefont {M.}~\bibnamefont
  {Porrati}},\ }\href {https://doi.org/10.1016/j.physletb.2004.06.090}
  {\bibfield  {journal} {\bibinfo  {journal} {Phys. Lett.}\ }\textbf {\bibinfo
  {volume} {B596}},\ \bibinfo {pages} {306} (\bibinfo {year} {2004})},\ \Eprint
  {https://arxiv.org/abs/hep-th/0402038} {arXiv:hep-th/0402038 [hep-th]}
  \BibitemShut {NoStop}%
\bibitem [{\citenamefont {Greene}\ \emph {et~al.}(2005)\citenamefont {Greene},
  \citenamefont {Schalm}, \citenamefont {Shiu},\ and\ \citenamefont {van~der
  Schaar}}]{Greene04}%
  \BibitemOpen
  \bibfield  {author} {\bibinfo {author} {\bibfnamefont {B.~R.}\ \bibnamefont
  {Greene}}, \bibinfo {author} {\bibfnamefont {K.}~\bibnamefont {Schalm}},
  \bibinfo {author} {\bibfnamefont {G.}~\bibnamefont {Shiu}},\ and\ \bibinfo
  {author} {\bibfnamefont {J.~P.}\ \bibnamefont {van~der Schaar}},\ }\href
  {https://doi.org/10.1088/1475-7516/2005/02/001} {\bibfield  {journal}
  {\bibinfo  {journal} {J. Cosmol. Astropart. Phys.}\ }\textbf {\bibinfo
  {volume} {0502}},\ \bibinfo {pages} {001}},\ \Eprint
  {https://arxiv.org/abs/hep-th/0411217} {arXiv:hep-th/0411217 [hep-th]}
  \BibitemShut {NoStop}%
\bibitem [{\citenamefont {Brandenberger}\ and\ \citenamefont
  {Martin}(2005)}]{BDM04}%
  \BibitemOpen
  \bibfield  {author} {\bibinfo {author} {\bibfnamefont {R.~H.}\ \bibnamefont
  {Brandenberger}}\ and\ \bibinfo {author} {\bibfnamefont {J.}~\bibnamefont
  {Martin}},\ }\href {https://doi.org/10.1103/PhysRevD.71.023504} {\bibfield
  {journal} {\bibinfo  {journal} {Phys. Rev.}\ }\textbf {\bibinfo {volume}
  {D71}},\ \bibinfo {pages} {023504} (\bibinfo {year} {2005})},\ \Eprint
  {https://arxiv.org/abs/hep-th/0410223} {arXiv:hep-th/0410223 [hep-th]}
  \BibitemShut {NoStop}%
\bibitem [{\citenamefont {Danielsson}(2005)}]{Daniel04}%
  \BibitemOpen
  \bibfield  {author} {\bibinfo {author} {\bibfnamefont {U.~H.}\ \bibnamefont
  {Danielsson}},\ }\href {https://doi.org/10.1103/PhysRevD.71.023516}
  {\bibfield  {journal} {\bibinfo  {journal} {Phys. Rev.}\ }\textbf {\bibinfo
  {volume} {D71}},\ \bibinfo {pages} {023516} (\bibinfo {year} {2005})},\
  \Eprint {https://arxiv.org/abs/hep-th/0411172} {arXiv:hep-th/0411172
  [hep-th]} \BibitemShut {NoStop}%
\bibitem [{\citenamefont {Goldstein}\ and\ \citenamefont
  {Lowe}(2003)}]{Goldstein02}%
  \BibitemOpen
  \bibfield  {author} {\bibinfo {author} {\bibfnamefont {K.}~\bibnamefont
  {Goldstein}}\ and\ \bibinfo {author} {\bibfnamefont {D.~A.}\ \bibnamefont
  {Lowe}},\ }\href {https://doi.org/10.1103/PhysRevD.67.063502} {\bibfield
  {journal} {\bibinfo  {journal} {Phys. Rev.}\ }\textbf {\bibinfo {volume}
  {D67}},\ \bibinfo {pages} {063502} (\bibinfo {year} {2003})},\ \Eprint
  {https://arxiv.org/abs/hep-th/0208167} {arXiv:hep-th/0208167 [hep-th]}
  \BibitemShut {NoStop}%
\bibitem [{\citenamefont {Easther}\ \emph
  {et~al.}(2005{\natexlab{b}})\citenamefont {Easther}, \citenamefont {Kinney},\
  and\ \citenamefont {Peiris}}]{Easther_Peir_04}%
  \BibitemOpen
  \bibfield  {author} {\bibinfo {author} {\bibfnamefont {R.}~\bibnamefont
  {Easther}}, \bibinfo {author} {\bibfnamefont {W.~H.}\ \bibnamefont
  {Kinney}},\ and\ \bibinfo {author} {\bibfnamefont {H.}~\bibnamefont
  {Peiris}},\ }\href {https://doi.org/10.1088/1475-7516/2005/05/009} {\bibfield
   {journal} {\bibinfo  {journal} {J. Cosmol. Astropart. Phys.}\ }\textbf
  {\bibinfo {volume} {0505}},\ \bibinfo {pages} {009}},\ \Eprint
  {https://arxiv.org/abs/astro-ph/0412613} {arXiv:astro-ph/0412613 [astro-ph]}
  \BibitemShut {NoStop}%
\bibitem [{\citenamefont {Okamoto}\ and\ \citenamefont
  {Lim}(2004)}]{Okamoto04}%
  \BibitemOpen
  \bibfield  {author} {\bibinfo {author} {\bibfnamefont {T.}~\bibnamefont
  {Okamoto}}\ and\ \bibinfo {author} {\bibfnamefont {E.~A.}\ \bibnamefont
  {Lim}},\ }\href {https://doi.org/10.1103/PhysRevD.69.083519} {\bibfield
  {journal} {\bibinfo  {journal} {Phys. Rev.}\ }\textbf {\bibinfo {volume}
  {D69}},\ \bibinfo {pages} {083519} (\bibinfo {year} {2004})},\ \Eprint
  {https://arxiv.org/abs/astro-ph/0312284} {arXiv:astro-ph/0312284 [astro-ph]}
  \BibitemShut {NoStop}%
\bibitem [{\citenamefont {Martin}\ and\ \citenamefont
  {Ringeval}(2005)}]{Martin_Ring04}%
  \BibitemOpen
  \bibfield  {author} {\bibinfo {author} {\bibfnamefont {J.}~\bibnamefont
  {Martin}}\ and\ \bibinfo {author} {\bibfnamefont {C.}~\bibnamefont
  {Ringeval}},\ }\href {https://doi.org/10.1088/1475-7516/2005/01/007}
  {\bibfield  {journal} {\bibinfo  {journal} {J. Cosmol. Astropart. Phys.}\
  }\textbf {\bibinfo {volume} {0501}},\ \bibinfo {pages} {007}},\ \Eprint
  {https://arxiv.org/abs/hep-ph/0405249} {arXiv:hep-ph/0405249 [hep-ph]}
  \BibitemShut {NoStop}%
\bibitem [{\citenamefont {Martin}\ and\ \citenamefont
  {Ringeval}(2004)}]{Martin_Ring04A}%
  \BibitemOpen
  \bibfield  {author} {\bibinfo {author} {\bibfnamefont {J.}~\bibnamefont
  {Martin}}\ and\ \bibinfo {author} {\bibfnamefont {C.}~\bibnamefont
  {Ringeval}},\ }\href {https://doi.org/10.1103/PhysRevD.69.127303} {\bibfield
  {journal} {\bibinfo  {journal} {Phys. Rev.}\ }\textbf {\bibinfo {volume}
  {D69}},\ \bibinfo {pages} {127303} (\bibinfo {year} {2004})},\ \Eprint
  {https://arxiv.org/abs/astro-ph/0402609} {arXiv:astro-ph/0402609 [astro-ph]}
  \BibitemShut {NoStop}%
\bibitem [{\citenamefont {{Chluba}}\ \emph
  {et~al.}(2012{\natexlab{a}})\citenamefont {{Chluba}}, \citenamefont
  {{Erickcek}},\ and\ \citenamefont {{Ben-Dayan}}}]{Chluba_D12}%
  \BibitemOpen
  \bibfield  {author} {\bibinfo {author} {\bibfnamefont {J.}~\bibnamefont
  {{Chluba}}}, \bibinfo {author} {\bibfnamefont {A.~L.}\ \bibnamefont
  {{Erickcek}}},\ and\ \bibinfo {author} {\bibfnamefont {I.}~\bibnamefont
  {{Ben-Dayan}}},\ }\href {https://doi.org/10.1088/0004-637X/758/2/76}
  {\bibfield  {journal} {\bibinfo  {journal} {\apj}\ }\textbf {\bibinfo
  {volume} {758}},\ \bibinfo {eid} {76} (\bibinfo {year}
  {2012}{\natexlab{a}})},\ \Eprint {https://arxiv.org/abs/1203.2681}
  {arXiv:1203.2681 [astro-ph.CO]} \BibitemShut {NoStop}%
\bibitem [{\citenamefont {{Chluba}}\ \emph
  {et~al.}(2012{\natexlab{b}})\citenamefont {{Chluba}}, \citenamefont
  {{Khatri}},\ and\ \citenamefont {{Sunyaev}}}]{Chluba_S12}%
  \BibitemOpen
  \bibfield  {author} {\bibinfo {author} {\bibfnamefont {J.}~\bibnamefont
  {{Chluba}}}, \bibinfo {author} {\bibfnamefont {R.}~\bibnamefont {{Khatri}}},\
  and\ \bibinfo {author} {\bibfnamefont {R.~A.}\ \bibnamefont {{Sunyaev}}},\
  }\href {https://doi.org/10.1111/j.1365-2966.2012.21474.x} {\bibfield
  {journal} {\bibinfo  {journal} {MNRAS}\ }\textbf {\bibinfo {volume} {425}},\
  \bibinfo {pages} {1129} (\bibinfo {year} {2012}{\natexlab{b}})},\ \Eprint
  {https://arxiv.org/abs/1202.0057} {arXiv:1202.0057 [astro-ph.CO]}
  \BibitemShut {NoStop}%
\bibitem [{\citenamefont {Emami}\ and\ \citenamefont
  {Smoot}(2018)}]{EmamiSmoot17}%
  \BibitemOpen
  \bibfield  {author} {\bibinfo {author} {\bibfnamefont {R.}~\bibnamefont
  {Emami}}\ and\ \bibinfo {author} {\bibfnamefont {G.}~\bibnamefont {Smoot}},\
  }\href {https://doi.org/10.1088/1475-7516/2018/01/007} {\bibfield  {journal}
  {\bibinfo  {journal} {J. Cosmol. Astropart. Phys.}\ }\textbf {\bibinfo
  {volume} {1801}}\bibfield  {number} {\bibinfo  {number} { (01)},\ \bibinfo
  {pages} {007}},\ }\Eprint {https://arxiv.org/abs/1705.09924}
  {arXiv:1705.09924 [astro-ph.CO]} \BibitemShut {NoStop}%
\bibitem [{\citenamefont {{Frenk}}\ and\ \citenamefont
  {{White}}(2012)}]{White_F12}%
  \BibitemOpen
  \bibfield  {author} {\bibinfo {author} {\bibfnamefont {C.~S.}\ \bibnamefont
  {{Frenk}}}\ and\ \bibinfo {author} {\bibfnamefont {S.~D.~M.}\ \bibnamefont
  {{White}}},\ }\href {https://doi.org/10.1002/andp.201200212} {\bibfield
  {journal} {\bibinfo  {journal} {Annalen der Physik}\ }\textbf {\bibinfo
  {volume} {524}},\ \bibinfo {pages} {507} (\bibinfo {year} {2012})},\ \Eprint
  {https://arxiv.org/abs/1210.0544} {arXiv:1210.0544 [astro-ph.CO]}
  \BibitemShut {NoStop}%
\bibitem [{\citenamefont {{Del Popolo}}\ and\ \citenamefont {{Le
  Delliou}}(2017)}]{delPopolo17}%
  \BibitemOpen
  \bibfield  {author} {\bibinfo {author} {\bibfnamefont {A.}~\bibnamefont {{Del
  Popolo}}}\ and\ \bibinfo {author} {\bibfnamefont {M.}~\bibnamefont {{Le
  Delliou}}},\ }\href {https://doi.org/10.3390/galaxies5010017} {\bibfield
  {journal} {\bibinfo  {journal} {Galaxies}\ }\textbf {\bibinfo {volume} {5}},\
  \bibinfo {pages} {17} (\bibinfo {year} {2017})},\ \Eprint
  {https://arxiv.org/abs/1606.07790} {arXiv:1606.07790 [astro-ph.CO]}
  \BibitemShut {NoStop}%
\bibitem [{\citenamefont {{Bullock}}\ and\ \citenamefont
  {{Boylan-Kolchin}}(2017)}]{Bullock_B17}%
  \BibitemOpen
  \bibfield  {author} {\bibinfo {author} {\bibfnamefont {J.~S.}\ \bibnamefont
  {{Bullock}}}\ and\ \bibinfo {author} {\bibfnamefont {M.}~\bibnamefont
  {{Boylan-Kolchin}}},\ }\href
  {https://doi.org/10.1146/annurev-astro-091916-055313} {\bibfield  {journal}
  {\bibinfo  {journal} {Annual Review of Astronomy and Astrophysics}\ }\textbf
  {\bibinfo {volume} {55}},\ \bibinfo {pages} {343} (\bibinfo {year} {2017})},\
  \Eprint {https://arxiv.org/abs/1707.04256} {arXiv:1707.04256 [astro-ph.CO]}
  \BibitemShut {NoStop}%
\bibitem [{\citenamefont {De~Rosa}\ \emph {et~al.}(2014)\citenamefont {De~Rosa}
  \emph {et~al.}}]{DeRosa_SMBH14}%
  \BibitemOpen
  \bibfield  {author} {\bibinfo {author} {\bibfnamefont {G.}~\bibnamefont
  {De~Rosa}} \emph {et~al.},\ }\href
  {https://doi.org/10.1088/0004-637X/790/2/145} {\bibfield  {journal} {\bibinfo
   {journal} {Astrophys. J.}\ }\textbf {\bibinfo {volume} {790}},\ \bibinfo
  {pages} {145} (\bibinfo {year} {2014})},\ \Eprint
  {https://arxiv.org/abs/1311.3260} {arXiv:1311.3260 [astro-ph.CO]}
  \BibitemShut {NoStop}%
\bibitem [{\citenamefont {{Wu}}\ \emph {et~al.}(2015)\citenamefont {{Wu}},
  \citenamefont {{Wang}}, \citenamefont {{Fan}}, \citenamefont {{Yi}},
  \citenamefont {{Zuo}}, \citenamefont {{Bian}}, \citenamefont {{Jiang}},
  \citenamefont {{McGreer}}, \citenamefont {{Wang}}, \citenamefont {{Yang}},
  \citenamefont {{Yang}}, \citenamefont {{Thompson}},\ and\ \citenamefont
  {{Beletsky}}}]{Wu_SMBH_Natur15}%
  \BibitemOpen
  \bibfield  {author} {\bibinfo {author} {\bibfnamefont {X.-B.}\ \bibnamefont
  {{Wu}}}, \bibinfo {author} {\bibfnamefont {F.}~\bibnamefont {{Wang}}},
  \bibinfo {author} {\bibfnamefont {X.}~\bibnamefont {{Fan}}}, \bibinfo
  {author} {\bibfnamefont {W.}~\bibnamefont {{Yi}}}, \bibinfo {author}
  {\bibfnamefont {W.}~\bibnamefont {{Zuo}}}, \bibinfo {author} {\bibfnamefont
  {F.}~\bibnamefont {{Bian}}}, \bibinfo {author} {\bibfnamefont
  {L.}~\bibnamefont {{Jiang}}}, \bibinfo {author} {\bibfnamefont {I.~D.}\
  \bibnamefont {{McGreer}}}, \bibinfo {author} {\bibfnamefont {R.}~\bibnamefont
  {{Wang}}}, \bibinfo {author} {\bibfnamefont {J.}~\bibnamefont {{Yang}}},
  \bibinfo {author} {\bibfnamefont {Q.}~\bibnamefont {{Yang}}}, \bibinfo
  {author} {\bibfnamefont {D.}~\bibnamefont {{Thompson}}},\ and\ \bibinfo
  {author} {\bibfnamefont {Y.}~\bibnamefont {{Beletsky}}},\ }\href
  {https://doi.org/10.1038/nature14241} {\bibfield  {journal} {\bibinfo
  {journal} {\nat}\ }\textbf {\bibinfo {volume} {518}},\ \bibinfo {pages} {512}
  (\bibinfo {year} {2015})},\ \Eprint {https://arxiv.org/abs/1502.07418}
  {arXiv:1502.07418 [astro-ph.GA]} \BibitemShut {NoStop}%
\bibitem [{\citenamefont {{Watson}}\ \emph {et~al.}(2015)\citenamefont
  {{Watson}}, \citenamefont {{Christensen}}, \citenamefont {{Knudsen}},
  \citenamefont {{Richard}}, \citenamefont {{Gallazzi}},\ and\ \citenamefont
  {{Micha{\l}owski}}}]{Dusty_Gal_15}%
  \BibitemOpen
  \bibfield  {author} {\bibinfo {author} {\bibfnamefont {D.}~\bibnamefont
  {{Watson}}}, \bibinfo {author} {\bibfnamefont {L.}~\bibnamefont
  {{Christensen}}}, \bibinfo {author} {\bibfnamefont {K.~K.}\ \bibnamefont
  {{Knudsen}}}, \bibinfo {author} {\bibfnamefont {J.}~\bibnamefont
  {{Richard}}}, \bibinfo {author} {\bibfnamefont {A.}~\bibnamefont
  {{Gallazzi}}},\ and\ \bibinfo {author} {\bibfnamefont {M.~J.}\ \bibnamefont
  {{Micha{\l}owski}}},\ }\href {https://doi.org/10.1038/nature14164} {\bibfield
   {journal} {\bibinfo  {journal} {\nat}\ }\textbf {\bibinfo {volume} {519}},\
  \bibinfo {pages} {327} (\bibinfo {year} {2015})},\ \Eprint
  {https://arxiv.org/abs/1503.00002} {arXiv:1503.00002 [astro-ph.GA]}
  \BibitemShut {NoStop}%
\bibitem [{\citenamefont {Hirano}\ \emph {et~al.}(2015)\citenamefont {Hirano},
  \citenamefont {Zhu}, \citenamefont {Yoshida}, \citenamefont {Spergel},\ and\
  \citenamefont {Yorke}}]{Hirano_Spergel2015}%
  \BibitemOpen
  \bibfield  {author} {\bibinfo {author} {\bibfnamefont {S.}~\bibnamefont
  {Hirano}}, \bibinfo {author} {\bibfnamefont {N.}~\bibnamefont {Zhu}},
  \bibinfo {author} {\bibfnamefont {N.}~\bibnamefont {Yoshida}}, \bibinfo
  {author} {\bibfnamefont {D.}~\bibnamefont {Spergel}},\ and\ \bibinfo {author}
  {\bibfnamefont {H.~W.}\ \bibnamefont {Yorke}},\ }\href
  {https://doi.org/10.1088/0004-637X/814/1/18} {\bibfield  {journal} {\bibinfo
  {journal} {Astrophys. J.}\ }\textbf {\bibinfo {volume} {814}},\ \bibinfo
  {pages} {18} (\bibinfo {year} {2015})},\ \Eprint
  {https://arxiv.org/abs/1504.05186} {arXiv:1504.05186 [astro-ph.CO]}
  \BibitemShut {NoStop}%
\bibitem [{\citenamefont {{Steinhardt}}\ \emph {et~al.}(2016)\citenamefont
  {{Steinhardt}}, \citenamefont {{Capak}}, \citenamefont {{Masters}},\ and\
  \citenamefont {{Speagle}}}]{Imp_earl16}%
  \BibitemOpen
  \bibfield  {author} {\bibinfo {author} {\bibfnamefont {C.~L.}\ \bibnamefont
  {{Steinhardt}}}, \bibinfo {author} {\bibfnamefont {P.}~\bibnamefont
  {{Capak}}}, \bibinfo {author} {\bibfnamefont {D.}~\bibnamefont {{Masters}}},\
  and\ \bibinfo {author} {\bibfnamefont {J.~S.}\ \bibnamefont {{Speagle}}},\
  }\href {https://doi.org/10.3847/0004-637X/824/1/21} {\bibfield  {journal}
  {\bibinfo  {journal} {Astrophys. J.}\ }\textbf {\bibinfo {volume} {824}},\
  \bibinfo {eid} {21} (\bibinfo {year} {2016})},\ \Eprint
  {https://arxiv.org/abs/1506.01377} {arXiv:1506.01377 [astro-ph.GA]}
  \BibitemShut {NoStop}%
\bibitem [{\citenamefont {{Valiante}}\ \emph {et~al.}(2016)\citenamefont
  {{Valiante}}, \citenamefont {{Schneider}}, \citenamefont {{Volonteri}},\ and\
  \citenamefont {{Omukai}}}]{First_StarsBH_Volont16}%
  \BibitemOpen
  \bibfield  {author} {\bibinfo {author} {\bibfnamefont {R.}~\bibnamefont
  {{Valiante}}}, \bibinfo {author} {\bibfnamefont {R.}~\bibnamefont
  {{Schneider}}}, \bibinfo {author} {\bibfnamefont {M.}~\bibnamefont
  {{Volonteri}}},\ and\ \bibinfo {author} {\bibfnamefont {K.}~\bibnamefont
  {{Omukai}}},\ }\href {https://doi.org/10.1093/mnras/stw225} {\bibfield
  {journal} {\bibinfo  {journal} {MNRAS}\ }\textbf {\bibinfo {volume} {457}},\
  \bibinfo {pages} {3356} (\bibinfo {year} {2016})},\ \Eprint
  {https://arxiv.org/abs/1601.07915} {arXiv:1601.07915 [astro-ph.GA]}
  \BibitemShut {NoStop}%
\bibitem [{\citenamefont {{Stefanon}}\ \emph {et~al.}(2017)\citenamefont
  {{Stefanon}}, \citenamefont {{Bouwens}}, \citenamefont {{Labb{\'e}}},
  \citenamefont {{Muzzin}}, \citenamefont {{Marchesini}}, \citenamefont
  {{Oesch}},\ and\ \citenamefont {{Gonzalez}}}]{Bouwens_halostellar17}%
  \BibitemOpen
  \bibfield  {author} {\bibinfo {author} {\bibfnamefont {M.}~\bibnamefont
  {{Stefanon}}}, \bibinfo {author} {\bibfnamefont {R.~J.}\ \bibnamefont
  {{Bouwens}}}, \bibinfo {author} {\bibfnamefont {I.}~\bibnamefont
  {{Labb{\'e}}}}, \bibinfo {author} {\bibfnamefont {A.}~\bibnamefont
  {{Muzzin}}}, \bibinfo {author} {\bibfnamefont {D.}~\bibnamefont
  {{Marchesini}}}, \bibinfo {author} {\bibfnamefont {P.}~\bibnamefont
  {{Oesch}}},\ and\ \bibinfo {author} {\bibfnamefont {V.}~\bibnamefont
  {{Gonzalez}}},\ }\href {https://doi.org/10.3847/1538-4357/aa72d8} {\bibfield
  {journal} {\bibinfo  {journal} {Astrophys. J.}\ }\textbf {\bibinfo {volume}
  {843}},\ \bibinfo {eid} {36} (\bibinfo {year} {2017})},\ \Eprint
  {https://arxiv.org/abs/1611.09354} {arXiv:1611.09354 [astro-ph.GA]}
  \BibitemShut {NoStop}%
\bibitem [{\citenamefont {{Glazebrook}}\ \emph
  {et~al.}(2017{\natexlab{a}})\citenamefont {{Glazebrook}}, \citenamefont
  {{Schreiber}}, \citenamefont {{Labb{\'e}}}, \citenamefont {{Nanayakkara}},
  \citenamefont {{Kacprzak}}, \citenamefont {{Oesch}}, \citenamefont
  {{Papovich}}, \citenamefont {{Spitler}}, \citenamefont {{Straatman}},
  \citenamefont {{Tran}},\ and\ \citenamefont {{Yuan}}}]{Massive_Q_Glaze17}%
  \BibitemOpen
  \bibfield  {author} {\bibinfo {author} {\bibfnamefont {K.}~\bibnamefont
  {{Glazebrook}}}, \bibinfo {author} {\bibfnamefont {C.}~\bibnamefont
  {{Schreiber}}}, \bibinfo {author} {\bibfnamefont {I.}~\bibnamefont
  {{Labb{\'e}}}}, \bibinfo {author} {\bibfnamefont {T.}~\bibnamefont
  {{Nanayakkara}}}, \bibinfo {author} {\bibfnamefont {G.~G.}\ \bibnamefont
  {{Kacprzak}}}, \bibinfo {author} {\bibfnamefont {P.~A.}\ \bibnamefont
  {{Oesch}}}, \bibinfo {author} {\bibfnamefont {C.}~\bibnamefont {{Papovich}}},
  \bibinfo {author} {\bibfnamefont {L.~R.}\ \bibnamefont {{Spitler}}}, \bibinfo
  {author} {\bibfnamefont {C.~M.~S.}\ \bibnamefont {{Straatman}}}, \bibinfo
  {author} {\bibfnamefont {K.-V.~H.}\ \bibnamefont {{Tran}}},\ and\ \bibinfo
  {author} {\bibfnamefont {T.}~\bibnamefont {{Yuan}}},\ }\href
  {https://doi.org/10.1038/nature21680} {\bibfield  {journal} {\bibinfo
  {journal} {\nat}\ }\textbf {\bibinfo {volume} {544}},\ \bibinfo {pages} {71}
  (\bibinfo {year} {2017}{\natexlab{a}})},\ \Eprint
  {https://arxiv.org/abs/1702.01751} {arXiv:1702.01751 [astro-ph.GA]}
  \BibitemShut {NoStop}%
\bibitem [{\citenamefont {{Venemans}}\ \emph {et~al.}(2017)\citenamefont
  {{Venemans}}, \citenamefont {{Walter}}, \citenamefont {{Decarli}},
  \citenamefont {{Ba{\~n}ados}}, \citenamefont {{Carilli}}, \citenamefont
  {{Winters}}, \citenamefont {{Schuster}}, \citenamefont {{da Cunha}},
  \citenamefont {{Fan}}, \citenamefont {{Farina}}, \citenamefont
  {{Mazzucchelli}}, \citenamefont {{Rix}},\ and\ \citenamefont
  {{Weiss}}}]{Dust_quas17}%
  \BibitemOpen
  \bibfield  {author} {\bibinfo {author} {\bibfnamefont {B.~P.}\ \bibnamefont
  {{Venemans}}}, \bibinfo {author} {\bibfnamefont {F.}~\bibnamefont
  {{Walter}}}, \bibinfo {author} {\bibfnamefont {R.}~\bibnamefont {{Decarli}}},
  \bibinfo {author} {\bibfnamefont {E.}~\bibnamefont {{Ba{\~n}ados}}}, \bibinfo
  {author} {\bibfnamefont {C.}~\bibnamefont {{Carilli}}}, \bibinfo {author}
  {\bibfnamefont {J.~M.}\ \bibnamefont {{Winters}}}, \bibinfo {author}
  {\bibfnamefont {K.}~\bibnamefont {{Schuster}}}, \bibinfo {author}
  {\bibfnamefont {E.}~\bibnamefont {{da Cunha}}}, \bibinfo {author}
  {\bibfnamefont {X.}~\bibnamefont {{Fan}}}, \bibinfo {author} {\bibfnamefont
  {E.~P.}\ \bibnamefont {{Farina}}}, \bibinfo {author} {\bibfnamefont
  {C.}~\bibnamefont {{Mazzucchelli}}}, \bibinfo {author} {\bibfnamefont
  {H.-W.}\ \bibnamefont {{Rix}}},\ and\ \bibinfo {author} {\bibfnamefont
  {A.}~\bibnamefont {{Weiss}}},\ }\href
  {https://doi.org/10.3847/2041-8213/aa943a} {\bibfield  {journal} {\bibinfo
  {journal} {Astrophys. J. Lett.}\ }\textbf {\bibinfo {volume} {851}},\
  \bibinfo {eid} {L8} (\bibinfo {year} {2017})},\ \Eprint
  {https://arxiv.org/abs/1712.01886} {arXiv:1712.01886 [astro-ph.GA]}
  \BibitemShut {NoStop}%
\bibitem [{\citenamefont {{Behroozi}}\ and\ \citenamefont
  {{Silk}}(2018)}]{Silk_LCDM_SMBH18}%
  \BibitemOpen
  \bibfield  {author} {\bibinfo {author} {\bibfnamefont {P.}~\bibnamefont
  {{Behroozi}}}\ and\ \bibinfo {author} {\bibfnamefont {J.}~\bibnamefont
  {{Silk}}},\ }\href {https://doi.org/10.1093/mnras/sty945} {\bibfield
  {journal} {\bibinfo  {journal} {MNRAS}\ }\textbf {\bibinfo {volume} {477}},\
  \bibinfo {pages} {5382} (\bibinfo {year} {2018})},\ \Eprint
  {https://arxiv.org/abs/1609.04402} {arXiv:1609.04402 [astro-ph.GA]}
  \BibitemShut {NoStop}%
\bibitem [{\citenamefont {{Tang}}\ \emph {et~al.}(2019)\citenamefont {{Tang}},
  \citenamefont {{Goto}}, \citenamefont {{Ohyama}}, \citenamefont {{Jin}},
  \citenamefont {{Done}}, \citenamefont {{Lu}}, \citenamefont {{Hashimoto}},
  \citenamefont {{Kilerci Eser}}, \citenamefont {{Chiang}},\ and\ \citenamefont
  {{Kim}}}]{SMBH19}%
  \BibitemOpen
  \bibfield  {author} {\bibinfo {author} {\bibfnamefont {J.-J.}\ \bibnamefont
  {{Tang}}}, \bibinfo {author} {\bibfnamefont {T.}~\bibnamefont {{Goto}}},
  \bibinfo {author} {\bibfnamefont {Y.}~\bibnamefont {{Ohyama}}}, \bibinfo
  {author} {\bibfnamefont {C.}~\bibnamefont {{Jin}}}, \bibinfo {author}
  {\bibfnamefont {C.}~\bibnamefont {{Done}}}, \bibinfo {author} {\bibfnamefont
  {T.-Y.}\ \bibnamefont {{Lu}}}, \bibinfo {author} {\bibfnamefont
  {T.}~\bibnamefont {{Hashimoto}}}, \bibinfo {author} {\bibfnamefont
  {E.}~\bibnamefont {{Kilerci Eser}}}, \bibinfo {author} {\bibfnamefont
  {C.-Y.}\ \bibnamefont {{Chiang}}},\ and\ \bibinfo {author} {\bibfnamefont
  {S.~J.}\ \bibnamefont {{Kim}}},\ }\href
  {https://doi.org/10.1093/mnras/stz134} {\bibfield  {journal} {\bibinfo
  {journal} {MNRAS}\ }\textbf {\bibinfo {volume} {484}},\ \bibinfo {pages}
  {2575} (\bibinfo {year} {2019})},\ \Eprint {https://arxiv.org/abs/1901.02615}
  {arXiv:1901.02615 [astro-ph.GA]} \BibitemShut {NoStop}%
\bibitem [{\citenamefont {{Cecchi}}\ \emph {et~al.}(2019)\citenamefont
  {{Cecchi}}, \citenamefont {{Bolzonella}}, \citenamefont {{Cimatti}},\ and\
  \citenamefont {{Girelli}}}]{Queis_Obs_mod19}%
  \BibitemOpen
  \bibfield  {author} {\bibinfo {author} {\bibfnamefont {R.}~\bibnamefont
  {{Cecchi}}}, \bibinfo {author} {\bibfnamefont {M.}~\bibnamefont
  {{Bolzonella}}}, \bibinfo {author} {\bibfnamefont {A.}~\bibnamefont
  {{Cimatti}}},\ and\ \bibinfo {author} {\bibfnamefont {G.}~\bibnamefont
  {{Girelli}}},\ }\href {https://doi.org/10.3847/2041-8213/ab2c80} {\bibfield
  {journal} {\bibinfo  {journal} {Astrophys. J.}\ }\textbf {\bibinfo {volume}
  {880}},\ \bibinfo {eid} {L14} (\bibinfo {year} {2019})},\ \Eprint
  {https://arxiv.org/abs/1906.11842} {arXiv:1906.11842 [astro-ph.GA]}
  \BibitemShut {NoStop}%
\bibitem [{\citenamefont {{Smith}}\ and\ \citenamefont
  {{Bromm}}(2019)}]{Bromm_SMBH_Rev19}%
  \BibitemOpen
  \bibfield  {author} {\bibinfo {author} {\bibfnamefont {A.}~\bibnamefont
  {{Smith}}}\ and\ \bibinfo {author} {\bibfnamefont {V.}~\bibnamefont
  {{Bromm}}},\ }\href@noop {} {\bibfield  {journal} {\bibinfo  {journal} {arXiv
  e-prints}\ ,\ \bibinfo {eid} {arXiv:1904.12890}} (\bibinfo {year} {2019})},\
  \Eprint {https://arxiv.org/abs/1904.12890} {arXiv:1904.12890 [astro-ph.GA]}
  \BibitemShut {NoStop}%
\bibitem [{\citenamefont {Wang}\ \emph {et~al.}(2019)\citenamefont {Wang} \emph
  {et~al.}}]{Dominant_Yoshi2019}%
  \BibitemOpen
  \bibfield  {author} {\bibinfo {author} {\bibfnamefont {T.}~\bibnamefont
  {Wang}} \emph {et~al.},\ }\href {https://doi.org/10.1038/s41586-019-1452-4}
  {\bibfield  {journal} {\bibinfo  {journal} {Nature}\ }\textbf {\bibinfo
  {volume} {572}},\ \bibinfo {pages} {211} (\bibinfo {year} {2019})},\ \Eprint
  {https://arxiv.org/abs/1908.02372} {arXiv:1908.02372 [astro-ph.GA]}
  \BibitemShut {NoStop}%
\bibitem [{\citenamefont {{El-Zant}}\ \emph {et~al.}(2001)\citenamefont
  {{El-Zant}}, \citenamefont {{Shlosman}},\ and\ \citenamefont
  {{Hoffman}}}]{Zant2001}%
  \BibitemOpen
  \bibfield  {author} {\bibinfo {author} {\bibfnamefont {A.}~\bibnamefont
  {{El-Zant}}}, \bibinfo {author} {\bibfnamefont {I.}~\bibnamefont
  {{Shlosman}}},\ and\ \bibinfo {author} {\bibfnamefont {Y.}~\bibnamefont
  {{Hoffman}}},\ }\href {https://doi.org/10.1086/322516} {\bibfield  {journal}
  {\bibinfo  {journal} {Astrophys. Jour.}\ }\textbf {\bibinfo {volume} {560}},\
  \bibinfo {pages} {636} (\bibinfo {year} {2001})},\ \Eprint
  {https://arxiv.org/abs/astro-ph/0103386} {astro-ph/0103386} \BibitemShut
  {NoStop}%
\bibitem [{\citenamefont {{El-Zant}}\ \emph {et~al.}(2004)\citenamefont
  {{El-Zant}}, \citenamefont {{Hoffman}}, \citenamefont {{Primack}},
  \citenamefont {{Combes}},\ and\ \citenamefont {{Shlosman}}}]{Zant2004}%
  \BibitemOpen
  \bibfield  {author} {\bibinfo {author} {\bibfnamefont {A.~A.}\ \bibnamefont
  {{El-Zant}}}, \bibinfo {author} {\bibfnamefont {Y.}~\bibnamefont
  {{Hoffman}}}, \bibinfo {author} {\bibfnamefont {J.}~\bibnamefont
  {{Primack}}}, \bibinfo {author} {\bibfnamefont {F.}~\bibnamefont
  {{Combes}}},\ and\ \bibinfo {author} {\bibfnamefont {I.}~\bibnamefont
  {{Shlosman}}},\ }\href {https://doi.org/10.1086/421938} {\bibfield  {journal}
  {\bibinfo  {journal} {Astrophys. Jour.l}\ }\textbf {\bibinfo {volume}
  {607}},\ \bibinfo {pages} {L75} (\bibinfo {year} {2004})},\ \Eprint
  {https://arxiv.org/abs/astro-ph/0309412} {astro-ph/0309412} \BibitemShut
  {NoStop}%
\bibitem [{\citenamefont {{Romano-D{\'{\i}}az}}\ \emph
  {et~al.}(2008)\citenamefont {{Romano-D{\'{\i}}az}}, \citenamefont
  {{Shlosman}}, \citenamefont {{Hoffman}},\ and\ \citenamefont
  {{Heller}}}]{RomanoDiaz2008}%
  \BibitemOpen
  \bibfield  {author} {\bibinfo {author} {\bibfnamefont {E.}~\bibnamefont
  {{Romano-D{\'{\i}}az}}}, \bibinfo {author} {\bibfnamefont {I.}~\bibnamefont
  {{Shlosman}}}, \bibinfo {author} {\bibfnamefont {Y.}~\bibnamefont
  {{Hoffman}}},\ and\ \bibinfo {author} {\bibfnamefont {C.}~\bibnamefont
  {{Heller}}},\ }\href {https://doi.org/10.1086/592687} {\bibfield  {journal}
  {\bibinfo  {journal} {Astrophys. Jour.l}\ }\textbf {\bibinfo {volume}
  {685}},\ \bibinfo {pages} {L105} (\bibinfo {year} {2008})},\ \Eprint
  {https://arxiv.org/abs/0808.0195} {arXiv:0808.0195} \BibitemShut {NoStop}%
\bibitem [{\citenamefont {{Goerdt}}\ \emph {et~al.}(2010)\citenamefont
  {{Goerdt}}, \citenamefont {{Moore}}, \citenamefont {{Read}},\ and\
  \citenamefont {{Stadel}}}]{Goerdt2010}%
  \BibitemOpen
  \bibfield  {author} {\bibinfo {author} {\bibfnamefont {T.}~\bibnamefont
  {{Goerdt}}}, \bibinfo {author} {\bibfnamefont {B.}~\bibnamefont {{Moore}}},
  \bibinfo {author} {\bibfnamefont {J.~I.}\ \bibnamefont {{Read}}},\ and\
  \bibinfo {author} {\bibfnamefont {J.}~\bibnamefont {{Stadel}}},\ }\href
  {https://doi.org/10.1088/0004-637X/725/2/1707} {\bibfield  {journal}
  {\bibinfo  {journal} {Astrophys. Jour.}\ }\textbf {\bibinfo {volume} {725}},\
  \bibinfo {pages} {1707} (\bibinfo {year} {2010})},\ \Eprint
  {https://arxiv.org/abs/0806.1951} {arXiv:0806.1951} \BibitemShut {NoStop}%
\bibitem [{\citenamefont {{Cole}}\ \emph {et~al.}(2011)\citenamefont {{Cole}},
  \citenamefont {{Dehnen}},\ and\ \citenamefont {{Wilkinson}}}]{Cole2011}%
  \BibitemOpen
  \bibfield  {author} {\bibinfo {author} {\bibfnamefont {D.~R.}\ \bibnamefont
  {{Cole}}}, \bibinfo {author} {\bibfnamefont {W.}~\bibnamefont {{Dehnen}}},\
  and\ \bibinfo {author} {\bibfnamefont {M.~I.}\ \bibnamefont {{Wilkinson}}},\
  }\href {https://doi.org/10.1111/j.1365-2966.2011.19110.x} {\bibfield
  {journal} {\bibinfo  {journal} {Mon. Not. Roy. Ast. Soc.}\ }\textbf {\bibinfo
  {volume} {416}},\ \bibinfo {pages} {1118} (\bibinfo {year} {2011})},\ \Eprint
  {https://arxiv.org/abs/1105.4050} {arXiv:1105.4050} \BibitemShut {NoStop}%
\bibitem [{\citenamefont {{Inoue}}\ and\ \citenamefont
  {{Saitoh}}(2011)}]{Inoue2011}%
  \BibitemOpen
  \bibfield  {author} {\bibinfo {author} {\bibfnamefont {S.}~\bibnamefont
  {{Inoue}}}\ and\ \bibinfo {author} {\bibfnamefont {T.~R.}\ \bibnamefont
  {{Saitoh}}},\ }\href {https://doi.org/10.1111/j.1365-2966.2011.19873.x}
  {\bibfield  {journal} {\bibinfo  {journal} {Mon. Not. Roy. Ast. Soc.}\
  }\textbf {\bibinfo {volume} {418}},\ \bibinfo {pages} {2527} (\bibinfo {year}
  {2011})},\ \Eprint {https://arxiv.org/abs/1108.0906} {arXiv:1108.0906}
  \BibitemShut {NoStop}%
\bibitem [{\citenamefont {{Laporte}}\ and\ \citenamefont
  {{White}}(2015)}]{Laporte2015}%
  \BibitemOpen
  \bibfield  {author} {\bibinfo {author} {\bibfnamefont {C.~F.~P.}\
  \bibnamefont {{Laporte}}}\ and\ \bibinfo {author} {\bibfnamefont {S.~D.~M.}\
  \bibnamefont {{White}}},\ }\href {https://doi.org/10.1093/mnras/stv112}
  {\bibfield  {journal} {\bibinfo  {journal} {Mon. Not. Roy. Ast. Soc.}\
  }\textbf {\bibinfo {volume} {451}},\ \bibinfo {pages} {1177} (\bibinfo {year}
  {2015})},\ \Eprint {https://arxiv.org/abs/1409.1924} {arXiv:1409.1924
  [astro-ph.GA]} \BibitemShut {NoStop}%
\bibitem [{\citenamefont {{Nipoti}}\ and\ \citenamefont
  {{Binney}}(2015)}]{Nipoti2015}%
  \BibitemOpen
  \bibfield  {author} {\bibinfo {author} {\bibfnamefont {C.}~\bibnamefont
  {{Nipoti}}}\ and\ \bibinfo {author} {\bibfnamefont {J.}~\bibnamefont
  {{Binney}}},\ }\href {https://doi.org/10.1093/mnras/stu2217} {\bibfield
  {journal} {\bibinfo  {journal} {Mon. Not. Roy. Ast. Soc.}\ }\textbf {\bibinfo
  {volume} {446}},\ \bibinfo {pages} {1820} (\bibinfo {year} {2015})},\ \Eprint
  {https://arxiv.org/abs/1410.6169} {arXiv:1410.6169} \BibitemShut {NoStop}%
\bibitem [{\citenamefont {{Del Popolo}}\ \emph {et~al.}(2018)\citenamefont
  {{Del Popolo}}, \citenamefont {{Pace}}, \citenamefont {{Le Delliou}},\ and\
  \citenamefont {{Lee}}}]{delPopolo2018}%
  \BibitemOpen
  \bibfield  {author} {\bibinfo {author} {\bibfnamefont {A.}~\bibnamefont {{Del
  Popolo}}}, \bibinfo {author} {\bibfnamefont {F.}~\bibnamefont {{Pace}}},
  \bibinfo {author} {\bibfnamefont {M.}~\bibnamefont {{Le Delliou}}},\ and\
  \bibinfo {author} {\bibfnamefont {X.}~\bibnamefont {{Lee}}},\ }\href
  {https://doi.org/10.1103/PhysRevD.98.063517} {\bibfield  {journal} {\bibinfo
  {journal} {Phys. Rev. D}\ }\textbf {\bibinfo {volume} {98}},\ \bibinfo {eid}
  {063517} (\bibinfo {year} {2018})},\ \Eprint
  {https://arxiv.org/abs/1809.10609} {arXiv:1809.10609} \BibitemShut {NoStop}%
\bibitem [{\citenamefont {{Read}}\ and\ \citenamefont
  {{Gilmore}}(2005)}]{Read2005}%
  \BibitemOpen
  \bibfield  {author} {\bibinfo {author} {\bibfnamefont {J.~I.}\ \bibnamefont
  {{Read}}}\ and\ \bibinfo {author} {\bibfnamefont {G.}~\bibnamefont
  {{Gilmore}}},\ }\href {https://doi.org/10.1111/j.1365-2966.2004.08424.x}
  {\bibfield  {journal} {\bibinfo  {journal} {Mon. Not. Roy. Ast. Soc.}\
  }\textbf {\bibinfo {volume} {356}},\ \bibinfo {pages} {107} (\bibinfo {year}
  {2005})},\ \Eprint {https://arxiv.org/abs/astro-ph/0409565}
  {astro-ph/0409565} \BibitemShut {NoStop}%
\bibitem [{\citenamefont {{Mashchenko}}\ \emph {et~al.}(2008)\citenamefont
  {{Mashchenko}}, \citenamefont {{Wadsley}},\ and\ \citenamefont
  {{Couchman}}}]{Mashchenko2008}%
  \BibitemOpen
  \bibfield  {author} {\bibinfo {author} {\bibfnamefont {S.}~\bibnamefont
  {{Mashchenko}}}, \bibinfo {author} {\bibfnamefont {J.}~\bibnamefont
  {{Wadsley}}},\ and\ \bibinfo {author} {\bibfnamefont {H.~M.~P.}\ \bibnamefont
  {{Couchman}}},\ }\href {https://doi.org/10.1126/science.1148666} {\bibfield
  {journal} {\bibinfo  {journal} {Science}\ }\textbf {\bibinfo {volume}
  {319}},\ \bibinfo {pages} {174} (\bibinfo {year} {2008})},\ \Eprint
  {https://arxiv.org/abs/0711.4803} {arXiv:0711.4803} \BibitemShut {NoStop}%
\bibitem [{\citenamefont {{Peirani}}\ \emph {et~al.}(2008)\citenamefont
  {{Peirani}}, \citenamefont {{Kay}},\ and\ \citenamefont
  {{Silk}}}]{Peirani2008}%
  \BibitemOpen
  \bibfield  {author} {\bibinfo {author} {\bibfnamefont {S.}~\bibnamefont
  {{Peirani}}}, \bibinfo {author} {\bibfnamefont {S.}~\bibnamefont {{Kay}}},\
  and\ \bibinfo {author} {\bibfnamefont {J.}~\bibnamefont {{Silk}}},\ }\href
  {https://doi.org/10.1051/0004-6361:20077956} {\bibfield  {journal} {\bibinfo
  {journal} {Atron. \& Astrophys.}\ }\textbf {\bibinfo {volume} {479}},\
  \bibinfo {pages} {123} (\bibinfo {year} {2008})},\ \Eprint
  {https://arxiv.org/abs/astro-ph/0612468} {astro-ph/0612468} \BibitemShut
  {NoStop}%
\bibitem [{\citenamefont {{Pontzen}}\ and\ \citenamefont
  {{Governato}}(2012)}]{Pontzen2012}%
  \BibitemOpen
  \bibfield  {author} {\bibinfo {author} {\bibfnamefont {A.}~\bibnamefont
  {{Pontzen}}}\ and\ \bibinfo {author} {\bibfnamefont {F.}~\bibnamefont
  {{Governato}}},\ }\href {https://doi.org/10.1111/j.1365-2966.2012.20571.x}
  {\bibfield  {journal} {\bibinfo  {journal} {Mon. Not. Roy. Ast. Soc.}\
  }\textbf {\bibinfo {volume} {421}},\ \bibinfo {pages} {3464} (\bibinfo {year}
  {2012})},\ \Eprint {https://arxiv.org/abs/1106.0499} {arXiv:1106.0499}
  \BibitemShut {NoStop}%
\bibitem [{\citenamefont {{Teyssier}}\ \emph {et~al.}(2013)\citenamefont
  {{Teyssier}}, \citenamefont {{Pontzen}}, \citenamefont {{Dubois}},\ and\
  \citenamefont {{Read}}}]{Teyssier2013}%
  \BibitemOpen
  \bibfield  {author} {\bibinfo {author} {\bibfnamefont {R.}~\bibnamefont
  {{Teyssier}}}, \bibinfo {author} {\bibfnamefont {A.}~\bibnamefont
  {{Pontzen}}}, \bibinfo {author} {\bibfnamefont {Y.}~\bibnamefont
  {{Dubois}}},\ and\ \bibinfo {author} {\bibfnamefont {J.~I.}\ \bibnamefont
  {{Read}}},\ }\href {https://doi.org/10.1093/mnras/sts563} {\bibfield
  {journal} {\bibinfo  {journal} {Mon. Not. Roy. Ast. Soc.}\ }\textbf {\bibinfo
  {volume} {429}},\ \bibinfo {pages} {3068} (\bibinfo {year} {2013})},\ \Eprint
  {https://arxiv.org/abs/1206.4895} {arXiv:1206.4895} \BibitemShut {NoStop}%
\bibitem [{\citenamefont {{Pontzen}}\ and\ \citenamefont
  {{Governato}}(2014)}]{Pontzen2014}%
  \BibitemOpen
  \bibfield  {author} {\bibinfo {author} {\bibfnamefont {A.}~\bibnamefont
  {{Pontzen}}}\ and\ \bibinfo {author} {\bibfnamefont {F.}~\bibnamefont
  {{Governato}}},\ }\href {https://doi.org/10.1038/nature12953} {\bibfield
  {journal} {\bibinfo  {journal} {Nature}\ }\textbf {\bibinfo {volume} {506}},\
  \bibinfo {pages} {171} (\bibinfo {year} {2014})},\ \Eprint
  {https://arxiv.org/abs/1402.1764} {arXiv:1402.1764 [astro-ph.CO]}
  \BibitemShut {NoStop}%
\bibitem [{\citenamefont {{El-Zant}}\ \emph {et~al.}(2016)\citenamefont
  {{El-Zant}}, \citenamefont {{Freundlich}},\ and\ \citenamefont
  {{Combes}}}]{EZFC}%
  \BibitemOpen
  \bibfield  {author} {\bibinfo {author} {\bibfnamefont {A.~A.}\ \bibnamefont
  {{El-Zant}}}, \bibinfo {author} {\bibfnamefont {J.}~\bibnamefont
  {{Freundlich}}},\ and\ \bibinfo {author} {\bibfnamefont {F.}~\bibnamefont
  {{Combes}}},\ }\href {https://doi.org/10.1093/mnras/stw1398} {\bibfield
  {journal} {\bibinfo  {journal} {Mon. Not. Roy. Astron. Soc.}\ }\textbf
  {\bibinfo {volume} {461}},\ \bibinfo {pages} {1745} (\bibinfo {year}
  {2016})},\ \Eprint {https://arxiv.org/abs/1603.00526} {arXiv:1603.00526
  [astro-ph.GA]} \BibitemShut {NoStop}%
\bibitem [{\citenamefont {{Col{\'{\i}}n}}\ \emph {et~al.}(2000)\citenamefont
  {{Col{\'{\i}}n}}, \citenamefont {{Avila-Reese}},\ and\ \citenamefont
  {{Valenzuela}}}]{Colin2000}%
  \BibitemOpen
  \bibfield  {author} {\bibinfo {author} {\bibfnamefont {P.}~\bibnamefont
  {{Col{\'{\i}}n}}}, \bibinfo {author} {\bibfnamefont {V.}~\bibnamefont
  {{Avila-Reese}}},\ and\ \bibinfo {author} {\bibfnamefont {O.}~\bibnamefont
  {{Valenzuela}}},\ }\href {https://doi.org/10.1086/317057} {\bibfield
  {journal} {\bibinfo  {journal} {Astrophys. Jour.}\ }\textbf {\bibinfo
  {volume} {542}},\ \bibinfo {pages} {622} (\bibinfo {year} {2000})},\ \Eprint
  {https://arxiv.org/abs/astro-ph/0004115} {astro-ph/0004115} \BibitemShut
  {NoStop}%
\bibitem [{\citenamefont {{Bode}}\ \emph {et~al.}(2001)\citenamefont {{Bode}},
  \citenamefont {{Ostriker}},\ and\ \citenamefont {{Turok}}}]{Bode2001}%
  \BibitemOpen
  \bibfield  {author} {\bibinfo {author} {\bibfnamefont {P.}~\bibnamefont
  {{Bode}}}, \bibinfo {author} {\bibfnamefont {J.~P.}\ \bibnamefont
  {{Ostriker}}},\ and\ \bibinfo {author} {\bibfnamefont {N.}~\bibnamefont
  {{Turok}}},\ }\href {https://doi.org/10.1086/321541} {\bibfield  {journal}
  {\bibinfo  {journal} {Astrophys. Jour.}\ }\textbf {\bibinfo {volume} {556}},\
  \bibinfo {pages} {93} (\bibinfo {year} {2001})},\ \Eprint
  {https://arxiv.org/abs/astro-ph/0010389} {astro-ph/0010389} \BibitemShut
  {NoStop}%
\bibitem [{\citenamefont {{Macci{\`o}}}\ \emph {et~al.}(2012)\citenamefont
  {{Macci{\`o}}}, \citenamefont {{Paduroiu}}, \citenamefont {{Anderhalden}},
  \citenamefont {{Schneider}},\ and\ \citenamefont {{Moore}}}]{Maccio2012b}%
  \BibitemOpen
  \bibfield  {author} {\bibinfo {author} {\bibfnamefont {A.~V.}\ \bibnamefont
  {{Macci{\`o}}}}, \bibinfo {author} {\bibfnamefont {S.}~\bibnamefont
  {{Paduroiu}}}, \bibinfo {author} {\bibfnamefont {D.}~\bibnamefont
  {{Anderhalden}}}, \bibinfo {author} {\bibfnamefont {A.}~\bibnamefont
  {{Schneider}}},\ and\ \bibinfo {author} {\bibfnamefont {B.}~\bibnamefont
  {{Moore}}},\ }\href {https://doi.org/10.1111/j.1365-2966.2012.21284.x}
  {\bibfield  {journal} {\bibinfo  {journal} {Mon. Not. Roy. Ast. Soc.}\
  }\textbf {\bibinfo {volume} {424}},\ \bibinfo {pages} {1105} (\bibinfo {year}
  {2012})},\ \Eprint {https://arxiv.org/abs/1202.1282} {arXiv:1202.1282}
  \BibitemShut {NoStop}%
\bibitem [{\citenamefont {{Lovell}}\ \emph {et~al.}(2014)\citenamefont
  {{Lovell}}, \citenamefont {{Frenk}}, \citenamefont {{Eke}}, \citenamefont
  {{Jenkins}}, \citenamefont {{Gao}},\ and\ \citenamefont
  {{Theuns}}}]{Lovell2014}%
  \BibitemOpen
  \bibfield  {author} {\bibinfo {author} {\bibfnamefont {M.~R.}\ \bibnamefont
  {{Lovell}}}, \bibinfo {author} {\bibfnamefont {C.~S.}\ \bibnamefont
  {{Frenk}}}, \bibinfo {author} {\bibfnamefont {V.~R.}\ \bibnamefont {{Eke}}},
  \bibinfo {author} {\bibfnamefont {A.}~\bibnamefont {{Jenkins}}}, \bibinfo
  {author} {\bibfnamefont {L.}~\bibnamefont {{Gao}}},\ and\ \bibinfo {author}
  {\bibfnamefont {T.}~\bibnamefont {{Theuns}}},\ }\href
  {https://doi.org/10.1093/mnras/stt2431} {\bibfield  {journal} {\bibinfo
  {journal} {Mon. Not. Roy. Ast. Soc.}\ }\textbf {\bibinfo {volume} {439}},\
  \bibinfo {pages} {300} (\bibinfo {year} {2014})},\ \Eprint
  {https://arxiv.org/abs/1308.1399} {arXiv:1308.1399} \BibitemShut {NoStop}%
\bibitem [{\citenamefont {{El-Zant}}\ \emph {et~al.}(2015)\citenamefont
  {{El-Zant}}, \citenamefont {{Khalil}},\ and\ \citenamefont
  {{Sil}}}]{El-Zant2015}%
  \BibitemOpen
  \bibfield  {author} {\bibinfo {author} {\bibfnamefont {A.}~\bibnamefont
  {{El-Zant}}}, \bibinfo {author} {\bibfnamefont {S.}~\bibnamefont
  {{Khalil}}},\ and\ \bibinfo {author} {\bibfnamefont {A.}~\bibnamefont
  {{Sil}}},\ }\href {https://doi.org/10.1103/PhysRevD.91.035030} {\bibfield
  {journal} {\bibinfo  {journal} {Phys. Rev. D}\ }\textbf {\bibinfo {volume}
  {91}},\ \bibinfo {eid} {035030} (\bibinfo {year} {2015})},\ \Eprint
  {https://arxiv.org/abs/1308.0836} {arXiv:1308.0836 [hep-ph]} \BibitemShut
  {NoStop}%
\bibitem [{\citenamefont {{Spergel}}\ and\ \citenamefont
  {{Steinhardt}}(2000)}]{Spergel2000}%
  \BibitemOpen
  \bibfield  {author} {\bibinfo {author} {\bibfnamefont {D.~N.}\ \bibnamefont
  {{Spergel}}}\ and\ \bibinfo {author} {\bibfnamefont {P.~J.}\ \bibnamefont
  {{Steinhardt}}},\ }\href {https://doi.org/10.1103/PhysRevLett.84.3760}
  {\bibfield  {journal} {\bibinfo  {journal} {Physical Review Letters}\
  }\textbf {\bibinfo {volume} {84}},\ \bibinfo {pages} {3760} (\bibinfo {year}
  {2000})},\ \Eprint {https://arxiv.org/abs/astro-ph/9909386}
  {astro-ph/9909386} \BibitemShut {NoStop}%
\bibitem [{\citenamefont {{Burkert}}(2000)}]{Burkert2000}%
  \BibitemOpen
  \bibfield  {author} {\bibinfo {author} {\bibfnamefont {A.}~\bibnamefont
  {{Burkert}}},\ }\href {https://doi.org/10.1086/312674} {\bibfield  {journal}
  {\bibinfo  {journal} {Astrophys. Jour.l}\ }\textbf {\bibinfo {volume}
  {534}},\ \bibinfo {pages} {L143} (\bibinfo {year} {2000})},\ \Eprint
  {https://arxiv.org/abs/astro-ph/0002409} {astro-ph/0002409} \BibitemShut
  {NoStop}%
\bibitem [{\citenamefont {{Kochanek}}\ and\ \citenamefont
  {{White}}(2000)}]{Kochanek2000}%
  \BibitemOpen
  \bibfield  {author} {\bibinfo {author} {\bibfnamefont {C.~S.}\ \bibnamefont
  {{Kochanek}}}\ and\ \bibinfo {author} {\bibfnamefont {M.}~\bibnamefont
  {{White}}},\ }\href {https://doi.org/10.1086/317149} {\bibfield  {journal}
  {\bibinfo  {journal} {Astrophys. Jour.}\ }\textbf {\bibinfo {volume} {543}},\
  \bibinfo {pages} {514} (\bibinfo {year} {2000})},\ \Eprint
  {https://arxiv.org/abs/astro-ph/0003483} {astro-ph/0003483} \BibitemShut
  {NoStop}%
\bibitem [{\citenamefont {{Miralda-Escud{\'e}}}(2002)}]{Miralda2002}%
  \BibitemOpen
  \bibfield  {author} {\bibinfo {author} {\bibfnamefont {J.}~\bibnamefont
  {{Miralda-Escud{\'e}}}},\ }\href {https://doi.org/10.1086/324138} {\bibfield
  {journal} {\bibinfo  {journal} {Astrophys. Jour.}\ }\textbf {\bibinfo
  {volume} {564}},\ \bibinfo {pages} {60} (\bibinfo {year} {2002})},\ \Eprint
  {https://arxiv.org/abs/astro-ph/0002050} {astro-ph/0002050} \BibitemShut
  {NoStop}%
\bibitem [{\citenamefont {{Zavala}}\ \emph {et~al.}(2013)\citenamefont
  {{Zavala}}, \citenamefont {{Vogelsberger}},\ and\ \citenamefont
  {{Walker}}}]{Zavala2013}%
  \BibitemOpen
  \bibfield  {author} {\bibinfo {author} {\bibfnamefont {J.}~\bibnamefont
  {{Zavala}}}, \bibinfo {author} {\bibfnamefont {M.}~\bibnamefont
  {{Vogelsberger}}},\ and\ \bibinfo {author} {\bibfnamefont {M.~G.}\
  \bibnamefont {{Walker}}},\ }\href {https://doi.org/10.1093/mnrasl/sls053}
  {\bibfield  {journal} {\bibinfo  {journal} {Mon. Not. Roy. Ast. Soc.}\
  }\textbf {\bibinfo {volume} {431}},\ \bibinfo {pages} {L20} (\bibinfo {year}
  {2013})},\ \Eprint {https://arxiv.org/abs/1211.6426} {arXiv:1211.6426
  [astro-ph.CO]} \BibitemShut {NoStop}%
\bibitem [{\citenamefont {{Elbert}}\ \emph {et~al.}(2015)\citenamefont
  {{Elbert}}, \citenamefont {{Bullock}}, \citenamefont {{Garrison-Kimmel}},
  \citenamefont {{Rocha}}, \citenamefont {{O{\~n}orbe}},\ and\ \citenamefont
  {{Peter}}}]{Elbert2015}%
  \BibitemOpen
  \bibfield  {author} {\bibinfo {author} {\bibfnamefont {O.~D.}\ \bibnamefont
  {{Elbert}}}, \bibinfo {author} {\bibfnamefont {J.~S.}\ \bibnamefont
  {{Bullock}}}, \bibinfo {author} {\bibfnamefont {S.}~\bibnamefont
  {{Garrison-Kimmel}}}, \bibinfo {author} {\bibfnamefont {M.}~\bibnamefont
  {{Rocha}}}, \bibinfo {author} {\bibfnamefont {J.}~\bibnamefont
  {{O{\~n}orbe}}},\ and\ \bibinfo {author} {\bibfnamefont {A.~H.~G.}\
  \bibnamefont {{Peter}}},\ }\href {https://doi.org/10.1093/mnras/stv1470}
  {\bibfield  {journal} {\bibinfo  {journal} {Mon. Not. Roy. Ast. Soc.}\
  }\textbf {\bibinfo {volume} {453}},\ \bibinfo {pages} {29} (\bibinfo {year}
  {2015})},\ \Eprint {https://arxiv.org/abs/1412.1477} {arXiv:1412.1477}
  \BibitemShut {NoStop}%
\bibitem [{\citenamefont {{Goodman}}(2000)}]{Goodman2000}%
  \BibitemOpen
  \bibfield  {author} {\bibinfo {author} {\bibfnamefont {J.}~\bibnamefont
  {{Goodman}}},\ }\href {https://doi.org/10.1016/S1384-1076(00)00015-4}
  {\bibfield  {journal} {\bibinfo  {journal} {New Astronomy}\ }\textbf
  {\bibinfo {volume} {5}},\ \bibinfo {pages} {103} (\bibinfo {year} {2000})},\
  \Eprint {https://arxiv.org/abs/astro-ph/0003018} {astro-ph/0003018}
  \BibitemShut {NoStop}%
\bibitem [{\citenamefont {{Hu}}\ \emph {et~al.}(2000)\citenamefont {{Hu}},
  \citenamefont {{Barkana}},\ and\ \citenamefont {{Gruzinov}}}]{Hu2000}%
  \BibitemOpen
  \bibfield  {author} {\bibinfo {author} {\bibfnamefont {W.}~\bibnamefont
  {{Hu}}}, \bibinfo {author} {\bibfnamefont {R.}~\bibnamefont {{Barkana}}},\
  and\ \bibinfo {author} {\bibfnamefont {A.}~\bibnamefont {{Gruzinov}}},\
  }\href {https://doi.org/10.1103/PhysRevLett.85.1158} {\bibfield  {journal}
  {\bibinfo  {journal} {Physical Review Letters}\ }\textbf {\bibinfo {volume}
  {85}},\ \bibinfo {pages} {1158} (\bibinfo {year} {2000})},\ \Eprint
  {https://arxiv.org/abs/astro-ph/0003365} {astro-ph/0003365} \BibitemShut
  {NoStop}%
\bibitem [{\citenamefont {{Schive}}\ \emph {et~al.}(2014)\citenamefont
  {{Schive}}, \citenamefont {{Liao}}, \citenamefont {{Woo}}, \citenamefont
  {{Wong}}, \citenamefont {{Chiueh}}, \citenamefont {{Broadhurst}},\ and\
  \citenamefont {{Hwang}}}]{Schive2014}%
  \BibitemOpen
  \bibfield  {author} {\bibinfo {author} {\bibfnamefont {H.-Y.}\ \bibnamefont
  {{Schive}}}, \bibinfo {author} {\bibfnamefont {M.-H.}\ \bibnamefont
  {{Liao}}}, \bibinfo {author} {\bibfnamefont {T.-P.}\ \bibnamefont {{Woo}}},
  \bibinfo {author} {\bibfnamefont {S.-K.}\ \bibnamefont {{Wong}}}, \bibinfo
  {author} {\bibfnamefont {T.}~\bibnamefont {{Chiueh}}}, \bibinfo {author}
  {\bibfnamefont {T.}~\bibnamefont {{Broadhurst}}},\ and\ \bibinfo {author}
  {\bibfnamefont {W.-Y.~P.}\ \bibnamefont {{Hwang}}},\ }\href
  {https://doi.org/10.1103/PhysRevLett.113.261302} {\bibfield  {journal}
  {\bibinfo  {journal} {Physical Review Letters}\ }\textbf {\bibinfo {volume}
  {113}},\ \bibinfo {eid} {261302} (\bibinfo {year} {2014})},\ \Eprint
  {https://arxiv.org/abs/1407.7762} {arXiv:1407.7762} \BibitemShut {NoStop}%
\bibitem [{\citenamefont {{Marsh}}\ and\ \citenamefont
  {{Silk}}(2014)}]{Marsh2014}%
  \BibitemOpen
  \bibfield  {author} {\bibinfo {author} {\bibfnamefont {D.~J.~E.}\
  \bibnamefont {{Marsh}}}\ and\ \bibinfo {author} {\bibfnamefont
  {J.}~\bibnamefont {{Silk}}},\ }\href {https://doi.org/10.1093/mnras/stt2079}
  {\bibfield  {journal} {\bibinfo  {journal} {Mon. Not. Roy. Ast. Soc.}\
  }\textbf {\bibinfo {volume} {437}},\ \bibinfo {pages} {2652} (\bibinfo {year}
  {2014})},\ \Eprint {https://arxiv.org/abs/1307.1705} {arXiv:1307.1705
  [astro-ph.CO]} \BibitemShut {NoStop}%
\bibitem [{\citenamefont {{Hui}}\ \emph {et~al.}(2017)\citenamefont {{Hui}},
  \citenamefont {{Ostriker}}, \citenamefont {{Tremaine}},\ and\ \citenamefont
  {{Witten}}}]{Hui_etal2017}%
  \BibitemOpen
  \bibfield  {author} {\bibinfo {author} {\bibfnamefont {L.}~\bibnamefont
  {{Hui}}}, \bibinfo {author} {\bibfnamefont {J.~P.}\ \bibnamefont
  {{Ostriker}}}, \bibinfo {author} {\bibfnamefont {S.}~\bibnamefont
  {{Tremaine}}},\ and\ \bibinfo {author} {\bibfnamefont {E.}~\bibnamefont
  {{Witten}}},\ }\href {https://doi.org/10.1103/PhysRevD.95.043541} {\bibfield
  {journal} {\bibinfo  {journal} {Phys. Rev. D}\ }\textbf {\bibinfo {volume}
  {95}},\ \bibinfo {eid} {043541} (\bibinfo {year} {2017})},\ \Eprint
  {https://arxiv.org/abs/1610.08297} {arXiv:1610.08297 [astro-ph.CO]}
  \BibitemShut {NoStop}%
\bibitem [{\citenamefont {{Genzel}}\ \emph {et~al.}(2011)\citenamefont
  {{Genzel}}, \citenamefont {{Newman}}, \citenamefont {{Jones}}, \citenamefont
  {{F{\"o}rster Schreiber}}, \citenamefont {{Shapiro}}, \citenamefont
  {{Genel}}, \citenamefont {{Lilly}}, \citenamefont {{Renzini}}, \citenamefont
  {{Tacconi}}, \citenamefont {{Bouch{\'e}}}, \citenamefont {{Burkert}},
  \citenamefont {{Cresci}}, \citenamefont {{Buschkamp}}, \citenamefont
  {{Carollo}}, \citenamefont {{Ceverino}}, \citenamefont {{Davies}},
  \citenamefont {{Dekel}}, \citenamefont {{Eisenhauer}}, \citenamefont
  {{Hicks}}, \citenamefont {{Kurk}}, \citenamefont {{Lutz}}, \citenamefont
  {{Mancini}}, \citenamefont {{Naab}}, \citenamefont {{Peng}}, \citenamefont
  {{Sternberg}}, \citenamefont {{Vergani}},\ and\ \citenamefont
  {{Zamorani}}}]{Genzzel_Clumps11}%
  \BibitemOpen
  \bibfield  {author} {\bibinfo {author} {\bibfnamefont {R.}~\bibnamefont
  {{Genzel}}}, \bibinfo {author} {\bibfnamefont {S.}~\bibnamefont {{Newman}}},
  \bibinfo {author} {\bibfnamefont {T.}~\bibnamefont {{Jones}}}, \bibinfo
  {author} {\bibfnamefont {N.~M.}\ \bibnamefont {{F{\"o}rster Schreiber}}},
  \bibinfo {author} {\bibfnamefont {K.}~\bibnamefont {{Shapiro}}}, \bibinfo
  {author} {\bibfnamefont {S.}~\bibnamefont {{Genel}}}, \bibinfo {author}
  {\bibfnamefont {S.~J.}\ \bibnamefont {{Lilly}}}, \bibinfo {author}
  {\bibfnamefont {A.}~\bibnamefont {{Renzini}}}, \bibinfo {author}
  {\bibfnamefont {L.~J.}\ \bibnamefont {{Tacconi}}}, \bibinfo {author}
  {\bibfnamefont {N.}~\bibnamefont {{Bouch{\'e}}}}, \bibinfo {author}
  {\bibfnamefont {A.}~\bibnamefont {{Burkert}}}, \bibinfo {author}
  {\bibfnamefont {G.}~\bibnamefont {{Cresci}}}, \bibinfo {author}
  {\bibfnamefont {P.}~\bibnamefont {{Buschkamp}}}, \bibinfo {author}
  {\bibfnamefont {C.~M.}\ \bibnamefont {{Carollo}}}, \bibinfo {author}
  {\bibfnamefont {D.}~\bibnamefont {{Ceverino}}}, \bibinfo {author}
  {\bibfnamefont {R.}~\bibnamefont {{Davies}}}, \bibinfo {author}
  {\bibfnamefont {A.}~\bibnamefont {{Dekel}}}, \bibinfo {author} {\bibfnamefont
  {F.}~\bibnamefont {{Eisenhauer}}}, \bibinfo {author} {\bibfnamefont
  {E.}~\bibnamefont {{Hicks}}}, \bibinfo {author} {\bibfnamefont
  {J.}~\bibnamefont {{Kurk}}}, \bibinfo {author} {\bibfnamefont
  {D.}~\bibnamefont {{Lutz}}}, \bibinfo {author} {\bibfnamefont
  {C.}~\bibnamefont {{Mancini}}}, \bibinfo {author} {\bibfnamefont
  {T.}~\bibnamefont {{Naab}}}, \bibinfo {author} {\bibfnamefont
  {Y.}~\bibnamefont {{Peng}}}, \bibinfo {author} {\bibfnamefont
  {A.}~\bibnamefont {{Sternberg}}}, \bibinfo {author} {\bibfnamefont
  {D.}~\bibnamefont {{Vergani}}},\ and\ \bibinfo {author} {\bibfnamefont
  {G.}~\bibnamefont {{Zamorani}}},\ }\href
  {https://doi.org/10.1088/0004-637X/733/2/101} {\bibfield  {journal} {\bibinfo
   {journal} {Astrophys, J.}\ }\textbf {\bibinfo {volume} {733}},\ \bibinfo
  {eid} {101} (\bibinfo {year} {2011})},\ \Eprint
  {https://arxiv.org/abs/1011.5360} {arXiv:1011.5360 [astro-ph.CO]}
  \BibitemShut {NoStop}%
\bibitem [{\citenamefont {{Cava}}\ \emph {et~al.}(2018)\citenamefont {{Cava}},
  \citenamefont {{Schaerer}}, \citenamefont {{Richard}}, \citenamefont
  {{P{\'e}rez-Gonz{\'a}lez}}, \citenamefont {{Dessauges-Zavadsky}},
  \citenamefont {{Mayer}},\ and\ \citenamefont
  {{Tamburello}}}]{Caval_Clumps18}%
  \BibitemOpen
  \bibfield  {author} {\bibinfo {author} {\bibfnamefont {A.}~\bibnamefont
  {{Cava}}}, \bibinfo {author} {\bibfnamefont {D.}~\bibnamefont {{Schaerer}}},
  \bibinfo {author} {\bibfnamefont {J.}~\bibnamefont {{Richard}}}, \bibinfo
  {author} {\bibfnamefont {P.~G.}\ \bibnamefont {{P{\'e}rez-Gonz{\'a}lez}}},
  \bibinfo {author} {\bibfnamefont {M.}~\bibnamefont {{Dessauges-Zavadsky}}},
  \bibinfo {author} {\bibfnamefont {L.}~\bibnamefont {{Mayer}}},\ and\ \bibinfo
  {author} {\bibfnamefont {V.}~\bibnamefont {{Tamburello}}},\ }\href
  {https://doi.org/10.1038/s41550-017-0295-x} {\bibfield  {journal} {\bibinfo
  {journal} {Nature Astronomy}\ }\textbf {\bibinfo {volume} {2}},\ \bibinfo
  {pages} {76} (\bibinfo {year} {2018})},\ \Eprint
  {https://arxiv.org/abs/1711.03977} {arXiv:1711.03977 [astro-ph.GA]}
  \BibitemShut {NoStop}%
\bibitem [{\citenamefont {Mukhanov}\ \emph {et~al.}(1992)\citenamefont
  {Mukhanov}, \citenamefont {Feldman},\ and\ \citenamefont
  {Brandenberger}}]{mukhanov1992theory}%
  \BibitemOpen
  \bibfield  {author} {\bibinfo {author} {\bibfnamefont {V.~F.}\ \bibnamefont
  {Mukhanov}}, \bibinfo {author} {\bibfnamefont {H.~A.}\ \bibnamefont
  {Feldman}},\ and\ \bibinfo {author} {\bibfnamefont {R.~H.}\ \bibnamefont
  {Brandenberger}},\ }\href@noop {} {\bibfield  {journal} {\bibinfo  {journal}
  {Physics Reports}\ }\textbf {\bibinfo {volume} {215}},\ \bibinfo {pages}
  {203} (\bibinfo {year} {1992})}\BibitemShut {NoStop}%
\bibitem [{\citenamefont {Armendariz-Picon}\ \emph {et~al.}(1999)\citenamefont
  {Armendariz-Picon}, \citenamefont {Damour},\ and\ \citenamefont
  {Mukhanov}}]{armendariz1999k}%
  \BibitemOpen
  \bibfield  {author} {\bibinfo {author} {\bibfnamefont {C.}~\bibnamefont
  {Armendariz-Picon}}, \bibinfo {author} {\bibfnamefont {T.}~\bibnamefont
  {Damour}},\ and\ \bibinfo {author} {\bibfnamefont {V.-i.}\ \bibnamefont
  {Mukhanov}},\ }\href@noop {} {\bibfield  {journal} {\bibinfo  {journal}
  {Physics Letters B}\ }\textbf {\bibinfo {volume} {458}},\ \bibinfo {pages}
  {209} (\bibinfo {year} {1999})}\BibitemShut {NoStop}%
\bibitem [{\citenamefont {Garriga}\ and\ \citenamefont
  {Mukhanov}(1999)}]{garriga1999perturbations}%
  \BibitemOpen
  \bibfield  {author} {\bibinfo {author} {\bibfnamefont {J.}~\bibnamefont
  {Garriga}}\ and\ \bibinfo {author} {\bibfnamefont {V.~F.}\ \bibnamefont
  {Mukhanov}},\ }\href@noop {} {\bibfield  {journal} {\bibinfo  {journal}
  {Physics Letters B}\ }\textbf {\bibinfo {volume} {458}},\ \bibinfo {pages}
  {219} (\bibinfo {year} {1999})}\BibitemShut {NoStop}%
\bibitem [{\citenamefont {Salopek}\ and\ \citenamefont
  {Bond}(1990)}]{salopek1990nonlinear}%
  \BibitemOpen
  \bibfield  {author} {\bibinfo {author} {\bibfnamefont {D.}~\bibnamefont
  {Salopek}}\ and\ \bibinfo {author} {\bibfnamefont {J.}~\bibnamefont {Bond}},\
  }\href@noop {} {\bibfield  {journal} {\bibinfo  {journal} {Physical Review
  D}\ }\textbf {\bibinfo {volume} {42}},\ \bibinfo {pages} {3936} (\bibinfo
  {year} {1990})}\BibitemShut {NoStop}%
\bibitem [{\citenamefont {Martin}\ and\ \citenamefont
  {Brandenberger}(2001{\natexlab{b}})}]{martin2001trans}%
  \BibitemOpen
  \bibfield  {author} {\bibinfo {author} {\bibfnamefont {J.}~\bibnamefont
  {Martin}}\ and\ \bibinfo {author} {\bibfnamefont {R.~H.}\ \bibnamefont
  {Brandenberger}},\ }\href@noop {} {\bibfield  {journal} {\bibinfo  {journal}
  {Physical Review D}\ }\textbf {\bibinfo {volume} {63}},\ \bibinfo {pages}
  {123501} (\bibinfo {year} {2001}{\natexlab{b}})}\BibitemShut {NoStop}%
\bibitem [{\citenamefont {Lemoine}\ \emph {et~al.}(2001)\citenamefont
  {Lemoine}, \citenamefont {Lubo}, \citenamefont {Martin},\ and\ \citenamefont
  {Uzan}}]{lemoine2001stress}%
  \BibitemOpen
  \bibfield  {author} {\bibinfo {author} {\bibfnamefont {M.}~\bibnamefont
  {Lemoine}}, \bibinfo {author} {\bibfnamefont {M.}~\bibnamefont {Lubo}},
  \bibinfo {author} {\bibfnamefont {J.}~\bibnamefont {Martin}},\ and\ \bibinfo
  {author} {\bibfnamefont {J.-P.}\ \bibnamefont {Uzan}},\ }\href@noop {}
  {\bibfield  {journal} {\bibinfo  {journal} {Physical Review D}\ }\textbf
  {\bibinfo {volume} {65}},\ \bibinfo {pages} {023510} (\bibinfo {year}
  {2001})}\BibitemShut {NoStop}%
\bibitem [{\citenamefont {Martin}\ and\ \citenamefont
  {Brandenberger}(2002)}]{martin2002corley}%
  \BibitemOpen
  \bibfield  {author} {\bibinfo {author} {\bibfnamefont {J.}~\bibnamefont
  {Martin}}\ and\ \bibinfo {author} {\bibfnamefont {R.~H.}\ \bibnamefont
  {Brandenberger}},\ }\href@noop {} {\bibfield  {journal} {\bibinfo  {journal}
  {Physical Review D}\ }\textbf {\bibinfo {volume} {65}},\ \bibinfo {pages}
  {103514} (\bibinfo {year} {2002})}\BibitemShut {NoStop}%
\bibitem [{Note1()}]{Note1}%
  \BibitemOpen
  \bibinfo {note} {Indeed, as discussed in Section~\ref {sec:jumpc}, in order
  to get modification of the matter power spectrum and halo mass function at
  scales of the order of a comoving Mpc, which are of particular interest, one
  needs $H/k_c \lesssim 10^{-4}$, much smaller than $c_s =0.01$, which is the
  smallest we consider.}\BibitemShut {Stop}%
\bibitem [{\citenamefont {Lucchin}\ and\ \citenamefont
  {Matarrese}(1985)}]{lucchin1985power}%
  \BibitemOpen
  \bibfield  {author} {\bibinfo {author} {\bibfnamefont {F.}~\bibnamefont
  {Lucchin}}\ and\ \bibinfo {author} {\bibfnamefont {S.}~\bibnamefont
  {Matarrese}},\ }\href@noop {} {\bibfield  {journal} {\bibinfo  {journal}
  {Physical Review D}\ }\textbf {\bibinfo {volume} {32}},\ \bibinfo {pages}
  {1316} (\bibinfo {year} {1985})}\BibitemShut {NoStop}%
\bibitem [{\citenamefont {Lyth}\ and\ \citenamefont
  {Stewart}(1992)}]{lyth1992curvature}%
  \BibitemOpen
  \bibfield  {author} {\bibinfo {author} {\bibfnamefont {D.~H.}\ \bibnamefont
  {Lyth}}\ and\ \bibinfo {author} {\bibfnamefont {E.~D.}\ \bibnamefont
  {Stewart}},\ }\href@noop {} {\bibfield  {journal} {\bibinfo  {journal}
  {Physics Letters B}\ }\textbf {\bibinfo {volume} {274}},\ \bibinfo {pages}
  {168} (\bibinfo {year} {1992})}\BibitemShut {NoStop}%
\bibitem [{\citenamefont {{Arkani-Hamed}}\ \emph {et~al.}(1998)\citenamefont
  {{Arkani-Hamed}}, \citenamefont {{Dimopoulos}},\ and\ \citenamefont
  {{Dvali}}}]{ArkTeV981}%
  \BibitemOpen
  \bibfield  {author} {\bibinfo {author} {\bibfnamefont {N.}~\bibnamefont
  {{Arkani-Hamed}}}, \bibinfo {author} {\bibfnamefont {S.}~\bibnamefont
  {{Dimopoulos}}},\ and\ \bibinfo {author} {\bibfnamefont {G.}~\bibnamefont
  {{Dvali}}},\ }\href {https://doi.org/10.1016/S0370-2693(98)00466-3}
  {\bibfield  {journal} {\bibinfo  {journal} {Physics Letters B}\ }\textbf
  {\bibinfo {volume} {429}},\ \bibinfo {pages} {263} (\bibinfo {year}
  {1998})},\ \Eprint {https://arxiv.org/abs/hep-ph/9803315}
  {arXiv:hep-ph/9803315 [hep-ph]} \BibitemShut {NoStop}%
\bibitem [{\citenamefont {{Antoniadis}}\ \emph {et~al.}(1998)\citenamefont
  {{Antoniadis}}, \citenamefont {{Arkani-Hamed}}, \citenamefont
  {{Dimopoulos}},\ and\ \citenamefont {{Dvali}}}]{ArkTev982}%
  \BibitemOpen
  \bibfield  {author} {\bibinfo {author} {\bibfnamefont {I.}~\bibnamefont
  {{Antoniadis}}}, \bibinfo {author} {\bibfnamefont {N.}~\bibnamefont
  {{Arkani-Hamed}}}, \bibinfo {author} {\bibfnamefont {S.}~\bibnamefont
  {{Dimopoulos}}},\ and\ \bibinfo {author} {\bibfnamefont {G.}~\bibnamefont
  {{Dvali}}},\ }\href {https://doi.org/10.1016/S0370-2693(98)00860-0}
  {\bibfield  {journal} {\bibinfo  {journal} {Physics Letters B}\ }\textbf
  {\bibinfo {volume} {436}},\ \bibinfo {pages} {257} (\bibinfo {year}
  {1998})},\ \Eprint {https://arxiv.org/abs/hep-ph/9804398}
  {arXiv:hep-ph/9804398 [hep-ph]} \BibitemShut {NoStop}%
\bibitem [{\citenamefont {{Randall}}\ and\ \citenamefont
  {{Sundrum}}(1999)}]{RanSun99}%
  \BibitemOpen
  \bibfield  {author} {\bibinfo {author} {\bibfnamefont {L.}~\bibnamefont
  {{Randall}}}\ and\ \bibinfo {author} {\bibfnamefont {R.}~\bibnamefont
  {{Sundrum}}},\ }\href {https://doi.org/10.1103/PhysRevLett.83.3370}
  {\bibfield  {journal} {\bibinfo  {journal} {\prl}\ }\textbf {\bibinfo
  {volume} {83}},\ \bibinfo {pages} {3370} (\bibinfo {year} {1999})},\ \Eprint
  {https://arxiv.org/abs/hep-ph/9905221} {arXiv:hep-ph/9905221 [hep-ph]}
  \BibitemShut {NoStop}%
\bibitem [{\citenamefont {{Abel}}\ and\ \citenamefont
  {{Santiago}}(2004)}]{StrinScale04}%
  \BibitemOpen
  \bibfield  {author} {\bibinfo {author} {\bibfnamefont {S.}~\bibnamefont
  {{Abel}}}\ and\ \bibinfo {author} {\bibfnamefont {J.}~\bibnamefont
  {{Santiago}}},\ }\href {https://doi.org/10.1088/0954-3899/30/3/R01}
  {\bibfield  {journal} {\bibinfo  {journal} {Journal of Physics G Nuclear
  Physics}\ }\textbf {\bibinfo {volume} {30}},\ \bibinfo {pages} {R83}
  (\bibinfo {year} {2004})},\ \Eprint {https://arxiv.org/abs/hep-ph/0404237}
  {arXiv:hep-ph/0404237 [hep-ph]} \BibitemShut {NoStop}%
\bibitem [{Note2()}]{Note2}%
  \BibitemOpen
  \bibinfo {note} {Cf. Ref \cite {Greene04}. Their more rigorous formulation
  using effective field theory invokes an 'earliest time', which is defined as
  the time the smallest CMB scale leaves the cutoff scale. The earliest time
  here would correspond to that when the first scales cross the high energy
  cutoff. In both situations the origin of the time variation of the
  backreaction lies in the same change in number of excited states in the
  interval between $k_c$ and $H$.}\BibitemShut {Stop}%
\bibitem [{\citenamefont {{Bardeen}}\ \emph {et~al.}(1986)\citenamefont
  {{Bardeen}}, \citenamefont {{Bond}}, \citenamefont {{Kaiser}},\ and\
  \citenamefont {{Szalay}}}]{BBKS}%
  \BibitemOpen
  \bibfield  {author} {\bibinfo {author} {\bibfnamefont {J.~M.}\ \bibnamefont
  {{Bardeen}}}, \bibinfo {author} {\bibfnamefont {J.~R.}\ \bibnamefont
  {{Bond}}}, \bibinfo {author} {\bibfnamefont {N.}~\bibnamefont {{Kaiser}}},\
  and\ \bibinfo {author} {\bibfnamefont {A.~S.}\ \bibnamefont {{Szalay}}},\
  }\href {https://doi.org/10.1086/164143} {\bibfield  {journal} {\bibinfo
  {journal} {\apj}\ }\textbf {\bibinfo {volume} {304}},\ \bibinfo {pages} {15}
  (\bibinfo {year} {1986})}\BibitemShut {NoStop}%
\bibitem [{\citenamefont {Luki{\'c}}\ \emph {et~al.}(2007)\citenamefont
  {Luki{\'c}}, \citenamefont {Heitmann}, \citenamefont {Habib}, \citenamefont
  {Bashinsky},\ and\ \citenamefont {Ricker}}]{lukic2007halo}%
  \BibitemOpen
  \bibfield  {author} {\bibinfo {author} {\bibfnamefont {Z.}~\bibnamefont
  {Luki{\'c}}}, \bibinfo {author} {\bibfnamefont {K.}~\bibnamefont {Heitmann}},
  \bibinfo {author} {\bibfnamefont {S.}~\bibnamefont {Habib}}, \bibinfo
  {author} {\bibfnamefont {S.}~\bibnamefont {Bashinsky}},\ and\ \bibinfo
  {author} {\bibfnamefont {P.~M.}\ \bibnamefont {Ricker}},\ }\href@noop {}
  {\bibfield  {journal} {\bibinfo  {journal} {The Astrophysical Journal}\
  }\textbf {\bibinfo {volume} {671}},\ \bibinfo {pages} {1160} (\bibinfo {year}
  {2007})}\BibitemShut {NoStop}%
\bibitem [{\citenamefont {{Boyanovsky}}\ \emph {et~al.}(2006)\citenamefont
  {{Boyanovsky}}, \citenamefont {{de Vega}},\ and\ \citenamefont
  {{Sanchez}}}]{NormaS06}%
  \BibitemOpen
  \bibfield  {author} {\bibinfo {author} {\bibfnamefont {D.}~\bibnamefont
  {{Boyanovsky}}}, \bibinfo {author} {\bibfnamefont {H.~J.}\ \bibnamefont {{de
  Vega}}},\ and\ \bibinfo {author} {\bibfnamefont {N.~G.}\ \bibnamefont
  {{Sanchez}}},\ }\href {https://doi.org/10.1103/PhysRevD.74.123006} {\bibfield
   {journal} {\bibinfo  {journal} {\prd}\ }\textbf {\bibinfo {volume} {74}},\
  \bibinfo {eid} {123006} (\bibinfo {year} {2006})},\ \Eprint
  {https://arxiv.org/abs/astro-ph/0607508} {arXiv:astro-ph/0607508 [astro-ph]}
  \BibitemShut {NoStop}%
\bibitem [{\citenamefont {Press}\ and\ \citenamefont
  {Schechter}(1974)}]{press1974formation}%
  \BibitemOpen
  \bibfield  {author} {\bibinfo {author} {\bibfnamefont {W.~H.}\ \bibnamefont
  {Press}}\ and\ \bibinfo {author} {\bibfnamefont {P.}~\bibnamefont
  {Schechter}},\ }\href@noop {} {\bibfield  {journal} {\bibinfo  {journal} {The
  Astrophysical Journal}\ }\textbf {\bibinfo {volume} {187}},\ \bibinfo {pages}
  {425} (\bibinfo {year} {1974})}\BibitemShut {NoStop}%
\bibitem [{\citenamefont {Bond}\ \emph {et~al.}(1991)\citenamefont {Bond},
  \citenamefont {Cole}, \citenamefont {Efstathiou},\ and\ \citenamefont
  {Kaiser}}]{bond1991excursion}%
  \BibitemOpen
  \bibfield  {author} {\bibinfo {author} {\bibfnamefont {J.}~\bibnamefont
  {Bond}}, \bibinfo {author} {\bibfnamefont {S.}~\bibnamefont {Cole}}, \bibinfo
  {author} {\bibfnamefont {G.}~\bibnamefont {Efstathiou}},\ and\ \bibinfo
  {author} {\bibfnamefont {N.}~\bibnamefont {Kaiser}},\ }\href@noop {}
  {\bibfield  {journal} {\bibinfo  {journal} {The Astrophysical Journal}\
  }\textbf {\bibinfo {volume} {379}},\ \bibinfo {pages} {440} (\bibinfo {year}
  {1991})}\BibitemShut {NoStop}%
\bibitem [{\citenamefont {{Sheth}}\ \emph {et~al.}(2001)\citenamefont
  {{Sheth}}, \citenamefont {{Mo}},\ and\ \citenamefont {{Tormen}}}]{SMT01}%
  \BibitemOpen
  \bibfield  {author} {\bibinfo {author} {\bibfnamefont {R.~K.}\ \bibnamefont
  {{Sheth}}}, \bibinfo {author} {\bibfnamefont {H.~J.}\ \bibnamefont {{Mo}}},\
  and\ \bibinfo {author} {\bibfnamefont {G.}~\bibnamefont {{Tormen}}},\ }\href
  {https://doi.org/10.1046/j.1365-8711.2001.04006.x} {\bibfield  {journal}
  {\bibinfo  {journal} {MNRAS}\ }\textbf {\bibinfo {volume} {323}},\ \bibinfo
  {pages} {1} (\bibinfo {year} {2001})},\ \Eprint
  {https://arxiv.org/abs/astro-ph/9907024} {arXiv:astro-ph/9907024 [astro-ph]}
  \BibitemShut {NoStop}%
\bibitem [{\citenamefont {{Sheth}}\ and\ \citenamefont
  {{Tormen}}(2002)}]{ST02}%
  \BibitemOpen
  \bibfield  {author} {\bibinfo {author} {\bibfnamefont {R.~K.}\ \bibnamefont
  {{Sheth}}}\ and\ \bibinfo {author} {\bibfnamefont {G.}~\bibnamefont
  {{Tormen}}},\ }\href {https://doi.org/10.1046/j.1365-8711.2002.04950.x}
  {\bibfield  {journal} {\bibinfo  {journal} {MNRAS}\ }\textbf {\bibinfo
  {volume} {329}},\ \bibinfo {pages} {61} (\bibinfo {year} {2002})},\ \Eprint
  {https://arxiv.org/abs/astro-ph/0105113} {arXiv:astro-ph/0105113 [astro-ph]}
  \BibitemShut {NoStop}%
\bibitem [{\citenamefont {{Despali}}\ \emph {et~al.}(2016)\citenamefont
  {{Despali}}, \citenamefont {{Giocoli}}, \citenamefont {{Angulo}},
  \citenamefont {{Tormen}}, \citenamefont {{Sheth}}, \citenamefont {{Baso}},\
  and\ \citenamefont {{Moscardini}}}]{Despali16}%
  \BibitemOpen
  \bibfield  {author} {\bibinfo {author} {\bibfnamefont {G.}~\bibnamefont
  {{Despali}}}, \bibinfo {author} {\bibfnamefont {C.}~\bibnamefont
  {{Giocoli}}}, \bibinfo {author} {\bibfnamefont {R.~E.}\ \bibnamefont
  {{Angulo}}}, \bibinfo {author} {\bibfnamefont {G.}~\bibnamefont {{Tormen}}},
  \bibinfo {author} {\bibfnamefont {R.~K.}\ \bibnamefont {{Sheth}}}, \bibinfo
  {author} {\bibfnamefont {G.}~\bibnamefont {{Baso}}},\ and\ \bibinfo {author}
  {\bibfnamefont {L.}~\bibnamefont {{Moscardini}}},\ }\href
  {https://doi.org/10.1093/mnras/stv2842} {\bibfield  {journal} {\bibinfo
  {journal} {MNRAS}\ }\textbf {\bibinfo {volume} {456}},\ \bibinfo {pages}
  {2486} (\bibinfo {year} {2016})},\ \Eprint {https://arxiv.org/abs/1507.05627}
  {arXiv:1507.05627 [astro-ph.CO]} \BibitemShut {NoStop}%
\bibitem [{\citenamefont {{Comparat}}\ \emph {et~al.}(2017)\citenamefont
  {{Comparat}}, \citenamefont {{Prada}}, \citenamefont {{Yepes}},\ and\
  \citenamefont {{Klypin}}}]{Comparat17}%
  \BibitemOpen
  \bibfield  {author} {\bibinfo {author} {\bibfnamefont {J.}~\bibnamefont
  {{Comparat}}}, \bibinfo {author} {\bibfnamefont {F.}~\bibnamefont {{Prada}}},
  \bibinfo {author} {\bibfnamefont {G.}~\bibnamefont {{Yepes}}},\ and\ \bibinfo
  {author} {\bibfnamefont {A.}~\bibnamefont {{Klypin}}},\ }\href
  {https://doi.org/10.1093/mnras/stx1183} {\bibfield  {journal} {\bibinfo
  {journal} {MNRAS}\ }\textbf {\bibinfo {volume} {469}},\ \bibinfo {pages}
  {4157} (\bibinfo {year} {2017})},\ \Eprint {https://arxiv.org/abs/1702.01628}
  {arXiv:1702.01628 [astro-ph.CO]} \BibitemShut {NoStop}%
\bibitem [{\citenamefont {{Macci{\`o}}}\ \emph {et~al.}(2020)\citenamefont
  {{Macci{\`o}}}, \citenamefont {{Courteau}}, \citenamefont {{Ouellette}},\
  and\ \citenamefont {{Dutton}}}]{Maccio_abunmatch20}%
  \BibitemOpen
  \bibfield  {author} {\bibinfo {author} {\bibfnamefont {A.~V.}\ \bibnamefont
  {{Macci{\`o}}}}, \bibinfo {author} {\bibfnamefont {S.}~\bibnamefont
  {{Courteau}}}, \bibinfo {author} {\bibfnamefont {N.~N.~Q.}\ \bibnamefont
  {{Ouellette}}},\ and\ \bibinfo {author} {\bibfnamefont {A.~A.}\ \bibnamefont
  {{Dutton}}},\ }\href {https://doi.org/10.1093/mnrasl/slaa094} {\bibfield
  {journal} {\bibinfo  {journal} {MNRAS}\ }\textbf {\bibinfo {volume} {496}},\
  \bibinfo {pages} {L101} (\bibinfo {year} {2020})},\ \Eprint
  {https://arxiv.org/abs/2006.00818} {arXiv:2006.00818 [astro-ph.GA]}
  \BibitemShut {NoStop}%
\bibitem [{\citenamefont {{Miller}}\ \emph {et~al.}(2014)\citenamefont
  {{Miller}}, \citenamefont {{Ellis}}, \citenamefont {{Newman}},\ and\
  \citenamefont {{Benson}}}]{Ellis14}%
  \BibitemOpen
  \bibfield  {author} {\bibinfo {author} {\bibfnamefont {S.~H.}\ \bibnamefont
  {{Miller}}}, \bibinfo {author} {\bibfnamefont {R.~S.}\ \bibnamefont
  {{Ellis}}}, \bibinfo {author} {\bibfnamefont {A.~B.}\ \bibnamefont
  {{Newman}}},\ and\ \bibinfo {author} {\bibfnamefont {A.}~\bibnamefont
  {{Benson}}},\ }\href {https://doi.org/10.1088/0004-637X/782/2/115} {\bibfield
   {journal} {\bibinfo  {journal} {\apj}\ }\textbf {\bibinfo {volume} {782}},\
  \bibinfo {eid} {115} (\bibinfo {year} {2014})},\ \Eprint
  {https://arxiv.org/abs/1310.1079} {arXiv:1310.1079 [astro-ph.CO]}
  \BibitemShut {NoStop}%
\bibitem [{\citenamefont {{Somerville}}\ and\ \citenamefont
  {{Dav{\'e}}}(2015)}]{Somerdown}%
  \BibitemOpen
  \bibfield  {author} {\bibinfo {author} {\bibfnamefont {R.~S.}\ \bibnamefont
  {{Somerville}}}\ and\ \bibinfo {author} {\bibfnamefont {R.}~\bibnamefont
  {{Dav{\'e}}}},\ }\href {https://doi.org/10.1146/annurev-astro-082812-140951}
  {\bibfield  {journal} {\bibinfo  {journal} {ARAA}\ }\textbf {\bibinfo
  {volume} {53}},\ \bibinfo {pages} {51} (\bibinfo {year} {2015})},\ \Eprint
  {https://arxiv.org/abs/1412.2712} {arXiv:1412.2712 [astro-ph.GA]}
  \BibitemShut {NoStop}%
\bibitem [{Note3()}]{Note3}%
  \BibitemOpen
  \bibinfo {note} {Note that in~\cite {Silk_LCDM_SMBH18} the more conservative
  cumulative stellar mass function, which averages over the increasing
  $\protect \frac {M_*}{M_h} (M_*)$, is employed. The data for $M_* = 10^8
  {M_\odot }$ to $10^{10} {M_\odot }$ are consistent with $\Lambda $CDM.
  Tension still arises for the data point shown for $M_* \ge 10^{11.7} {M_\odot
  }$ at $z \approx 5.5$ (their Fig.~2 top-left). The results in Fig.~18
  of~\cite {COSMOS15} show data consistent with $M_*/M_h = 1/6.3$, or even
  larger, already at $z \sim 5$, for $M_* \gtrsim 10^{11} {M_\odot }$. In
  general, it seems that galaxies can be accommodated into halos with $M_*/M_h
  \ll 1/6.3$ for $z \gtrsim 5$ if one keeps to $M_* \lesssim 10^{10.5} {M_\odot
  }$ (e.g., \cite {IncFBehrooz13}, and Fig.~9 of~\cite {UMach}). On the other
  hand, the most extreme cases in the data presented by Steinhardt et.
  al.~\cite {Imp_earl16} (where $M_*/M_h \approx 1/7$ seems required), are just
  barely consistent with standard $\Lambda $CDM.}\BibitemShut {Stop}%
\bibitem [{\citenamefont {Mancuso}\ \emph {et~al.}(2016)\citenamefont
  {Mancuso}, \citenamefont {Lapi}, \citenamefont {Shi}, \citenamefont
  {Gonzalez-Nuevo}, \citenamefont {Aversa},\ and\ \citenamefont
  {Danese}}]{Mancuso16}%
  \BibitemOpen
  \bibfield  {author} {\bibinfo {author} {\bibfnamefont {C.}~\bibnamefont
  {Mancuso}}, \bibinfo {author} {\bibfnamefont {A.}~\bibnamefont {Lapi}},
  \bibinfo {author} {\bibfnamefont {J.}~\bibnamefont {Shi}}, \bibinfo {author}
  {\bibfnamefont {J.}~\bibnamefont {Gonzalez-Nuevo}}, \bibinfo {author}
  {\bibfnamefont {R.}~\bibnamefont {Aversa}},\ and\ \bibinfo {author}
  {\bibfnamefont {L.}~\bibnamefont {Danese}},\ }\href
  {https://doi.org/10.3847/0004-637X/823/2/128} {\bibfield  {journal} {\bibinfo
   {journal} {Astrophys. J.}\ }\textbf {\bibinfo {volume} {823}},\ \bibinfo
  {pages} {128} (\bibinfo {year} {2016})},\ \Eprint
  {https://arxiv.org/abs/1604.02507} {arXiv:1604.02507 [astro-ph.GA]}
  \BibitemShut {NoStop}%
\bibitem [{\citenamefont {{Alcalde Pampliega}}\ \emph
  {et~al.}(2019)\citenamefont {{Alcalde Pampliega}}, \citenamefont
  {{P{\'e}rez-Gonz{\'a}lez}}, \citenamefont {{Barro}}, \citenamefont
  {{Dom{\'\i}nguez S{\'a}nchez}}, \citenamefont {{Eliche-Moral}}, \citenamefont
  {{Cardiel}}, \citenamefont {{Hern{\'a}n-Caballero}}, \citenamefont
  {{Rodriguez-Mu{\~n}oz}}, \citenamefont {{S{\'a}nchez Bl{\'a}zquez}},\ and\
  \citenamefont {{Esquej}}}]{Balmer19}%
  \BibitemOpen
  \bibfield  {author} {\bibinfo {author} {\bibfnamefont {B.}~\bibnamefont
  {{Alcalde Pampliega}}}, \bibinfo {author} {\bibfnamefont {P.~G.}\
  \bibnamefont {{P{\'e}rez-Gonz{\'a}lez}}}, \bibinfo {author} {\bibfnamefont
  {G.}~\bibnamefont {{Barro}}}, \bibinfo {author} {\bibfnamefont
  {H.}~\bibnamefont {{Dom{\'\i}nguez S{\'a}nchez}}}, \bibinfo {author}
  {\bibfnamefont {M.~C.}\ \bibnamefont {{Eliche-Moral}}}, \bibinfo {author}
  {\bibfnamefont {N.}~\bibnamefont {{Cardiel}}}, \bibinfo {author}
  {\bibfnamefont {A.}~\bibnamefont {{Hern{\'a}n-Caballero}}}, \bibinfo {author}
  {\bibfnamefont {L.}~\bibnamefont {{Rodriguez-Mu{\~n}oz}}}, \bibinfo {author}
  {\bibfnamefont {P.}~\bibnamefont {{S{\'a}nchez Bl{\'a}zquez}}},\ and\
  \bibinfo {author} {\bibfnamefont {P.}~\bibnamefont {{Esquej}}},\ }\href
  {https://doi.org/10.3847/1538-4357/ab14f2} {\bibfield  {journal} {\bibinfo
  {journal} {Astrophys. Jour.}\ }\textbf {\bibinfo {volume} {876}},\ \bibinfo
  {eid} {135} (\bibinfo {year} {2019})},\ \Eprint
  {https://arxiv.org/abs/1806.04152} {arXiv:1806.04152 [astro-ph.GA]}
  \BibitemShut {NoStop}%
\bibitem [{\citenamefont {{Glazebrook}}\ \emph
  {et~al.}(2017{\natexlab{b}})\citenamefont {{Glazebrook}}, \citenamefont
  {{Schreiber}}, \citenamefont {{Labb{\'e}}}, \citenamefont {{Nanayakkara}},
  \citenamefont {{Kacprzak}}, \citenamefont {{Oesch}}, \citenamefont
  {{Papovich}}, \citenamefont {{Spitler}}, \citenamefont {{Straatman}},
  \citenamefont {{Tran}},\ and\ \citenamefont {{Yuan}}}]{NatImp}%
  \BibitemOpen
  \bibfield  {author} {\bibinfo {author} {\bibfnamefont {K.}~\bibnamefont
  {{Glazebrook}}}, \bibinfo {author} {\bibfnamefont {C.}~\bibnamefont
  {{Schreiber}}}, \bibinfo {author} {\bibfnamefont {I.}~\bibnamefont
  {{Labb{\'e}}}}, \bibinfo {author} {\bibfnamefont {T.}~\bibnamefont
  {{Nanayakkara}}}, \bibinfo {author} {\bibfnamefont {G.~G.}\ \bibnamefont
  {{Kacprzak}}}, \bibinfo {author} {\bibfnamefont {P.~A.}\ \bibnamefont
  {{Oesch}}}, \bibinfo {author} {\bibfnamefont {C.}~\bibnamefont {{Papovich}}},
  \bibinfo {author} {\bibfnamefont {L.~R.}\ \bibnamefont {{Spitler}}}, \bibinfo
  {author} {\bibfnamefont {C.~M.~S.}\ \bibnamefont {{Straatman}}}, \bibinfo
  {author} {\bibfnamefont {K.-V.~H.}\ \bibnamefont {{Tran}}},\ and\ \bibinfo
  {author} {\bibfnamefont {T.}~\bibnamefont {{Yuan}}},\ }\href
  {https://doi.org/10.1038/nature21680} {\bibfield  {journal} {\bibinfo
  {journal} {\nat}\ }\textbf {\bibinfo {volume} {544}},\ \bibinfo {pages} {71}
  (\bibinfo {year} {2017}{\natexlab{b}})},\ \Eprint
  {https://arxiv.org/abs/1702.01751} {arXiv:1702.01751 [astro-ph.GA]}
  \BibitemShut {NoStop}%
\bibitem [{\citenamefont {{Mawatari}}\ \emph {et~al.}(2019)\citenamefont
  {{Mawatari}}, \citenamefont {{Inoue}}, \citenamefont {{Hashimoto}},
  \citenamefont {{Silverman}}, \citenamefont {{Kajisawa}}, \citenamefont
  {{Yamanaka}}, \citenamefont {{Yamada}}, \citenamefont {{Davidzon}},
  \citenamefont {{Capak}}, \citenamefont {{Lin}}, \citenamefont {{Hsieh}},
  \citenamefont {{Taniguchi}}, \citenamefont {{Tanaka}}, \citenamefont {{Ono}},
  \citenamefont {{Harikane}}, \citenamefont {{Sugahara}}, \citenamefont
  {{Fujimoto}},\ and\ \citenamefont {{Nagao}}}]{Capak19}%
  \BibitemOpen
  \bibfield  {author} {\bibinfo {author} {\bibfnamefont {K.}~\bibnamefont
  {{Mawatari}}}, \bibinfo {author} {\bibfnamefont {A.~K.}\ \bibnamefont
  {{Inoue}}}, \bibinfo {author} {\bibfnamefont {T.}~\bibnamefont
  {{Hashimoto}}}, \bibinfo {author} {\bibfnamefont {J.}~\bibnamefont
  {{Silverman}}}, \bibinfo {author} {\bibfnamefont {M.}~\bibnamefont
  {{Kajisawa}}}, \bibinfo {author} {\bibfnamefont {S.}~\bibnamefont
  {{Yamanaka}}}, \bibinfo {author} {\bibfnamefont {T.}~\bibnamefont
  {{Yamada}}}, \bibinfo {author} {\bibfnamefont {I.}~\bibnamefont
  {{Davidzon}}}, \bibinfo {author} {\bibfnamefont {P.}~\bibnamefont {{Capak}}},
  \bibinfo {author} {\bibfnamefont {L.}~\bibnamefont {{Lin}}}, \bibinfo
  {author} {\bibfnamefont {B.-C.}\ \bibnamefont {{Hsieh}}}, \bibinfo {author}
  {\bibfnamefont {Y.}~\bibnamefont {{Taniguchi}}}, \bibinfo {author}
  {\bibfnamefont {M.}~\bibnamefont {{Tanaka}}}, \bibinfo {author}
  {\bibfnamefont {Y.}~\bibnamefont {{Ono}}}, \bibinfo {author} {\bibfnamefont
  {Y.}~\bibnamefont {{Harikane}}}, \bibinfo {author} {\bibfnamefont
  {Y.}~\bibnamefont {{Sugahara}}}, \bibinfo {author} {\bibfnamefont
  {S.}~\bibnamefont {{Fujimoto}}},\ and\ \bibinfo {author} {\bibfnamefont
  {T.}~\bibnamefont {{Nagao}}},\ }\href@noop {} {\bibfield  {journal} {\bibinfo
   {journal} {arXiv e-prints}\ ,\ \bibinfo {eid} {arXiv:1912.10954}} (\bibinfo
  {year} {2019})},\ \Eprint {https://arxiv.org/abs/1912.10954}
  {arXiv:1912.10954 [astro-ph.GA]} \BibitemShut {NoStop}%
\bibitem [{\citenamefont {Straatman}\ \emph {et~al.}(2014)\citenamefont
  {Straatman} \emph {et~al.}}]{subqueis14}%
  \BibitemOpen
  \bibfield  {author} {\bibinfo {author} {\bibfnamefont {C.~M.~S.}\
  \bibnamefont {Straatman}} \emph {et~al.},\ }\href
  {https://doi.org/10.1088/2041-8205/783/1/L14} {\bibfield  {journal} {\bibinfo
   {journal} {Astrophys. J.}\ }\textbf {\bibinfo {volume} {783}},\ \bibinfo
  {pages} {L14} (\bibinfo {year} {2014})},\ \Eprint
  {https://arxiv.org/abs/1312.4952} {arXiv:1312.4952 [astro-ph.GA]}
  \BibitemShut {NoStop}%
\bibitem [{\citenamefont {{Schreiber}}\ \emph {et~al.}(2018)\citenamefont
  {{Schreiber}}, \citenamefont {{Glazebrook}}, \citenamefont {{Nanayakkara}},
  \citenamefont {{Kacprzak}}, \citenamefont {{Labb{\'e}}}, \citenamefont
  {{Oesch}}, \citenamefont {{Yuan}}, \citenamefont {{Tran}}, \citenamefont
  {{Papovich}}, \citenamefont {{Spitler}},\ and\ \citenamefont
  {{Straatman}}}]{SchreiSAM18}%
  \BibitemOpen
  \bibfield  {author} {\bibinfo {author} {\bibfnamefont {C.}~\bibnamefont
  {{Schreiber}}}, \bibinfo {author} {\bibfnamefont {K.}~\bibnamefont
  {{Glazebrook}}}, \bibinfo {author} {\bibfnamefont {T.}~\bibnamefont
  {{Nanayakkara}}}, \bibinfo {author} {\bibfnamefont {G.~G.}\ \bibnamefont
  {{Kacprzak}}}, \bibinfo {author} {\bibfnamefont {I.}~\bibnamefont
  {{Labb{\'e}}}}, \bibinfo {author} {\bibfnamefont {P.}~\bibnamefont
  {{Oesch}}}, \bibinfo {author} {\bibfnamefont {T.}~\bibnamefont {{Yuan}}},
  \bibinfo {author} {\bibfnamefont {K.~V.}\ \bibnamefont {{Tran}}}, \bibinfo
  {author} {\bibfnamefont {C.}~\bibnamefont {{Papovich}}}, \bibinfo {author}
  {\bibfnamefont {L.}~\bibnamefont {{Spitler}}},\ and\ \bibinfo {author}
  {\bibfnamefont {C.}~\bibnamefont {{Straatman}}},\ }\href
  {https://doi.org/10.1051/0004-6361/201833070} {\bibfield  {journal} {\bibinfo
   {journal} {Astron. \& Astrophys.}\ }\textbf {\bibinfo {volume} {618}},\
  \bibinfo {eid} {A85} (\bibinfo {year} {2018})},\ \Eprint
  {https://arxiv.org/abs/1807.02523} {arXiv:1807.02523 [astro-ph.GA]}
  \BibitemShut {NoStop}%
\bibitem [{\citenamefont {{Santini}}\ \emph {et~al.}(2019)\citenamefont
  {{Santini}}, \citenamefont {{Merlin}}, \citenamefont {{Fontana}},
  \citenamefont {{Magnelli}}, \citenamefont {{Paris}}, \citenamefont
  {{Castellano}}, \citenamefont {{Grazian}}, \citenamefont {{Pentericci}},
  \citenamefont {{Pilo}},\ and\ \citenamefont {{Torelli}}}]{PALMA19}%
  \BibitemOpen
  \bibfield  {author} {\bibinfo {author} {\bibfnamefont {P.}~\bibnamefont
  {{Santini}}}, \bibinfo {author} {\bibfnamefont {E.}~\bibnamefont {{Merlin}}},
  \bibinfo {author} {\bibfnamefont {A.}~\bibnamefont {{Fontana}}}, \bibinfo
  {author} {\bibfnamefont {B.}~\bibnamefont {{Magnelli}}}, \bibinfo {author}
  {\bibfnamefont {D.}~\bibnamefont {{Paris}}}, \bibinfo {author} {\bibfnamefont
  {M.}~\bibnamefont {{Castellano}}}, \bibinfo {author} {\bibfnamefont
  {A.}~\bibnamefont {{Grazian}}}, \bibinfo {author} {\bibfnamefont
  {L.}~\bibnamefont {{Pentericci}}}, \bibinfo {author} {\bibfnamefont
  {S.}~\bibnamefont {{Pilo}}},\ and\ \bibinfo {author} {\bibfnamefont
  {M.}~\bibnamefont {{Torelli}}},\ }\href
  {https://doi.org/10.1093/mnras/stz801} {\bibfield  {journal} {\bibinfo
  {journal} {Mon. Not. Roy. Ast. Soc.}\ }\textbf {\bibinfo {volume} {486}},\
  \bibinfo {pages} {560} (\bibinfo {year} {2019})},\ \Eprint
  {https://arxiv.org/abs/1902.09548} {arXiv:1902.09548 [astro-ph.GA]}
  \BibitemShut {NoStop}%
\bibitem [{\citenamefont {{Merlin}}\ \emph {et~al.}(2019)\citenamefont
  {{Merlin}}, \citenamefont {{Fortuni}}, \citenamefont {{Torelli}},
  \citenamefont {{Santini}}, \citenamefont {{Castellano}}, \citenamefont
  {{Fontana}}, \citenamefont {{Grazian}}, \citenamefont {{Pentericci}},
  \citenamefont {{Pilo}},\ and\ \citenamefont {{Schmidt}}}]{R+D19}%
  \BibitemOpen
  \bibfield  {author} {\bibinfo {author} {\bibfnamefont {E.}~\bibnamefont
  {{Merlin}}}, \bibinfo {author} {\bibfnamefont {F.}~\bibnamefont {{Fortuni}}},
  \bibinfo {author} {\bibfnamefont {M.}~\bibnamefont {{Torelli}}}, \bibinfo
  {author} {\bibfnamefont {P.}~\bibnamefont {{Santini}}}, \bibinfo {author}
  {\bibfnamefont {M.}~\bibnamefont {{Castellano}}}, \bibinfo {author}
  {\bibfnamefont {A.}~\bibnamefont {{Fontana}}}, \bibinfo {author}
  {\bibfnamefont {A.}~\bibnamefont {{Grazian}}}, \bibinfo {author}
  {\bibfnamefont {L.}~\bibnamefont {{Pentericci}}}, \bibinfo {author}
  {\bibfnamefont {S.}~\bibnamefont {{Pilo}}},\ and\ \bibinfo {author}
  {\bibfnamefont {K.~B.}\ \bibnamefont {{Schmidt}}},\ }\href
  {https://doi.org/10.1093/mnras/stz2615} {\bibfield  {journal} {\bibinfo
  {journal} {Mon. Not. Roy. Ast. Soc.}\ }\textbf {\bibinfo {volume} {490}},\
  \bibinfo {pages} {3309} (\bibinfo {year} {2019})},\ \Eprint
  {https://arxiv.org/abs/1909.07996} {arXiv:1909.07996 [astro-ph.GA]}
  \BibitemShut {NoStop}%
\bibitem [{\citenamefont {{Valentino}}\ \emph {et~al.}(2020)\citenamefont
  {{Valentino}}, \citenamefont {{Tanaka}}, \citenamefont {{Davidzon}},
  \citenamefont {{Toft}}, \citenamefont {{G{\'o}mez-Guijarro}}, \citenamefont
  {{Stockmann}}, \citenamefont {{Onodera}}, \citenamefont {{Brammer}},
  \citenamefont {{Ceverino}}, \citenamefont {{Faisst}}, \citenamefont
  {{Gallazzi}}, \citenamefont {{Hayward}}, \citenamefont {{Ilbert}},
  \citenamefont {{Kubo}}, \citenamefont {{Magdis}}, \citenamefont {{Selsing}},
  \citenamefont {{Shimakawa}}, \citenamefont {{Sparre}}, \citenamefont
  {{Steinhardt}}, \citenamefont {{Yabe}},\ and\ \citenamefont
  {{Zabl}}}]{Quiescent1.5Gyr19}%
  \BibitemOpen
  \bibfield  {author} {\bibinfo {author} {\bibfnamefont {F.}~\bibnamefont
  {{Valentino}}}, \bibinfo {author} {\bibfnamefont {M.}~\bibnamefont
  {{Tanaka}}}, \bibinfo {author} {\bibfnamefont {I.}~\bibnamefont
  {{Davidzon}}}, \bibinfo {author} {\bibfnamefont {S.}~\bibnamefont {{Toft}}},
  \bibinfo {author} {\bibfnamefont {C.}~\bibnamefont {{G{\'o}mez-Guijarro}}},
  \bibinfo {author} {\bibfnamefont {M.}~\bibnamefont {{Stockmann}}}, \bibinfo
  {author} {\bibfnamefont {M.}~\bibnamefont {{Onodera}}}, \bibinfo {author}
  {\bibfnamefont {G.}~\bibnamefont {{Brammer}}}, \bibinfo {author}
  {\bibfnamefont {D.}~\bibnamefont {{Ceverino}}}, \bibinfo {author}
  {\bibfnamefont {A.~L.}\ \bibnamefont {{Faisst}}}, \bibinfo {author}
  {\bibfnamefont {A.}~\bibnamefont {{Gallazzi}}}, \bibinfo {author}
  {\bibfnamefont {C.~C.}\ \bibnamefont {{Hayward}}}, \bibinfo {author}
  {\bibfnamefont {O.}~\bibnamefont {{Ilbert}}}, \bibinfo {author}
  {\bibfnamefont {M.}~\bibnamefont {{Kubo}}}, \bibinfo {author} {\bibfnamefont
  {G.~E.}\ \bibnamefont {{Magdis}}}, \bibinfo {author} {\bibfnamefont
  {J.}~\bibnamefont {{Selsing}}}, \bibinfo {author} {\bibfnamefont
  {R.}~\bibnamefont {{Shimakawa}}}, \bibinfo {author} {\bibfnamefont
  {M.}~\bibnamefont {{Sparre}}}, \bibinfo {author} {\bibfnamefont
  {C.}~\bibnamefont {{Steinhardt}}}, \bibinfo {author} {\bibfnamefont
  {K.}~\bibnamefont {{Yabe}}},\ and\ \bibinfo {author} {\bibfnamefont
  {J.}~\bibnamefont {{Zabl}}},\ }\href
  {https://doi.org/10.3847/1538-4357/ab64dc} {\bibfield  {journal} {\bibinfo
  {journal} {\apj}\ }\textbf {\bibinfo {volume} {889}},\ \bibinfo {eid} {93}
  (\bibinfo {year} {2020})},\ \Eprint {https://arxiv.org/abs/1909.10540}
  {arXiv:1909.10540 [astro-ph.GA]} \BibitemShut {NoStop}%
\bibitem [{\citenamefont {{Girelli}}\ \emph {et~al.}(2019)\citenamefont
  {{Girelli}}, \citenamefont {{Bolzonella}},\ and\ \citenamefont
  {{Cimatti}}}]{GirelliSAM19}%
  \BibitemOpen
  \bibfield  {author} {\bibinfo {author} {\bibfnamefont {G.}~\bibnamefont
  {{Girelli}}}, \bibinfo {author} {\bibfnamefont {M.}~\bibnamefont
  {{Bolzonella}}},\ and\ \bibinfo {author} {\bibfnamefont {A.}~\bibnamefont
  {{Cimatti}}},\ }\href@noop {} {\bibfield  {journal} {\bibinfo  {journal}
  {arXiv e-prints}\ ,\ \bibinfo {eid} {arXiv:1910.07544}} (\bibinfo {year}
  {2019})},\ \Eprint {https://arxiv.org/abs/1910.07544} {arXiv:1910.07544
  [astro-ph.GA]} \BibitemShut {NoStop}%
\bibitem [{\citenamefont {{Forrest}}\ \emph {et~al.}(2019)\citenamefont
  {{Forrest}} \emph {et~al.}}]{XMassQ19}%
  \BibitemOpen
  \bibfield  {author} {\bibinfo {author} {\bibfnamefont {B.}~\bibnamefont
  {{Forrest}}} \emph {et~al.},\ }\href@noop {} {\bibfield  {journal} {\bibinfo
  {journal} {arXiv e-prints}\ ,\ \bibinfo {eid} {arXiv:1910.10158}} (\bibinfo
  {year} {2019})},\ \Eprint {https://arxiv.org/abs/1910.10158}
  {arXiv:1910.10158 [astro-ph.GA]} \BibitemShut {NoStop}%
\bibitem [{\citenamefont {{Song}}\ \emph {et~al.}(2016)\citenamefont {{Song}}
  \emph {et~al.}}]{Song16}%
  \BibitemOpen
  \bibfield  {author} {\bibinfo {author} {\bibfnamefont {M.}~\bibnamefont
  {{Song}}} \emph {et~al.},\ }\href {https://doi.org/10.3847/0004-637X/825/1/5}
  {\bibfield  {journal} {\bibinfo  {journal} {Astrophys. Jour.}\ }\textbf
  {\bibinfo {volume} {825}},\ \bibinfo {eid} {5} (\bibinfo {year} {2016})},\
  \Eprint {https://arxiv.org/abs/1507.05636} {arXiv:1507.05636 [astro-ph.GA]}
  \BibitemShut {NoStop}%
\bibitem [{\citenamefont {{Peacock}}(1999)}]{Peacock_B}%
  \BibitemOpen
  \bibfield  {author} {\bibinfo {author} {\bibfnamefont {J.~A.}\ \bibnamefont
  {{Peacock}}},\ }\href@noop {} {\emph {\bibinfo {title} {Cosmological Physics,
  by John A.~Peacock, pp.~704.~ISBN 052141072X.~Cambridge, UK: Cambridge
  University Press, January 1999.}}}\ (\bibinfo {year} {1999})\ p.\ \bibinfo
  {pages} {704}\BibitemShut {NoStop}%
\bibitem [{\citenamefont {Harikane}\ \emph {et~al.}(2016)\citenamefont
  {Harikane} \emph {et~al.}}]{Harikane16}%
  \BibitemOpen
  \bibfield  {author} {\bibinfo {author} {\bibfnamefont {Y.}~\bibnamefont
  {Harikane}} \emph {et~al.},\ }\href
  {https://doi.org/10.3847/0004-637X/821/2/123} {\bibfield  {journal} {\bibinfo
   {journal} {Astrophys. J.}\ }\textbf {\bibinfo {volume} {821}},\ \bibinfo
  {pages} {123} (\bibinfo {year} {2016})},\ \Eprint
  {https://arxiv.org/abs/1511.07873} {arXiv:1511.07873 [astro-ph.GA]}
  \BibitemShut {NoStop}%
\bibitem [{\citenamefont {{Leauthaud}}\ \emph {et~al.}(2017)\citenamefont
  {{Leauthaud}}, \citenamefont {{Saito}}, \citenamefont {{Hilbert}},
  \citenamefont {{Barreira}}, \citenamefont {{More}}, \citenamefont {{White}},
  \citenamefont {{Alam}}, \citenamefont {{Behroozi}}, \citenamefont {{Bundy}},
  \citenamefont {{Coupon}}, \citenamefont {{Erben}}, \citenamefont {{Heymans}},
  \citenamefont {{Hildebrandt}}, \citenamefont {{Mandelbaum}}, \citenamefont
  {{Miller}}, \citenamefont {{Moraes}}, \citenamefont {{Pereira}},
  \citenamefont {{Rodr{\'\i}guez-Torres}}, \citenamefont {{Schmidt}},
  \citenamefont {{Shan}}, \citenamefont {{Viel}},\ and\ \citenamefont
  {{Villaescusa-Navarro}}}]{Leauthaud_Lens17}%
  \BibitemOpen
  \bibfield  {author} {\bibinfo {author} {\bibfnamefont {A.}~\bibnamefont
  {{Leauthaud}}}, \bibinfo {author} {\bibfnamefont {S.}~\bibnamefont
  {{Saito}}}, \bibinfo {author} {\bibfnamefont {S.}~\bibnamefont {{Hilbert}}},
  \bibinfo {author} {\bibfnamefont {A.}~\bibnamefont {{Barreira}}}, \bibinfo
  {author} {\bibfnamefont {S.}~\bibnamefont {{More}}}, \bibinfo {author}
  {\bibfnamefont {M.}~\bibnamefont {{White}}}, \bibinfo {author} {\bibfnamefont
  {S.}~\bibnamefont {{Alam}}}, \bibinfo {author} {\bibfnamefont
  {P.}~\bibnamefont {{Behroozi}}}, \bibinfo {author} {\bibfnamefont
  {K.}~\bibnamefont {{Bundy}}}, \bibinfo {author} {\bibfnamefont
  {J.}~\bibnamefont {{Coupon}}}, \bibinfo {author} {\bibfnamefont
  {T.}~\bibnamefont {{Erben}}}, \bibinfo {author} {\bibfnamefont
  {C.}~\bibnamefont {{Heymans}}}, \bibinfo {author} {\bibfnamefont
  {H.}~\bibnamefont {{Hildebrandt}}}, \bibinfo {author} {\bibfnamefont
  {R.}~\bibnamefont {{Mandelbaum}}}, \bibinfo {author} {\bibfnamefont
  {L.}~\bibnamefont {{Miller}}}, \bibinfo {author} {\bibfnamefont
  {B.}~\bibnamefont {{Moraes}}}, \bibinfo {author} {\bibfnamefont {M.~E.~S.}\
  \bibnamefont {{Pereira}}}, \bibinfo {author} {\bibfnamefont {S.~A.}\
  \bibnamefont {{Rodr{\'\i}guez-Torres}}}, \bibinfo {author} {\bibfnamefont
  {F.}~\bibnamefont {{Schmidt}}}, \bibinfo {author} {\bibfnamefont {H.-Y.}\
  \bibnamefont {{Shan}}}, \bibinfo {author} {\bibfnamefont {M.}~\bibnamefont
  {{Viel}}},\ and\ \bibinfo {author} {\bibfnamefont {F.}~\bibnamefont
  {{Villaescusa-Navarro}}},\ }\href {https://doi.org/10.1093/mnras/stx258}
  {\bibfield  {journal} {\bibinfo  {journal} {MNRAS}\ }\textbf {\bibinfo
  {volume} {467}},\ \bibinfo {pages} {3024} (\bibinfo {year} {2017})},\ \Eprint
  {https://arxiv.org/abs/1611.08606} {arXiv:1611.08606 [astro-ph.CO]}
  \BibitemShut {NoStop}%
\bibitem [{\citenamefont {Arinyo-i Prats}\ \emph {et~al.}(2015)\citenamefont
  {Arinyo-i Prats}, \citenamefont {Miralda-Escudé}, \citenamefont {Viel},\
  and\ \citenamefont {Cen}}]{NonlinLyman}%
  \BibitemOpen
  \bibfield  {author} {\bibinfo {author} {\bibfnamefont {A.}~\bibnamefont
  {Arinyo-i Prats}}, \bibinfo {author} {\bibfnamefont {J.}~\bibnamefont
  {Miralda-Escudé}}, \bibinfo {author} {\bibfnamefont {M.}~\bibnamefont
  {Viel}},\ and\ \bibinfo {author} {\bibfnamefont {R.}~\bibnamefont {Cen}},\
  }\href {https://doi.org/10.1088/1475-7516/2015/12/017} {\bibfield  {journal}
  {\bibinfo  {journal} {J. Cosmol. Astropart. Phys.}\ }\textbf {\bibinfo
  {volume} {1512}}\bibfield  {number} {\bibinfo  {number} { (12)},\ \bibinfo
  {pages} {017}},\ }\Eprint {https://arxiv.org/abs/1506.04519}
  {arXiv:1506.04519 [astro-ph.CO]} \BibitemShut {NoStop}%
\bibitem [{\citenamefont {{Binney}}\ and\ \citenamefont
  {{Tremaine}}(2008)}]{BT2008}%
  \BibitemOpen
  \bibfield  {author} {\bibinfo {author} {\bibfnamefont {J.}~\bibnamefont
  {{Binney}}}\ and\ \bibinfo {author} {\bibfnamefont {S.}~\bibnamefont
  {{Tremaine}}},\ }\href@noop {} {\emph {\bibinfo {title} {Galactic Dynamics:
  Second Edition, by James Binney and Scott Tremaine.~ISBN 978-0-691-13026-2
  (HB).~Published by Princeton University Press, Princeton, NJ USA, 2008.}}}\
  (\bibinfo  {publisher} {Princeton University Press},\ \bibinfo {year}
  {2008})\BibitemShut {NoStop}%
\bibitem [{Note4()}]{Note4}%
  \BibitemOpen
  \bibinfo {note} {Note also, in the latter case of $c_s =0.01$, the small bump
  on the right, visible in the low $\gamma $ spectra beyond the cutoff scale.
  These are due to non-adibaticity at the crossing of the high energy
  transition at $\eta _c$. They are much less prominent for smaller $k_c/H \ge
  10^{4}$ values, used in the rest of this work. They nevertheless represent
  another potentially interesting effect associated with non-adiabatic
  evolution}\BibitemShut {NoStop}%
\bibitem [{\citenamefont {{Nakama}}\ \emph {et~al.}(2017)\citenamefont
  {{Nakama}}, \citenamefont {{Chluba}},\ and\ \citenamefont
  {{Kamionkowski}}}]{JensKam17}%
  \BibitemOpen
  \bibfield  {author} {\bibinfo {author} {\bibfnamefont {T.}~\bibnamefont
  {{Nakama}}}, \bibinfo {author} {\bibfnamefont {J.}~\bibnamefont {{Chluba}}},\
  and\ \bibinfo {author} {\bibfnamefont {M.}~\bibnamefont {{Kamionkowski}}},\
  }\href {https://doi.org/10.1103/PhysRevD.95.121302} {\bibfield  {journal}
  {\bibinfo  {journal} {\prd}\ }\textbf {\bibinfo {volume} {95}},\ \bibinfo
  {eid} {121302} (\bibinfo {year} {2017})},\ \Eprint
  {https://arxiv.org/abs/1703.10559} {arXiv:1703.10559 [astro-ph.CO]}
  \BibitemShut {NoStop}%
\bibitem [{\citenamefont {Davidzon}\ \emph {et~al.}(2017)\citenamefont
  {Davidzon} \emph {et~al.}}]{COSMOS15}%
  \BibitemOpen
  \bibfield  {author} {\bibinfo {author} {\bibfnamefont {I.}~\bibnamefont
  {Davidzon}} \emph {et~al.},\ }\href
  {https://doi.org/10.1051/0004-6361/201730419} {\bibfield  {journal} {\bibinfo
   {journal} {Astron. Astrophys.}\ }\textbf {\bibinfo {volume} {605}},\
  \bibinfo {pages} {A70} (\bibinfo {year} {2017})},\ \Eprint
  {https://arxiv.org/abs/1701.02734} {arXiv:1701.02734 [astro-ph.GA]}
  \BibitemShut {NoStop}%
\bibitem [{\citenamefont {{Behroozi}}\ \emph {et~al.}(2013)\citenamefont
  {{Behroozi}}, \citenamefont {{Wechsler}},\ and\ \citenamefont
  {{Conroy}}}]{IncFBehrooz13}%
  \BibitemOpen
  \bibfield  {author} {\bibinfo {author} {\bibfnamefont {P.~S.}\ \bibnamefont
  {{Behroozi}}}, \bibinfo {author} {\bibfnamefont {R.~H.}\ \bibnamefont
  {{Wechsler}}},\ and\ \bibinfo {author} {\bibfnamefont {C.}~\bibnamefont
  {{Conroy}}},\ }\href {https://doi.org/10.1088/0004-637X/770/1/57} {\bibfield
  {journal} {\bibinfo  {journal} {Astrophys. Jour.}\ }\textbf {\bibinfo
  {volume} {770}},\ \bibinfo {eid} {57} (\bibinfo {year} {2013})},\ \Eprint
  {https://arxiv.org/abs/1207.6105} {arXiv:1207.6105 [astro-ph.CO]}
  \BibitemShut {NoStop}%
\bibitem [{\citenamefont {Behroozi}\ \emph {et~al.}(2019)\citenamefont
  {Behroozi}, \citenamefont {Wechsler}, \citenamefont {Hearin},\ and\
  \citenamefont {Conroy}}]{UMach}%
  \BibitemOpen
  \bibfield  {author} {\bibinfo {author} {\bibfnamefont {P.}~\bibnamefont
  {Behroozi}}, \bibinfo {author} {\bibfnamefont {R.~H.}\ \bibnamefont
  {Wechsler}}, \bibinfo {author} {\bibfnamefont {A.~P.}\ \bibnamefont
  {Hearin}},\ and\ \bibinfo {author} {\bibfnamefont {C.}~\bibnamefont
  {Conroy}},\ }\href {https://doi.org/10.1093/mnras/stz1182} {\bibfield
  {journal} {\bibinfo  {journal} {Mon. Not. Roy. Astron. Soc.}\ }\textbf
  {\bibinfo {volume} {488}},\ \bibinfo {pages} {3143} (\bibinfo {year}
  {2019})},\ \Eprint {https://arxiv.org/abs/1806.07893} {arXiv:1806.07893}
  \BibitemShut {NoStop}%
\end{thebibliography}%

\end{document}